\documentclass[a4paper,fleqn,usenatbib]{mnras}

\usepackage{newtxtext,newtxmath}
\usepackage[T1]{fontenc}
\usepackage{ae,aecompl}
\usepackage{bm}
\usepackage{xcolor}
\usepackage{scalerel}
\usepackage{tikz}
\usetikzlibrary{svg.path}

\definecolor{orcidlogocol}{HTML}{A6CE39}
\tikzset{
  orcidlogo/.pic={
    \fill[orcidlogocol] svg{M256,128c0,70.7-57.3,128-128,128C57.3,256,0,198.7,0,128C0,57.3,57.3,0,128,0C198.7,0,256,57.3,256,128z};
    \fill[white] svg{M86.3,186.2H70.9V79.1h15.4v48.4V186.2z}
                 svg{M108.9,79.1h41.6c39.6,0,57,28.3,57,53.6c0,27.5-21.5,53.6-56.8,53.6h-41.8V79.1z M124.3,172.4h24.5c34.9,0,42.9-26.5,42.9-39.7c0-21.5-13.7-39.7-43.7-39.7h-23.7V172.4z}
                 svg{M88.7,56.8c0,5.5-4.5,10.1-10.1,10.1c-5.6,0-10.1-4.6-10.1-10.1c0-5.6,4.5-10.1,10.1-10.1C84.2,46.7,88.7,51.3,88.7,56.8z};
  }
}

\newcommand\orcidicon[1]{\href{https://orcid.org/#1}{\mbox{\scalerel*{
\begin{tikzpicture}[yscale=-1,transform shape]
\pic{orcidlogo};
\end{tikzpicture}
}{|}}}}

\usepackage{float}
\usepackage[export]{adjustbox}
\usepackage{graphicx}	% Including figure files

\usepackage{amsmath}	% Advanced maths commands
\usepackage{amssymb}	% Extra maths symbols

\usepackage{multicol}
\usepackage{verbatim}   % For commenting out multiple lines. --JR
\usepackage{threeparttable}
\usepackage{upgreek}
\usepackage{rotating}

\usepackage{subfig}
\usepackage{comment}
\usepackage{longtable} 
\usepackage{caption}
\usepackage{threeparttable}
\newcommand{\lcdm}{$\Lambda$CDM}

\newcommand{\om}{\Omega_{m0}}
\newcommand{\ol}{\Omega_{\Lambda}}
\newcommand{\ok}{\Omega_{k0}}
\newcommand{\FT}[1]{}
\newcommand{\rfe}{${\cal R}_{\rm{Fe\textsc{ii}}}$}
\newcommand{\Feii}{Fe\,\textsc{ii}}
\newcommand{\Mgii}{Mg\,\textsc{ii}}
\newcommand{\hb}{{\sc{H}}$\beta$\/}

\definecolor{MZ}{RGB}{255,128,0}

\title[H$\beta$ QSO study]{Do reverberation-measured H$\beta$ quasars provide a useful test of cosmology?}

\author[]{
Narayan Khadka$^{\orcidicon{0000-0001-5512-2716}1}$\thanks{E-mail: nkhadka@phys.ksu.edu},
Mary Loli Mart\'inez-Aldama$^{\orcidicon{0000-0002-7843-7689}{2,3}}$\thanks{E-mail: mmary@cft.edu.pl},
Michal Zaja\v{c}ek$^{\orcidicon{0000-0001-6450-1187}4}$\thanks{E-mail: zajacek@mail.muni.cz},
\newauthor \hspace{0.1mm}
Bo\.{z}ena Czerny$^{\orcidicon{0000-0001-5848-4333}2}$\thanks{E-mail: bcz@cft.edu.pl},
Bharat Ratra$^{\orcidicon{0000-0002-7307-0726}1}$\thanks{E-mail: ratra@phys.ksu.edu}\\
$^{1}$Department of Physics, Kansas State University, 116 Cardwell Hall, Manhattan, KS 66506, USA\\
$^{2}$Center for Theoretical Physics, Polish Academy of Sciences, Al.\ Lotnik\'{o}w 32/46, 02-668 Warsaw, Poland\\
$^{3}$Departamento de Astronomia, Universidad de Chile, Camino del Observatorio 1515, Santiago, Chile\\
$^{4}$Department of Theoretical Physics and Astrophysics, Faculty of Science, Masaryk University, Kotl\'a\v{r}sk\'a 2, 611 37 Brno, Czech Republic
}

\date{Accepted XXX. Received YYY; in original form ZZZ}

\pubyear{2021}

\begin{document}
\label{firstpage}
\pagerange{\pageref{firstpage}--\pageref{lastpage}}
\maketitle

\begin{abstract}
%%%
We use 118 H$\beta$ quasar (QSO) observations in the redshift range $0.0023 \leq z \leq 0.89$ to simultaneously constrain cosmological model parameters and QSO 2-parameter radius-luminosity ($R-L$) relation parameters in six different cosmological models. We find that the $R-L$ relation parameters for these QSOs are independent of the assumed cosmology so these QSOs seem to be standardizable through the $R-L$ relation (although there is a complication that might render this untrue). Cosmological constraints obtained using these QSOs are weak, more favor currently decelerated cosmological expansion, and typically are in $\sim 2\sigma$ tension with those obtained from a joint analysis of baryon acoustic oscillation and Hubble parameter measurements. Extending the $R-L$ relation to a 3-parameter one to try to correct for the accretion rate effect does not result in a reduction of the cosmological constraints discrepancy nor does it result in the hoped-for significant reduction of the intrinsic scatter of the $R-L$ relation.  
\end{abstract}


%%%%%%%%%%%%%%%%%%%%%%%%%%%%%%%%%%%%%%%%%%%%%%%%%%

%%%%%%%%%%%%%%%%% BODY OF PAPER %%%%%%%%%%%%%%%%%%

\section{Introduction}
\label{sec:Introduction}

The spatially-flat $\Lambda$CDM cosmological model \citep{Peebles1984}, with the dark energy assumed to be a time-independent cosmological constant $\Lambda$, accounts well for many observed properties of the Universe \citep[see e.g.][]{Farooqetal2017, Scolnicetal2018, PlanckCollaboration2020, eBOSSCollaboration2021}. There are however some discrepancies \citep[see e.g.][]{eleonora2021, PerivolaropoulosSkara2021}, and measurements do not strongly rule out a mildly spatially non-flat geometry or mildly dynamical dark energy models.

It is unclear whether current reports of discrepancies with spatially-flat $\Lambda$CDM implies new physics beyond the model, or whether they just reflect an underestimate of the measurement errors. Since the statistical errors in better-established cosmological probes are now under better control, a significant issue is whether the systematic errors have been underestimated. The best way to test this is to use alternate cosmological probes. Some work has already been done in this direction and it includes the use of HII starburst galaxy observations which extend to redshift $z \sim 2.4$ \citep{ManiaRatra2012, Chavezetal2014, GonzalezMoran2019, GonzalezMoranetal2021, Caoetal2020, Caoetal2021a, Caoetal_2021c, Johnsonetal2021, Mehrabietal2022}, quasar (QSO) angular size measurements which probe to $z \sim 2.7$ \citep{Caoetal2017, Ryanetal2019, Caoetal2020, Caoetal2021b, Zhengetal2021, Lianetal2021}, QSO X-ray and UV flux measurements which reach to $z \sim 7.5$ \citep{RisalitiLusso2015, RisalitiLusso2019, KhadkaRatra2020a, KhadkaRatra2020b, KhadkaRatra2021a, KhadkaRatra2021b, Yangetal2020, Lussoetal2020, ZhaoXia2021, Lietal2021, Lianetal2021, Rezaeietal2021, Luongoetal2021},\footnote{In the most recent \cite{Lussoetal2020} QSO flux compilation, their  assumed UV--X-ray correlation model is valid only to a significantly lower redshift, $z \sim 1.5-1.7$, meaning these QSOs can be used to derive only lower-$z$ cosmological constraints \citep{KhadkaRatra2021a, KhadkaRatra2021b}.} and gamma-ray burst (GRB) data that extend to $z \sim 8.2$ \citep{Wang_2016, Wangetal2021, Dirirsa2019, Amati2019, KhadkaRatra2020c, Khadkaetal2021a, Demianskietal_2021, Luongoetal2021, Huetal2021, OrlandoMarco2021, Caoetal2021d, Caoetal2022, CaoRatra2022}. Other interesting and potentially important quasar-based methods include the parallax distance measurements \citep{wang_parallax2020,GRAVITY2021} and super-Eddington quasars as standardizable candles \citep{wang2013,marziani2014,marziani_atom2019}.

In our previous work  \citep{khadka2021} we focused on reverberation-mapped active galactic nuclei (AGN) as alternate probes. This method was independently proposed by \citet{haas2011} and  \citet{watson2011}. Soon after, an optimized strategy for higher redshifts, based not on H$\beta$ but on the \Mgii\ line was outlined by \citet{czerny2013}.  The first cosmological constraints based on this method were obtained by \citet{Mary2019}, \citet{Michal2021}, and \citet{Czerny2021}. The method is based on the broad-line region (BLR) radius-luminosity correlation (hereafter $R-L$ relation) which allows one to convert the rest-frame time delay of the broad emission line with respect to the ionizing continuum to an absolute monochromatic luminosity of a given source. In \citet{khadka2021} we derived (weak) cosmological constraints from the currently most-complete set of 78 measurements of the \Mgii\ time delay with respect to the continuum for a sample of AGN covering the redshift range between 0.0033 and 1.89. We showed that this new probe was standardizable and so avoids the circularity problem. We used these \Mgii\ QSOs alone, and in combination with baryon acoustic oscillation (BAO) and Hubble parameter [H(z)] chronometric measurements, and tested different cosmological models. We did not detect any tension between the weak \Mgii\ cosmological constraints and the spatially-flat $\Lambda$CDM model. On the other hand, mild dark energy dynamics or a little spatial curvature could not be excluded based on the \Mgii\ quasar sample, which motivates further cosmological tests, including those based on the $R-L$ relation using other reverberation-measured AGN.

In the current paper we use a larger sample of 118 measurements of the time delay done using the broad H$\beta$ line. This sample covers a narrower redshift range, from 0.002 to 0.890, where the lower and the upper limits correspond to the detection of the H$\beta$ line (4861.35 \AA) in the optical and near-infrared bands ($\sim 4870-9190$ \AA). For these H$\beta$ QSOs, rest-frame time delays and luminosities are correlated through the power-law $R-L$ relation, $\tau\propto L^{\gamma}$ \citep{2000ApJ...533..631K,2005ApJ...629...61K,2013ApJ...767..149B}, where the mean BLR radius is given by $R=c\tau$ with $c$ being the speed of light. We can use this correlation to standardize these QSOs or at least we can study the $R-L$ relation to see if it can be used to standardize H$\beta$ QSOs. Initially, the $R-L$ relation exhibited a small scatter of $\sim 0.1-0.2$ around a slope of $\gamma\sim 0.5$ \citep{2013ApJ...767..149B} when the QSO sample consisted mostly of lower-accreting sources that exhibit a larger variability. When additional sources were included, sources with a larger Eddington ratio that are less variable, the scatter of the $R-L$ relation increased considerably \citep{du2015, du2016,2017ApJ...851...21G,2018ApJ...856....6D}, raising the question of whether the canonical 2-parameter $R-L$ relation with a power-law slope around $0.5$ is valid for all sources. Analyses of the H$\beta$ sample revealed that the scatter is largely driven by the accretion rate and/or the UV/optical spectral energy distribution shape \citep{du2015, du2016,2018ApJ...856....6D, Mary2019,2020ApJ...903..112D,2020ApJ...899...73F}. Hence extended $R-L$ relations were investigated with added independent observables associated with the accretion rate, such as the iron line relative strength or the fractional variability, to correct for the accretion-rate effect \citep{duwang_2019,Mary2020}.

Here we analyze these 118 \hb\ sources data using the 2-parameter $R-L$ relation in six different cosmological models. These models include both spatially-flat and non-flat geometry as well as both time-independent and dynamical dark energy densities. We fit cosmological model parameters and $R-L$ relation parameters simultaneously in a given model so our results are free from the circularity problem if the $R-L$ relation is independent of the cosmological model used in the analysis (as is the case). Current H$\beta$ QSO data are able to provide only weak cosmological constraints and these constraints are in $\sim 2\sigma$ tension with constraints obtained using other better-established cosmological probes. We also used \Feii\ measurements for these 118 sources in an extended 3-parameter $R-L$ relation hoping to find less discrepant \hb\ cosmological constraints as well as a significantly smaller value for the $R-L$ relation intrinsic dispersion, but our results show that inclusion of a third parameter in an extended $R-L$ relation does not accomplish these aims. While in the case of the full 118 \hb\ QSO sample, the 3-parameter $R-L$ relation is very strongly favored over the 2-parameter one, when we divide the sample into equal high and low \rfe\ subsets,\footnote{Here \rfe\ is the flux ratio parameter of optical \Feii\ to \hb.} these high and low \rfe\ data subsets do not provide significant evidence for or against the 3-parameter $R-L$ relation relative to the 2-parameter one and so it appears that it is currently premature to draw a strong conclusion about observational support for the 3-parameter $R-L$ relation.

The paper is structured as follows. In Sec.~2 we summarize the cosmological models we use. In Sec.~3 we describe the H$\beta$ quasar data we use to constrain cosmological model and $R-L$ relation parameters. In Sec.~4 we describe the data analysis methods we use. In Sec.~5 we present our results. In Sec.~6 we suggest possible explanations for the \hb\ cosmological constraints discrepancy, contrast our \hb\ results to the previous \Mgii\ ones, and discuss some future possibilities. We conclude in Sec.~7.

\section{Models}
\label{sec:models}

In this paper, we use six different cosmological models to predict model-dependent rest-frame time-delays of H$\beta$ reverberation-mapped quasars at known redshifts. Three of these cosmological models assume flat spatial hypersurfaces while the other three allow for non-zero spatial curvature.\footnote{Discussions of observational constraints on spatial curvature can be traced back through \citet{Chenetal2016}, \citet{Ranaetal2017}, \citet{Oobaetal2018a, Oobaetal2018b}, \citet{Yuetal2018}, \citet{ParkRatra2019a, ParkRatra2019b}, \citet{Wei2018}, \citet{DESCollaboration2019}, \citet{Lietal2020}, \citet{Handley2019}, \citet{EfstathiouGratton2020}, \citet{DiValentinoetal2021a}, \citet{VelasquezToribioFabris2020}, \citet{Vagnozzietal2020, Vagnozzietal2021}, \citet{KiDSCollaboration2021}, \citet{ArjonaNesseris2021}, and \citet{Dhawanetal2021}.} These time-delay predictions are computed from the expansion rate of the universe, $H(z)$, which depends on the values of the cosmological parameters of the model under study. By comparing the predicted time-delays with the observed time-delays, we can measure the cosmological parameters.

In the $\Lambda$CDM model the Hubble parameter is
\begin{equation}
\label{eq:friedLCDM}
    H(z) = H_0\sqrt{\Omega_{m0}(1+z)^3 + \Omega_{k0}(1+z)^2 + \Omega_{\Lambda}},
\end{equation}
where $H_0$ is the Hubble constant, $\Omega_{m0}$, $\Omega_{k0}$, and $\Omega_{\Lambda}$ are related through the equation $\Omega_{m0}$ + $\Omega_{k0}$ + $\Omega_{\Lambda}$ = 1, and are the present values of the non-relativistic matter density parameter, the spatial curvature energy density parameter, and the cosmological constant energy density parameter, respectively. In the spatially non-flat $\Lambda$CDM model, the conventional choice of free parameters is $\Omega_{m0}$, $\Omega_{k0}$, and $H_0$ while in the spatially-flat $\Lambda$CDM model we use the same set of free parameters but now with $\Omega_{k0} = 0$.\footnote{For the BAO + $H(z)$ data analyses, in all six cosmological models, we describe $\Omega_{m0}$ in terms of the present values of the CDM and baryonic matter (physical) energy density parameters, and instead of $\Omega_{m0}$ we use $\Omega_c h^2$ and $\Omega_b h^2$ as free parameters. Here $h$ is the Hubble constant in units of 100 km s$^{-1}$ Mpc$^{-1}$ and $\Omega_{m0} = \Omega_c + \Omega_b$.}

In the XCDM dynamical dark energy parametrization the Hubble parameter is
\begin{equation}
\label{eq:XCDM}
    H(z) = H_0\sqrt{\Omega_{m0}(1+z)^3 + \Omega_{k0}(1+z)^2 + \Omega_{X0}(1+z)^{3(1+\omega_X)}},
\end{equation}
where $\Omega_{m0}$, $\Omega_{k0}$, and $\Omega_{X0}$ are related through the equation $\Omega_{m0}$ + $\Omega_{k0}$ + $\Omega_{X0}$ = 1, and $\Omega_{X0}$ is the present value of the $X$-fluid dark energy density parameter. $\omega_X$ is the equation of state parameter of the $X$-fluid (the ratio of the pressure to the energy density). In the spatially non-flat XCDM parametrization, the conventional choice of free parameters is $\Omega_{m0}$, $\Omega_{k0}$, $\omega_X$, and $H_0$ while in the spatially-flat XCDM parametrization we use the same set of free parameters but now with $\Omega_{k0} = 0$. In the XCDM parameterization when $\omega_X = -1$ the $\Lambda$CDM model is recovered.

In the $\phi$CDM model the scalar field $\phi$ is the dynamical dark energy \citep{PeeblesRatra1988, RatraPeebles1988, Pavlovetal2013}.\footnote{Discussions of observational constraints on the $\phi$CDM model can be traced back through \cite{chen_etal_2017}, \citet{Zhaietal2017}, \citet{Oobaetal2018c, Oobaetal2019}, \citet{ParkRatra2018, ParkRatra2019c, ParkRatra2020}, \citet{Sangwanetal2018}, \citet{SolaPercaulaetal2019}, \citet{Singhetal2019}, \citet{UrenaLopezRoy2020}, \citet{SinhaBanerjee2021}, \citet{Xuetal2021}, and \citet{deCruzetal2021}.} The scalar field potential energy density which determines $\Omega_{\phi}(z, \alpha)$, the scalar field dynamical dark energy density parameter, is assumed to be an inverse power law of $\phi$,
\begin{equation}
\label{eq:phiCDMV}
    V(\phi) = \frac{1}{2}\kappa m_{p}^2 \phi^{-\alpha},
\end{equation}
where $m_{p}$ is the Planck mass, $\alpha$ is a positive parameter,  and $\kappa$ is a constant whose value is determined using the shooting method to guarantee that the current energy budget equation $\Omega_{m0} + \Omega_{k0} + \Omega_{\phi}(z = 0, \alpha) = 1$ is satisfied.

With this potential energy density, the dynamics of a spatially homogeneous scalar field and cosmological scale factor $a$ is governed by the scalar field equation of motion and the Friedmann equation
\begin{align}
\label{eq:field}
   & \ddot{\phi}  + 3\frac{\dot{a}}{a}\dot\phi - \frac{1}{2}\alpha \kappa m_{p}^2 \phi^{-\alpha - 1} = 0, \\
\label{eq:friedpCDM}
   & \left(\frac{\dot{a}}{a}\right)^2 = \frac{8 \pi}{3 m_{p}^2}\left(\rho_m + \rho_{\phi}\right) - \frac{k}{a^2}.
\end{align}
Here an overdot denotes a derivative with respect to time, $k$ is positive, zero, and negative for closed, flat, and open spatial hypersurfaces (corresponding to $\Omega_{k0} < 0, =0, {\rm and} >0$), $\rho_m$ is the non-relativistic matter energy density, and the scalar field energy density
\begin{equation}
    \rho_{\phi} = \frac{m^2_p}{32\pi}\left[\dot{\phi}^2 + \kappa m^2_p \phi^{-\alpha}\right].
\end{equation}
The numerical solution of the coupled differential equations (\ref{eq:field}) and (\ref{eq:friedpCDM}) is used to compute $\rho_{\phi}$ and then $\Omega_{\phi}(z, \alpha)$ is determined from 
\begin{equation}
    \Omega_{\phi}(z, \alpha) = \frac{8\pi \rho_{\phi}}{3 m^2_p H^2_0}.
\end{equation}

The Hubble parameter in the $\phi$CDM model is
\begin{equation}
    H(z) = H_0\sqrt{\Omega_{m0}(1+z)^3 + \Omega_{k0}(1+z)^2 + \Omega_{\phi}\left(z, \alpha\right)},
\end{equation}
where $\Omega_{m0}$, $\Omega_{k0}$, and $\Omega_{\phi}(0, \alpha)$ are related through the equation $\Omega_{m0}$ + $\Omega_{k0}$ + $\Omega_{\phi}(0, \alpha)$ = 1. In the spatially non-flat $\phi$CDM model, the conventional choice of free parameters is $\Omega_{m0}$, $\Omega_{k0}$, $\alpha$, and $H_0$ while in the spatially-flat $\phi$CDM model we use the same set of free parameters but now with $\Omega_{k0} = 0$. When $\alpha = 0$ the $\phi$CDM model becomes the $\Lambda$CDM model.

\section{Data}
\label{sec:data}

\begin{figure}
 \includegraphics[width=0.8\linewidth]{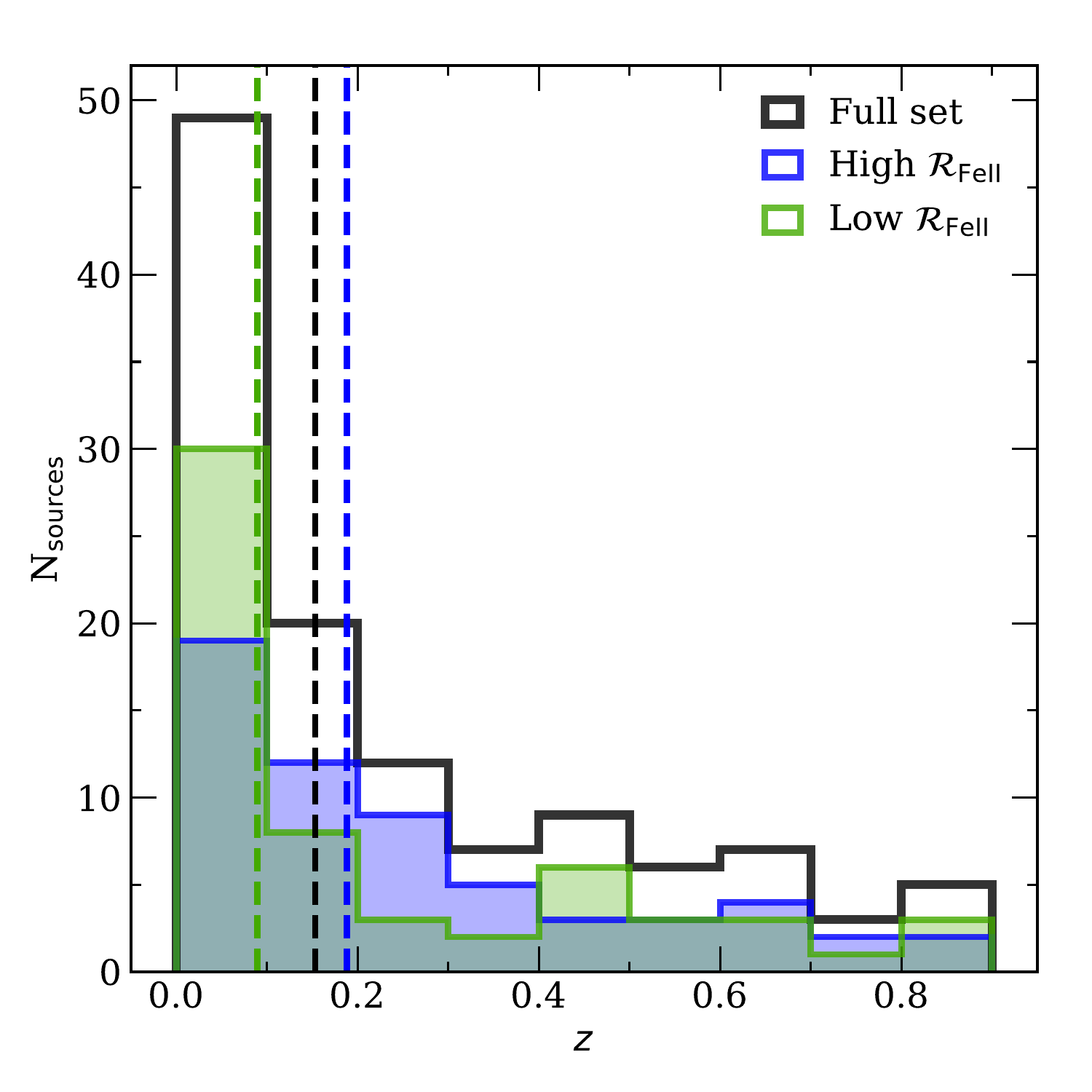}
\caption{Redshift distribution for the full (black), high-\rfe\ (blue), and low-\rfe\ (green) data sets, see Sec.~\ref{sec:data}. Vertical lines correspond to the median of each distribution.}
\label{fig:hist_z}
\end{figure}

In our analyses here we use 118 sources with well-established reverberation-measured time-delays of the H$\beta$ line with respect to the continuum and with measurements of the intensity of optical \Feii, expressed as the flux ratio parameter \rfe=F(Fe\textsc{ii}$_{4434-4684\AA}$)/F(\hb). This sample spans redshift and luminosity ranges of $0.002<z<0.89$ and $41.5< \log(L_{5100}\,[{\rm erg\,s^{-1}}])<45.9$, respectively. The redshift distribution is shown in Fig.\ \ref{fig:hist_z}. The \hb\ time delay measurements are taken from the compilation of \citet{Mary2019}, supplemented with measurements from \citet{zhang_2018} (3C 273), \citet{huang2019} (I Zw 1), \citet{rakshit2020} (PKS 1510-089), and \citet{li_2021} (PG 0923+201 and PG 1001+291). The flux at 5100\,\AA\ in the observed-frame (log~$F_{5100}$) was estimated from the luminosities collected by \citet{Mary2019}, assuming $H_0 = 70$ km s$^{-1}$ Mpc$^{-1}$, $\Omega_{m0} = 0.3$, and $\Omega_\Lambda = 0.7$. In some cases the cosmological parameter values differ from those used here, however the flux estimations are in agreement within the uncertainties and our results are not influenced by the small differences (see footnote 12 below). We have indicated in Table~\ref{tab:hbQSOdata} the  original cosmological parameters. References for each of the original measurements are given in Table~\ref{tab:hbQSOdata}.

Several reverberation studies of broad emission lines indicate that UV and optical time-delays show a systematic offset with the Eddington ratio\footnote{$\eta=L_{\rm bol}/ L_{\rm Edd}$, where $L_{\rm bol}=40(L_{5100}/1\times10^{42} {\rm erg~ s}^{-1})^{-0.2}$  and $L_{\rm Edd}=1.5\times10^{38}M_{\rm BH}/M_{\odot}$ \citep{netzer2019}.} $\eta$ \citep{du2015, duwang_2019,Mary2020}. The  first attempt to correct for this effect, and to try to reduce the dispersion, was proposed by \citet{du2016}. Later, \citet{Mary2019} proposed a correction based on the accretion rate, however, their correction introduces a correlation between the accretion rate and the time-delay that can bias the results. Here we consider the approach of \citet{duwang_2019} and \citet{Yu2020} who proposed using the observationally inferred \rfe\ measurements as a proxy for the Eddington ratio, replacing the usual 2-parameter $R-L$ relation with a 3-parameter one. In contrast to the Eddington ratio, \rfe\ is independent of the time delay. For this purpose we use the \rfe\ estimations of \citet{duwang_2019} and \citet{shen_2019}.\footnote{The equivalent widths of \hb\ and \Feii\ are taken from \citet{shen_2019} to estimate the \rfe\ parameter in the SDSS-RM sources, see Table~\ref{tab:hbQSOdata}.} Table~\ref{tab:hbQSOdata} describes the sample we use here, listing the source name, RA, DEC, redshift, flux at 5100\AA\ ($F_{5100}$), \hb\ rest-frame time-delay ($\tau$)\footnote{The \hb\ time-delay measurements have asymmetric error bars and in our analyses here we use the corresponding symmetrized error bar $\sigma = 0.5(2\sigma_1 \sigma_2/(\sigma_1 + \sigma_2) + \sqrt{\sigma_1\sigma_2})$ \citep{Barlow2004} where $\sigma_1$ and $\sigma_2$ are the asymmetric upper and lower error bars respectively. $\sigma_1$, $\sigma_2$, and $\sigma$ for all sources are listed in Table~\ref{tab:hbQSOdata}. We used the same symmetrization technique in \cite{khadka2021}.},  \rfe\ value, and literature reference. 

\begin{figure*}
 \includegraphics[width=0.8\linewidth]{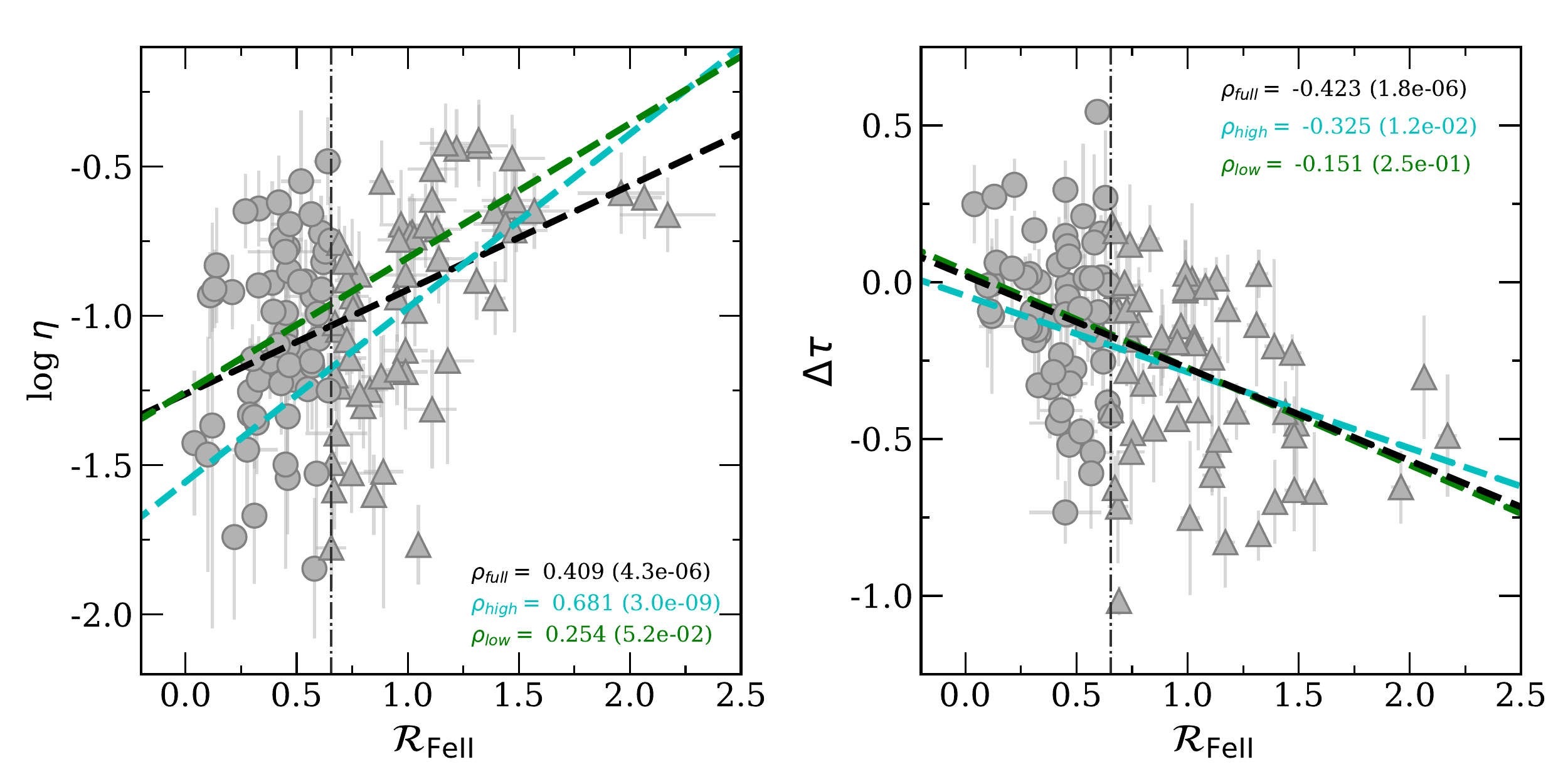}
\caption{Left panel: Correlation between the \rfe\ parameter and the Eddington ratio ($\eta$).  Right panel: offset of the observed time-delay with respect to $\tau$ from the usual 2-parameter $R-L$ relation ($\Delta \tau$ defined as $\Delta \tau\equiv \log(\tau/\tau_{\rm R-L})$, where $\tau$ is the observed rest-frame time delay and $\tau_{\rm R-L}$ is the one predicted from the 2-parameter $R-L$ relation). In both panels, black, green, and cyan dashed lines indicate the best-fit for the full, low-\rfe\ (circles symbols), and high-\rfe\ (triangle symbols) set, respectively. Vertical black dot-dashed lines indicate the median \rfe\ value and circles (triangles) indicate low-\rfe\ (high-\rfe) data points. Spearman's rank correlation coefficient ($\rho$) and the $p$-value for each population are also shown.}
\label{fig:delta_rfe}
\end{figure*}

The strong correlation between the Eddington ratio and \rfe\ \citep{borosongreen1992, marziani2003} motivates inclusion of \rfe\ in a 3-parameter $R-L$ relation. The left panel of Fig.~\ref{fig:delta_rfe} shows the moderate correlation between these parameters in our sample (with Spearman's rank correlation coefficient $\rho=0.409$ and $p$-value $=4.25\times10^{-6}$).\footnote{A value of $\rho > 0.5 (< - 0.5)$ would imply strong correlation (anticorrelation), while the probability $p < 5\times 10^{-2}$ implies that the correlation or anticorrelation, despite being not too strong, is nevertheless highly significant.} Another argument which supports the inclusion of \rfe\ in an extended 3-parameter $R-L$ relation is the moderate correlation ($\rho=-0.423$, $p$-value $=1.78\times10^{-6}$) between \rfe\ and the offset ($\Delta \tau$) of the observed time-delay with respect to $\tau$ predicted by the usual 2-parameter $R-L$ relation\footnote{Defined as $\Delta \tau\equiv \log(\tau/\tau_{\rm R-L})$, where $\tau$ is the observed rest-frame time delay and $\tau_{\rm R-L}$ is the time-delay predicted from the 2-parameter $R-L$ relation.} \citep{2013ApJ...767..149B} shown in the right panel of Fig.~\ref{fig:delta_rfe}. 

\citet{Mary2020} and \citet{Michal2021} divided the reverberation-measured \Mgii\ objects into two groups, with low and with high Eddington ratios, and obtained a lower scatter and a better fit for the 2-parameter $R-L$ relation in each subpopulation. Following these results, we have divided our \hb\ sample here into two equal subgroups at the median value of \rfe\ (\rfe = 0.655), where sources with higher \rfe\ values correspond to the high accretors and vice-versa; in what follows we call these subsamples the high-\rfe\ and the low-\rfe\ subgroups. Fig.~\ref{fig:hist_z} shows the redshift distribution of each subset. Table~\ref{tab:hbQSOdata} indicates the sources of each subgroup. The best-fit line and the correlation coefficient for each subgroup are shown in Fig.~\ref{fig:delta_rfe}. In the case of the high accretors there is a significant correlation between \rfe\ and $\eta$, reflecting the small influence of the orientation in high-accreting sources which  mostly show a face-on orientation \citep{panda2019,marziani2021}. In the subset of low accretors the correlation is weak, which most likely reflects the difficulties of measuring weak \Feii\ contribution to the spectrum and the effect caused by the large viewing angles ($\theta \sim 5-50^{\circ}$) found in the AGN population \citep{marziani2021}. On the other hand, the correlation between $\Delta \tau$ and \rfe\ is not significant in the low- and high-\rfe\ subsets according to Spearman's rank coefficient and the $p-$values (right panel of Fig.~\ref{fig:delta_rfe}).\footnote{Although there is a moderate correlation in the full set, the intercept of the best-fit line has a large uncertainty [$\Delta \tau=(0.02\pm0.05)+(-0.29\pm0.06){\cal R}_{\rm{Fe\textsc{ii}}}$], which is also observed in the best-fit lines of the high [$\Delta \tau=(-0.04\pm0.11)+(-0.24\pm0.10){\cal R}_{\rm{Fe\textsc{ii}}}$] and low [$\Delta \tau=(0.04\pm0.09)+(-0.31\pm0.20){\cal R}_{\rm{Fe\textsc{ii}}}$] \rfe\ data subsets, with fewer sources and so possibly weaker correlations.} The slope of the best-fit line in each subset is very similar to the one of the full sample ($a_{\rm full}=-0.295\pm0.057$, $a_{\rm high}=-0.243\pm0.101$, $a_{\rm low}=-0.309\pm0.199$). Taking as a reference the full sample ($\rho_{\rm full}=0.423$, $p-$value=$1.8\times10^{-6}$), the difference in the slope for the low- and high-\rfe\ subsets is $0.02\sigma$ and $0.45\sigma$, respectively, which are not statistically significant. This suggests that the relation given by the full sample is consistent with the $\Delta \tau$-\rfe\ relations in the low- and high-\rfe\ subsets, however, the weak correlations seen in the two 59-source subsets likely have an effect on the results of the analyses, see Sec.~\ref{sec:QSO} and \ref{sec:discussion}.

In addition to H$\beta$ QSO data, we also use 11 BAO observations and 31 $H(z)$ measurements in our analyses here. These BAO and $H(z)$ data are given in Table 1 of \cite{KhadkaRatra2021a} and Table 2 of \cite{Ryanetal2018}. We use BAO + $H(z)$ data results as a proxy for the better-established cosmological probe results and compare them with those from H$\beta$ QSO data to determine whether or not the QSO data results are consistent with better-established data results.

\section{Methods}
\label{sec:methods}

For an H$\beta$ QSO the rest-frame time-delay of the H$\beta$ line and the QSO luminosity are related through the $R-L$ relation \citep{2013ApJ...767..149B}
\begin{equation}
\label{eq:corr}
   \log \left({\frac{\tau} {\rm day}}\right) = \beta + \gamma \log\left({\frac{L_{5100}}{10^{44}\,{\rm erg\,s^{-1}}}}\right),
\end{equation}
where $\log$ = $\log_{10}$, $\tau$ is the rest-frame time-delay of the H$\beta$ line in units of day, $L_{5100}$ is the monochromatic luminosity of the quasar at 5100 {\AA} in units of erg s$^{-1}$, and the intercept $\beta$ and the slope $\gamma$ must be determined from data. In what follows we refer to H$\beta$ data analyses that are based on the 2-parameter $R-L$ relation as H$\beta$ QSO-118, or H$\beta$ high-\rfe, or H$\beta$ low-\rfe\ analyses, depending on the data subgroup used.

In our analyses, in addition to the 2-parameter $R-L$ relation given in eq.\ (\ref{eq:corr}), we consider an extended 3-parameter $R-L$ relation in an attempt to correct for the accretion rate effect observed in the 2-parameter $R-L$ relation (see Sec.~\ref{sec:data} and \citeauthor{Mary2020}, \citeyear{Mary2020} for further details), and thus to try to reduce the intrinsic dispersion and tighten the cosmological constraints. The 3-parameter $R-L$ relation is
\begin{equation}
\label{eq:corr3}
   \log \left({\frac{\tau} {\rm day}}\right) = \beta + \gamma \log\left({\frac{L_{5100}}{10^{44}\,{\rm erg\,s^{-1}}}}\right) + k {\cal R}_{\text{Fe{\sc II}}},
\end{equation}
\citep{duwang_2019} where $k$ is the third free parameter associated with the intensity of the optical \Feii\  flux ratio parameter \rfe\ and must be determined from data. Note an important difference between our older paper \citep{Mary2020} and the current one: eq.~(\ref{eq:corr3}), which is eq.~(5) in \citet{duwang_2019}, assumes a dependence $ \log \tau \propto {\cal R}_{\rm FeII}$, while \citet{Mary2020} assumed $ \log \tau \propto \log {\cal R}_{\rm FeII}$, or equivalently, $\tau=K L_{5100}^{\gamma}{\cal R}_{\rm FeII}^{k}$. In what follows, we refer to H$\beta$ data analyses that are based on the 3-parameter $R-L$ relation as H$\beta$ QSO-118$^\prime$, or H$\beta{}^\prime$ high-\rfe, or H$\beta{}^\prime$ low-\rfe\ analyses, depending on the data subgroup used.

The luminosity can be expressed in terms of flux as
\begin{equation}
\label{eq:lumi}
     L_{5100} = 4\pi D^{2}_L F_{5100}
\end{equation}
where $F_{5100}$ is the measured quasar flux at 5100 {\AA} in units of ${\rm erg\,cm^{-2}\,s^{-1}}$, and $D_L(z, p)$, a function of $z$ and cosmological parameters $p$, is the luminosity distance in units of cm. The luminosity distance is 
\begin{equation}
\label{eq:DM}
  \frac{H_0\sqrt{\left|\Omega_{k0}\right|}D_L(z, p)}{(1+z)} = 
    \begin{cases}
    {\rm sinh}\left[g(z)\right] & \text{if}\ \Omega_{k0} > 0, \\
    \vspace{1mm}
    g(z) & \text{if}\ \Omega_{k0} = 0,\\
    \vspace{1mm}
    {\rm sin}\left[g(z)\right] & \text{if}\ \Omega_{k0} < 0,
    \end{cases}   
\end{equation}
where
\begin{equation}
\label{eq:fun}
   g(z) = H_0\sqrt{\left|\Omega_{k0}\right|}\int^z_0 \frac{dz'}{H(z', p)},
\end{equation}
and the Hubble parameter $H(z, p)$ is given in Sec.\ 2 for each cosmological model.

For quasars at known redshifts, rest-frame time-delays in a given cosmological model can be predicted using eqs.\ (\ref{eq:corr}) or (\ref{eq:corr3}), as well as (\ref{eq:lumi}) and (\ref{eq:DM}). Cosmological model and correlation relation parameters can then be constrained by comparing these predicted time-delays with corresponding measured time-delays using the log likelihood function \citep{Dago2005}
\begin{equation}
\label{eq:chi2}
    \ln({\rm LF}) = -\frac{1}{2}\sum^{N}_{i = 1} \left[\frac{[\log(\tau^{\rm obs}_{X,i}) - \log(\tau^{\rm th}_{X,i})]^2}{s^2_i} + \ln(2\pi s^2_i)\right].
\end{equation}
Here $\ln$ = $\log_e$, $\tau^{\rm th}_{X,i}(p)$ and $\tau^{\rm obs}_{X,i}$ are the predicted and observed time-delays at measured redshift $z_i$. In the 2-parameter $R-L$ relation case,  $s^2_i = \sigma^2_{\log{\tau_{\rm obs},i}} + \gamma^2 \sigma^2_{\log{F_{5100},i}} + \sigma_{\rm ext}^2$,  while in the 3-parameter $R-L$ relation case,  $s^2_i = \sigma^2_{\log{\tau_{\rm obs},i}} + \gamma^2 \sigma^2_{\log{F_{5100},i}} + k^2 \sigma^2_{{\cal R}_{\text{Fe{\sc II}}},i} + \sigma_{\rm ext}^2$, where $\sigma_{\log{\tau_{\rm obs},i}}$, $\sigma_{\log{F_{5100},i}}$, and $\sigma_{{\cal R}_{\text{Fe{\sc II}}},i}$ are the measurement error on the observed time-delay ($\tau^{\rm obs}_{X,i}$), measured flux ($F_{5100,i}$), and measured \rfe$_{,i}$ respectively. $\sigma_{\rm ext}$ is the intrinsic dispersion of the $R-L$ relation.

Cosmological constraints from the BAO + $H(z)$ data are taken from \cite{KhadkaRatra2021a} and we refer the reader to that paper for a description of the derivation of these constraints and a detailed discussion of these constraints.

We maximize the log likelihood function given in eq.\ (\ref{eq:chi2}) by using the Markov chain Monte Carlo (MCMC) sampling method as implemented in the {\sc MontePython} code \citep{Brinckmann2019}. Best-fit value and corresponding uncertainty of each free parameter are determined from analyses of the MCMC chains by using the {\sc Python} package {\sc Getdist} \citep{Lewis_2019}, which we also use to plot the likelihoods. Convergence of the MCMC chains for each free parameter is confirmed by requring that the Gelman-Rubin criterion ($R - 1 < 0.05$) be satisfied. We use a flat prior for each parameter, with non-zero prior ranges listed in Table \ref{tab:prior}. The QSO data we use here cannot constrain $H_0$ because there is a degeneracy between $\beta$ and $H_0$, so in QSO data analyses here we set $H_0$ to $70$ ${\rm km}\hspace{1mm}{\rm s}^{-1}{\rm Mpc}^{-1}$.

\begin{table}
	\centering
	\caption{Summary of the non-zero flat prior parameter ranges.}
	\label{tab:prior}
	\begin{threeparttable}
	\begin{tabular}{l|c}
	\hline
	Parameter & Prior range \\
	\hline
	$\Omega_bh^2$ & $[0, 1]$ \\
	$\Omega_ch^2$ & $[0, 1]$ \\
    $\Omega_{m0}$ & $[0, 1]$ \\
    $\Omega_{k0}$ & $[-2, 1]$ \\
    $\omega_{X}$ & $[-5, 0.33]$ \\
    $\alpha$ & $[0, 10]$ \\
    $\sigma_{\rm ext}$ & $[0, 5]$ \\
    $\beta$ & $[0, 10]$ \\
    $\gamma$ & $[0, 5]$ \\
    $k$ & $[-10, 10]$ \\
	\hline
	\end{tabular}
    \end{threeparttable}
\end{table}

For the comparison of different cosmological models and for the comparison of the different $R-L$ relations, we compute the Akaike and the Bayesian information criterion ($AIC$ and $BIC$) values. The $AIC$ and the $BIC$ values are defined as
\begin{align}
\label{eq:AIC}
    AIC =& -2 \ln({\rm LF}_{\rm max}) + 2d,\\
\label{eq:BIC}
    BIC =& -2 \ln({\rm LF}_{\rm max}) + d\ln{N}\, ,
\end{align}
where $\rm LF_{\rm max}$ is the maximum likelihood value, $N$ is the number of measurements, and $d$ is the number of free parameters, with $dof = N - d$ being the degrees of freedom. We also compute $\Delta AIC$ and $\Delta BIC$ differences of the 3-parameter $R-L$ with respect to the corresponding 2-parameter $R-L$ reference model. $\Delta AIC(BIC) \in [0, 2]$ is weak evidence in favor of the 2-parameter reference model, $\Delta AIC(BIC) \in(2, 6]$ is positive evidence for the reference model, $\Delta AIC(BIC)>6$ is strong evidence for the reference model, and $\Delta AIC(BIC)>10$ is very strong evidence for the reference model.  Negative values of $\Delta AIC$ or $\Delta BIC$ indicate that the model under investigation fits the data better than the reference model.

\begin{table*}
	\centering
	\small\addtolength{\tabcolsep}{-5pt}
	\small
	\caption{Unmarginalized best-fit parameters for H$\beta$ data sets.$^{\rm a}$ $\Delta AIC$ and $\Delta BIC$ values are computed with respect to the $AIC$ and $BIC$ values of the corresponding 2-parameter $R-L$ relation computation. The QSO-$118^{\prime}$ and H$\beta^\prime$ results assume the 3-parameter $R-L$ relation.}
	\label{tab:BFP}
	\begin{threeparttable}
	\begin{tabular}{l|cccccccccccccccccc} % four columns, alignment for each
		\hline
		Model & Data set & \ \ $\Omega_{\rm m0}$ \ \ & \ \ $\Omega_{\rm k0}$ \ \ & \ \ $\omega_{X}$ \ \ & \ \ $\alpha$ \ \ & \ \ $\sigma_{\rm ext}$ \ \ & \ \ $\beta$ \ \ & \ \ \ \ $\gamma$ \ \ \ \ & $k$  &  $dof$ & $-2\ln({\rm LF}_{\rm max})$ & $AIC$ & \ \ $BIC$ \ \ & \ $\Delta AIC$ \ & \ $\Delta BIC$\\
		\hline
	    Flat $\Lambda$CDM & H$\beta$ QSO-118 & 0.998  & - & - &- & 0.231 & 1.361 & 0.422 & - & 114 & 17.52 & 25.52 & 36.60  & - & - \\
	    & H$\beta$ low-\rfe\ & 0.999 & - & - & - & 0.206 & 1.461 & 0.471 & - & 55 & $-5.20$ & 2.80 & 11.11  & - & - \\
	    & H$\beta$ high-\rfe\ & 0.999 & - & - & - & 0.220 & 1.266 & 0.383 & - & 55 & $5.32$ & 13.32 & 21.63  & - & - \\
		& H$\beta$ QSO-$118^{\prime}$ & 0.998 & - & - & - & 0.210 & 1.558 & 0.448 & $-0.264$ & 113 & $-1.20$ & 8.80 & 22.65  & $-16.72$ & $-13.95$\\
		& H$\beta^{\prime}$ low-\rfe\ & 0.998 & - & - & - & 0.198 & 1.583 & 0.479 & $-0.272$ & 54 & $-7.00$ & 3.00 & 13.39 & $0.20$ & 2.28 \\
		& H$\beta^{\prime}$ high-\rfe\ & 0.991 & - & - & - & 0.213 & 1.421 & 0.404 & $-0.150$ & 54 & 3.10 & 13.10 & 23.49 & $-0.22$ & 1.86 \\
		\hline
		Non-flat $\Lambda$CDM & H$\beta$ QSO-118 & 0.998 & $-0.015$ & - & - & 0.229 & 1.365 & 0.422 & - & 113 & 15.68 & 25.68 & 39.53  & - & - \\
		& H$\beta$ low-\rfe\ & 0.995 & $-0.015$ & - & - & 0.200 & 1.460 & 0.472 & - & 54 & $-5.13$ & 4.87 & 15.26 & - & -  \\
		& H$\beta$ high-\rfe\ & 0.877 & $-0.041$ & - & - & 0.221 & 1.264 & 0.386 & - & 54 & $5.34$ & 15.34 & 25.73 & - & -  \\
		& H$\beta$ QSO-$118^{\prime}$ & 0.988 & 0.000& - & - & 0.208 & 1.544 & 0.447 & $-0.247$ & 112 & $-2.86$ & 9.14 & 25.76 & $-16.54$ & $-13.77$ \\
		& H$\beta^{\prime}$ low-\rfe\ & 0.992 & 0.042& - & - & 0.198 & 1.574 & 0.479 & $-0.265$ & 53 & $-6.93$ & 5.07 & 17.54 & 0.20 & 2.28 \\
		& H$\beta^{\prime}$ high-\rfe\ & 0.993 & $-0.030$ & - & - & 0.214 & 1.431 & 0.407 & $-0.161$ & 53 & 3.11 & 15.11 & 27.58 & $-0.23$ & 1.85 \\
		\hline
		Flat XCDM & H$\beta$ QSO-118 & 0.046 & - & 0.140 & - & 0.232 & 1.367 & 0.421 & - & 113 & 16.74 & 26.74 & 40.59 & - & - \\
		& H$\beta$ low-\rfe\ & 0.062 & - & 0.138 & - & 0.202 & 1.473 & 0.474 & - & 54 & $-6.02$ & 5.98 & 18.45 & - & - \\
		& H$\beta$ high-\rfe\ & 0.389 & - & 0.143 & - & 0.221 & 1.271 & 0.386 & - & 54 & 5.26 & 15.26 & 25.65 & - & - \\
		& H$\beta$ QSO-$118^{\prime}$ & 0.062 & - & 0.139 & - & 0.205 & 1.558 & 0.443 & $-0.252$ & 112 & $-1.98$ & 10.02 & 26.64 & $-16.72$ & $-13.95$ \\
		& H$\beta^{\prime}$ low-\rfe\ & 0.166 & - & 0.139 & - & 0.198 & 1.582 & 0.469 & $-0.272$ & 53 & $-7.56$ & 4.44 & 16.91 & $-1.54$ & $-1.54$ \\
		& H$\beta^{\prime}$ high-\rfe\ & 0.567 & - & 0.127 & - & 0.220 & 1.449 & 0.407 & $-0.169$ & 53 & 3.00 & 15.00 & 27.47 & $-0.26$ & 1.82 \\
		\hline
		Non-flat XCDM & H$\beta$ QSO-118 & 0.323 & $-1.982$ & 0.090 & - & 0.230 & 1.411 & 0.440 & - & 112 & 14.70 & 26.70 & 43.32 & - & - \\
		& H$\beta$ low-\rfe\ & 0.513 & $-1.954$ & 0.127 & - & 0.202 & 1.511 & 0.494 & - & 53 & $-8.46$ & 3.54 & 16.01 & - & -\\
		& H$\beta$ high-\rfe\ & 0.332 & 1.990 & $-3.072$ & - & 0.218 & 1.323 & 0.399 & - & 53 & 5.10 & 17.10 & 29.67 & - & -\\
		& H$\beta$ QSO-$118^{\prime}$ & 0.988 & $-1.812$ & 0.093 & - & 0.213 & 1.558 & 0.457 & $-0.258$ & 111 & $-4.10$ & 9.90 & 29.29 & $-16.80$ & $-14.03$ \\
		& H$\beta^{\prime}$ low-\rfe\ & 0.314 & $-1.968$ & 0.115 & - & 0.198 & 1.575 & 0.488 & $-0.152$ & 52 & $-9.30$ & 4.67 & 19.24 & $1.13$ & $3.23$ \\
		& H$\beta^{\prime}$ high-\rfe\ & 0.848 & 1.574 & $-3.027$ & - & 0.214 & 1.523 & 0.424 & $-0.176$ & 52 & 2.54 & 16.54 & 31.08 & $-0.56$ & 1.41\\
		\hline
		Flat $\phi$CDM & H$\beta$ QSO-118 & 0.999 & - & - & 6.209 & 0.232 & 1.360 & 0.421 & - & 113 & 17.52 & 27.52 & 41.37 & - & - \\
		& H$\beta$ low-\rfe\ & 0.997 & - & - & 8.107 & 0.204 & 1.461 & 0.473 & - & 54 & $-5.20$ & 4.80 & 15.19 & - & - \\
		& H$\beta$ high-\rfe\ & 0.999 & - & - & 8.135 & 0.221 & 1.267 & 0.383 & - & 54 & 5.32 & 15.32 & 25.71 & - & - \\
		& H$\beta$ QSO-$118^{\prime}$ & 0.995 & - & - & 6.547& 0.210 & 1.552 & 0.448 & $-0.257$ & 112 & $-1.20$ & 10.80 & 27.42 & $-16.72$ & $-14.03$ \\
		& H$\beta^{\prime}$ low-\rfe\ & 0.999 & - & - & 8.807& 0.200 & 1.582 & 0.476 & $-0.281$ & 53 & $-7.02$ & 4.98 & 17.45 & $0.18$ & 2.26 \\
		& H$\beta^{\prime}$ high-\rfe\ & 0.981 & - & - & 5.38& 0.217 & 1.424 & 0.402 & $-0.153$ & 53 & 3.08 & 15.08 & 27.54 & $-0.24$ & 1.83\\
		\hline
		Non-flat $\phi$CDM & H$\beta$ QSO-118 & 0.999 & $-0.982$ & - & 9.910& 0.234 & 1.370 & 0.423 & - & 112 & 16.30 & 28.30 & 44.92 & - & - \\
		& H$\beta$ low-\rfe\ & 0.980 & $-0.975$ & - & 9.539 & 0.197 & 1.479 & 0.472 & - & 53 & $-6.54$ & 5.46 & 17.93 & - & - \\
		& H$\beta$ high-\rfe\ & 0.970 & $-0.837$ & - & 8.168 & 0.221 & 1.275 & 0.387 & - & 53 & 5.26 & 17.26 & 29.73 & - & - \\
		& H$\beta$ QSO-$118^{\prime}$ & 0.976 & $-0.909$ & - & 9.226 & 0.210 & 1.562 & 0.461 & $-0.253$ & 111 & $-2.42$ & 11.58 & 30.97 & $-16.72$ & $-13.95$\\
		& H$\beta^{\prime}$ low-\rfe\ & 0.964 & $-0.949$ & - & 9.733 & 0.197 & 1.591 & 0.486 & $-0.281$ & 52 & $-8.04$ & 5.96 & 20.50 & $0.50$ & 2.57\\
		& H$\beta^{\prime}$ high-\rfe\ & 0.998 & $-0.931$ & - & 8.211 & 0.211 & 1.441 & 0.411 & $-0.151$ & 52 & 2.90 & 16.90 & 31.44 & $-0.36$ & 1.71\\
		\hline
	\end{tabular}
	\begin{tablenotes}
    \item[a] $H_0$ is set to $70$ ${\rm km}\hspace{1mm}{\rm s}^{-1}{\rm Mpc}^{-1}$ for QSO-only data analyses.
    \end{tablenotes}
    \end{threeparttable}
\end{table*}

\begin{sidewaystable*}
\begin{adjustbox}{angle=0}
\centering
\small\addtolength{\tabcolsep}{0.0pt}
\begin{threeparttable}
\caption{Marginalized one-dimensional best-fit parameters with 1$\sigma$ confidence intervals, or 1$\sigma$ or 2$\sigma$ limits, for the H$\beta$ and BAO + $H(z)$ data sets. The QSO-$118^{\prime}$ and H$\beta^\prime$ results assume the 3-parameter $R-L$ relation.}
\label{tab:1d_BFP2}
\setlength{\tabcolsep}{1.3mm}{
\begin{tabular}{lcccccccccccccc}
\hline
Model & Data & $\Omega_{b}h^2$ & $\Omega_{c}h^2$ & $\om$ & $\ol$\tnote{a} & $\ok$ & $\omega_{X}$ & $\alpha$ & $H_0$\tnote{b} & $\sigma_{\rm ext}$ & $\beta$ & $\gamma$ & $k$ \\
\hline
Flat \lcdm\ & H$\beta$ QSO-118 &-&-& $> 0.336$ & $< 0.664$ & - & - & - & - & $0.236^{+0.020}_{-0.018}$ & $1.350^{-0.026}_{-0.028}$ & $0.415^{+0.030}_{-0.029}$ & -\\
& H$\beta$ low-\rfe\ &-&-& $> 0.325$ & --- & - & - & - & - & $0.212^{+0.026}_{-0.023}$ & $1.448^{-0.034}_{-0.036}$ & $0.465^{+0.039}_{-0.037}$ & -\\
& H$\beta$ high-\rfe\ &-&-& --- & --- & - & - & - & - & $0.228^{+0.029}_{-0.025}$ & $1.246^{-0.037}_{-0.038}$ & $0.374^{+0.043}_{-0.042}$ & -\\
& H$\beta$ QSO-$118^{\prime}$ &-&-& $> 0.388$ & $< 0.612$ & - & - & - & - & $0.216^{+0.019}_{-0.017}$ & $1.541^{-0.049}_{-0.051}$ & $0.441^{+0.029}_{-0.029}$ & $-0.259^{+0.060}_{-0.059}$\\
& H$\beta^{\prime}$ low-\rfe\ &-&-& $> 0.264$ & $< 0.750$ & - & - & - & - & $0.210^{+0.028}_{-0.025}$ & $1.577^{-0.100}_{-0.104}$ & $0.468^{+0.040}_{-0.039}$ & $-0.309^{+0.231}_{-0.222}$\\
& H$\beta^{\prime}$ high-\rfe\ &-&- & --- & $< 0.910$ & - & - & - & - & $0.228^{+0.030}_{-0.027}$ & $1.396^{-0.128}_{-0.125}$ & $0.394^{+0.047}_{-0.048}$ & $-0.144^{+0.114}_{-0.115}$\\
& BAO+$H(z)$& $0.024^{+0.003}_{-0.003}$ & $0.119^{+0.008}_{-0.008}$ & $0.299^{+0.015}_{-0.017}$ & - & - & - & - &$69.300^{+1.800}_{-1.800}$&-&-&-&-\\
\hline
Non-flat \lcdm\ & H$\beta$ QSO-118 &-&-& $> 0.190$ & $< 1.330$ & $-0.006^{+0.416}_{-0.549}$ & - & - & - & $0.237^{+0.020}_{-0.018}$ & $1.341^{-0.027}_{-0.029}$ & $0.411^{+0.030}_{-0.030}$ & -\\
& H$\beta$ low-\rfe\ &-&-& $> 0.193$ & $< 1.430$ & $-0.012^{+0.407}_{-0.614}$ & - & - & - & $0.213^{+0.027}_{-0.023}$ & $1.439^{-0.034}_{-0.037}$ & $0.460^{+0.039}_{-0.038}$ & -\\
& H$\beta$ high-\rfe\ &-&-& --- & $< 1.790$ & $-0.034^{+0.489}_{-0.835}$ & - & - & - & $0.229^{+0.028}_{-0.025}$ & $1.239^{-0.037}_{-0.040}$ & $0.369^{+0.043}_{-0.041}$ & -\\
& H$\beta$ QSO-$118^{\prime}$ &-&-& $> 0.230$ & $< 1.270$ & - & - & - & - & $0.217^{+0.019}_{-0.016}$ & $1.532^{-0.050}_{-0.052}$ & $0.437^{+0.029}_{-0.028}$ & $-0.259^{+0.060}_{-0.060}$\\
& H$\beta^{\prime}$ low-\rfe\ &-&-& $> 0.169$ & $< 1.560$ & $-0.026^{+0.441}_{-0.682}$ & - & - & - & $0.211^{+0.028}_{-0.025}$ & $1.570^{+0.103}_{-0.100}$ & $0.464^{+0.039}_{-0.040}$ & $-0.322^{+0.225}_{-0.224}$\\
& H$\beta^{\prime}$ high-\rfe\ &-&-& --- & $< 1.740$ & $-0.034^{+0.490}_{-0.793}$ & - & - & - & $0.229^{+0.030}_{-0.027}$ & $1.378^{+0.128}_{-0.122}$ & $0.387^{+0.048}_{-0.045}$ & $-0.136^{+0.110}_{-0.116}$\\
& BAO+$H(z)$& $0.025^{+0.004}_{-0.004}$ & $0.113^{+0.019}_{-0.019}$ & $0.292^{+0.023}_{-0.023}$ & $0.667^{+0.093}_{+0.081}$ & $-0.014^{+0.075}_{-0.075}$ & - & - &$68.700^{+2.300}_{-2.300}$&-&-&-&-\\
\hline
Flat XCDM & H$\beta$ QSO-118 &-&-& $> 0.217$ & - & - & $< 0.200$ & - & - & $0.236^{+0.020}_{-0.018}$ & $1.350^{-0.027}_{-0.028}$ & $0.415^{+0.030}_{-0.030}$ & -\\
& H$\beta$ low-\rfe\ &-&-& $> 0.233$ & - & - & $< 0.058$ & - & - & $0.211^{+0.027}_{-0.023}$ & $1.447^{-0.034}_{-0.037}$ & $0.465^{+0.038}_{-0.038}$ & -\\
& H$\beta$ high-\rfe\ &-&-& --- & - & - & $< 0.010$ & - & - & $0.228^{+0.029}_{-0.025}$ & $1.245^{-0.039}_{-0.042}$ & $0.373^{+0.043}_{-0.043}$ & -\\
& H$\beta$ QSO-$118^{\prime}$ &-&-& $> 0.207$ & - & - & $< 0.200$ & - & - & $0.217^{+0.019}_{-0.018}$ & $1.541^{-0.053}_{-0.054}$ & $0.442^{+0.030}_{-0.031}$ & $-0.259^{+0.062}_{-0.063}$\\
& H$\beta{\prime}$ low-\rfe\ &-&-& $> 0.212$ & - & - & $< 0.046$ & - & - & $0.209^{+0.028}_{-0.024}$ & $1.572^{-0.101}_{-0.101}$ & $0.468^{+0.041}_{-0.040}$ & $-0.303^{+0.226}_{-0.224}$\\
& H$\beta{\prime}$ high-\rfe\ &-&-& --- & - & - & $< 0.100$ & - & - & $0.227^{+0.030}_{-0.026}$ & $1.395^{-0.127}_{-0.128}$ & $0.392^{+0.048}_{-0.048}$ & $-0.147^{+0.115}_{-0.112}$\\
& BAO+$H(z)$ & $0.030^{+0.005}_{-0.005}$ & $0.093^{+0.019}_{-0.017}$ & $0.282^{+0.021}_{-0.021}$ & - & - & $-0.744^{+0.140}_{-0.097}$ & - &$65.800^{+2.200}_{-2.500}$& - & - & - &-\\
\hline
Non-flat XCDM & H$\beta$ QSO-118 &-&-& --- & - & --- & $< 0.100$ & - & - & $0.235^{+0.020}_{-0.018}$ & $1.377^{-0.044}_{-0.036}$ & $0.428^{+0.033}_{-0.032}$ & -\\
& H$\beta$ low-\rfe\ &-&-& --- & - & --- & $< 0.100$ & - & - & $0.210^{+0.026}_{-0.023}$ & $1.475^{-0.050}_{-0.044}$ & $0.478^{+0.041}_{-0.039}$ & -\\
& H$\beta$ high-\rfe\ &-&-& --- & - & $0.667^{+1.126}_{-0.704}$ & $< 0.100$ & - & - & $0.227^{+0.029}_{-0.025}$ & $1.263^{-0.050}_{-0.045}$ & $0.380^{+0.045}_{-0.044}$ & -\\
& H$\beta$ QSO-$118^{\prime}$ &-&-& --- & - & --- & $< 0.200$ & - & -  & $0.216^{+0.019}_{-0.018}$ & $1.571^{-0.063}_{-0.058}$ & $0.452^{+0.034}_{-0.031}$ & $-0.257^{+0.063}_{-0.062}$\\
& H$\beta^{\prime}$ low-\rfe\ &-&-& --- & - & --- & $< 0.200$ & - & - & $0.209^{+0.028}_{-0.024}$ & $1.589^{-0.102}_{-0.103}$ & $0.478^{+0.043}_{-0.041}$ & $-0.266^{+0.223}_{-0.229}$\\
& H$\beta^{\prime}$ high-\rfe\ &-&-& --- & - & $> -1.23$ & $< 0.100$ & - & - & $0.226^{+0.030}_{-0.026}$ & $1.425^{-0.135}_{-0.126}$ & $0.401^{+0.051}_{-0.047}$ & $-0.151^{+0.111}_{-0.113}$\\
& BAO+$H(z)$ & $0.029^{+0.005}_{-0.005}$ & $0.099^{+0.021}_{-0.021}$ & $0.293^{+0.027}_{-0.027}$ & - & $-0.120^{+0.130}_{-0.130}$ & $-0.693^{+0.130}_{-0.077}$ & - &$65.900^{+2.400}_{-2.400}$& - & - & - & -\\
\hline
Flat $\phi$CDM & H$\beta$ QSO-118 &-&-& $> 0.191$ & - & - & - & --- & -  & $0.236^{+0.020}_{-0.018}$ & $1.353^{-0.026}_{-0.027}$ & $0.418^{+0.030}_{-0.029}$ & -\\
& H$\beta$ low-\rfe\ &-&-& $> 0.191$ & - & - & - & --- & - & $0.211^{+0.026}_{-0.023}$ & $1.453^{-0.033}_{-0.034}$ & $0.468^{+0.038}_{-0.037}$ & -\\
& H$\beta$ high-\rfe\ &-&-& --- & - & - & - & --- & - & $0.227^{+0.029}_{-0.025}$ & $1.255^{-0.035}_{-0.037}$ & $0.378^{+0.043}_{-0.042}$ & -\\
& H$\beta$ QSO-$118^{\prime}$ &-&-& $> 0.208$ & - & - & - & --- & - & $0.216^{+0.019}_{-0.018}$ & $1.544^{-0.052}_{-0.051}$ & $0.443^{+0.029}_{-0.029}$ & $-0.259^{+0.061}_{-0.061}$\\
& H$\beta^{\prime}$ low-\rfe\ &-&-& $> 0.158$ & - & - & - & --- & - & $0.209^{+0.027}_{-0.025}$ & $1.577^{-0.101}_{-0.099}$ & $0.471^{+0.039}_{-0.039}$ & $-0.296^{+0.221}_{-0.221}$\\
& H$\beta^{\prime}$ high-\rfe\ &-&-& $0.678^{+0.312}_{-0.289}$ & - & - & - & --- & - & $0.225^{+0.030}_{-0.025}$ & $1.410^{-0.123}_{-0.120}$ & $0.398^{+0.047}_{-0.047}$ & $-0.152^{+0.113}_{-0.109}$\\
& BAO+$H(z)$ & $0.032^{+0.006}_{-0.003}$ & $0.081^{+0.017}_{-0.017}$ & $0.266^{+0.023}_{-0.023}$ & - & - & - & $1.530^{+0.620}_{-0.850}$ &$65.100^{+2.100}_{-2.100}$& - & - & - & -\\
\hline
Non-flat $\phi$CDM & H$\beta$ QSO-118 &-&-& $> 0.158$ & - & $-0.093^{+0.368}_{-0.366}$ & - & --- & - & $0.235^{+0.020}_{-0.018}$ & $1.354^{-0.026}_{-0.027}$ & $0.417^{+0.029}_{-0.030}$ & -\\
& H$\beta$ low-\rfe\ &-&-& $> 0.157$ & - & $-0.091^{+0.362}_{-0.371}$ & - & --- & - & $0.211^{+0.027}_{-0.024}$ & $1.453^{-0.034}_{-0.034}$ & $0.468^{+0.038}_{-0.037}$ & -\\
& H$\beta$ high-\rfe\ &-&-& $0.501^{+0.313}_{-0.310}$ & - & $0.009^{+0.410}_{-0.345}$ & - & --- & - & $0.227^{+0.030}_{-0.026}$ & $1.255^{-0.035}_{-0.036}$ & $0.378^{+0.043}_{-0.041}$ & -\\
& H$\beta$ QSO-$118^{\prime}$ &-&-& $> 0.178$ & - & $-0.085^{+0.353}_{-0.389}$ & - & --- & - & $0.216^{+0.019}_{-0.018}$ & $1.546^{-0.051}_{-0.053}$ & $0.443^{+0.029}_{-0.030}$ & $-0.258^{+0.060}_{-0.062}$\\
& H$\beta^{\prime}$ low-\rfe\ &-&-& --- & - & $-0.077^{+0.381}_{0.372}$ & - & --- & - & $0.208^{+0.028}_{-0.025}$ & $1.577^{-0.101}_{-0.099}$ & $0.471^{+0.039}_{-0.039}$ & $-0.294^{+0.221}_{-0.223}$\\
& H$\beta^{\prime}$ high-\rfe\ &-&-& $0.571^{+0.313}_{-0.309}$ & - & $-0.010^{+0.414}_{0.356}$ & - & --- & - & $0.226^{+0.031}_{-0.027}$ & $1.412^{-0.120}_{-0.121}$ & $0.397^{+0.047}_{-0.045}$ & $-0.149^{+0.108}_{-0.116}$\\
& BAO+$H(z)$ & $0.032^{+0.006}_{-0.004}$ & $0.085^{+0.017}_{-0.021}$ & $0.271^{+0.024}_{-0.028}$ & - & $-0.080^{+0.100}_{-0.100}$ & - & $1.660^{+0.670}_{-0.830}$ &$65.500^{+2.500}_{-2.500}$& - & - & - & -\\
\hline
\end{tabular}}
\begin{tablenotes}
\item[a]In our analyses $\Omega_{\Lambda}$ is a derived parameter and in each case $\Omega_{\Lambda}$ chains are derived using the current energy budget equation $\Omega_{\Lambda}= 1-\Omega_{m0}-\Omega_{k0}$ (where $\Omega_{k0}=0$ in the flat $\Lambda$CDM model). We determine best-fit values and uncertainties for $\Omega_{\Lambda}$ from these chains. 
\item[b]${\rm km}\hspace{1mm}{\rm s}^{-1}{\rm Mpc}^{-1}$. $H_0$ is set to $70$ ${\rm km}\hspace{1mm}{\rm s}^{-1}{\rm Mpc}^{-1}$ for the QSO-only data analyses.
\end{tablenotes}
\end{threeparttable}
\end{adjustbox}
\end{sidewaystable*}

\section{Results}
\label{sec:QSO}

Results from the 2-parameter $R-L$ relation H$\beta$ QSO-118, H$\beta$ low-\rfe, and H$\beta$ high-\rfe\ analyses and from the 3-parameter $R-L$ relation H$\beta$ QSO-118$^\prime$, H$\beta^\prime$ low-\rfe, and H$\beta^\prime$ high-\rfe\ analyses are given in Tables \ref{tab:BFP} and \ref{tab:1d_BFP2}. The unmarginalized best-fit parameter values are listed in Table \ref{tab:BFP} and the marginalized one-dimensional best-fit parameter values and limits are given in Table \ref{tab:1d_BFP2}.\footnote{We noted in the first paragraph of Sec.\ 3 that for some of the data points we assumed slightly different cosmological parameter values, when converting from monochromatic luminosities to flux densities, in comparison with the cosmological parameters used in the original papers. To quantify the differences that are caused by assuming slightly different flat $\Lambda$CDM parameters, we analyzed the flat $\Lambda$CDM model using the correct H$\beta$ QSO-118 data set (adopting the original cosmological parameters), finding $\Omega_{m0} > 0.334$, $\Omega_{\Lambda} < 0.666$, $\sigma_{\rm ext} = 0.237^{+0.018}_{-0.022}$, $\beta = 1.349 \pm 0.029$, and $\gamma = 0.421 \pm 0.031$, i.e. the differences with respect to the results shown in the first line of Table \ref{tab:1d_BFP2} are negligible.} Corresponding one-dimensional likelihood distributions and two-dimensional likelihood contours for different cosmological models are plotted in Figs.\ \ref{fig:Eiso-Ep1}--\ref{fig:Eiso-Ep5}. The H$\beta$ QSO-118 sample is currently the most complete sample of reliable reverberation-measured quasars with reliable \rfe\ measurements. There are three additional quasars with \hb\ time delays\footnote{We also performed 2-parameter $R-L$ analyses with 121 sources, now also including J141955, MCG~+06-26-012, and MCG~+06-30-015, in addition to the 118 sources in Table~\ref{tab:hbQSOdata}. We used both uncorrected time delays and time delays corrected with respect to the canonical $R-L$ relation; see \citet{Mary2019} for the methodology used for the correction which introduces a bias and so we do not present these results. The uncorrected 121 sources cosmological constraints are quite similar to those from the
H$\beta$ QSO-118 sample, as expected.}, however they are not considered in the analysis because the \rfe\ value for SDSS~J141955 is not unreliable, MCG+06-26-012 has a relatively low sampling cadence, and MCG+06-30-015 is affected by reddening \citep{duwang_2019}.

Results from BAO + $H(z)$ data are given in Table \ref{tab:1d_BFP2}. We use the BAO + $H(z)$ cosmological constraints to compare to those determined from H$\beta$ QSO data. This comparison allows us to draw a qualitative idea of whether or not the H$\beta$ QSO constraints  are consistent with those derived from better-established cosmological data which favor $\Omega_{m0}$ = 0.3. 

\subsection{$R-L$ correlation relation parameter measurements}

The derivation of cosmological constraints from these H$\beta$ QSO data depends on the validity of the assumed $R-L$ correlation relation. As discussed next, for both the 2-parameter and the 3-parameter $R-L$ relation, the $R-L$ relation parameter ($\beta$, $\gamma$, and $k$) values measured using the complete H$\beta$ data set, or measured using the high-\rfe\ or low-\rfe\ data subsets, are almost completely independent of the cosmological model used in the analysis. This indicates that the H$\beta$ QSOs are standardizable through the $R-L$ relation. There are, however, potential complications, to be discussed below.

From Table \ref{tab:1d_BFP2}, for the 2-parameter H$\beta$ QSO-118 data set, in all cosmological models, the values of $\beta$ lie in the range $1.341^{+0.027}_{-0.029}$ to $1.377^{+0.044}_{-0.036}$ and the values of $\gamma$ lie in the range $0.411^{+0.030}_{-0.030}$ to $0.428^{+0.033}_{-0.032}$. The difference between the largest and the smallest central values of $\beta$ is 0.80$\sigma$ (of the quadrature sum of the two error bars) while this difference for $\gamma$ values is 0.39$\sigma$, both of which are not statistically significant. For the 3-parameter H$\beta$ QSO-118$^{\prime}$ data set, in all cosmological models, the values of $\beta$ lie in the range $1.532^{+0.050}_{-0.052}$ to $1.571^{+0.063}_{-0.058}$, the values of $\gamma$ lie in the range $0.437^{+0.029}_{-0.028}$ to $0.452^{+0.034}_{-0.031}$, and the values of $k$ lie in the range $-0.259^{+0.062}_{-0.063}$ to $-0.257^{+0.063}_{-0.062}$. The difference between the largest and the smallest central values of $\beta$ is 0.51$\sigma$ while this difference for $\gamma$ values and $k$ values is 0.35$\sigma$ and 0.02$\sigma$, respectively, and again these are not statistically significant.

\begin{figure*}
\begin{multicols}{2}
    \includegraphics[width=\linewidth,height=7cm]{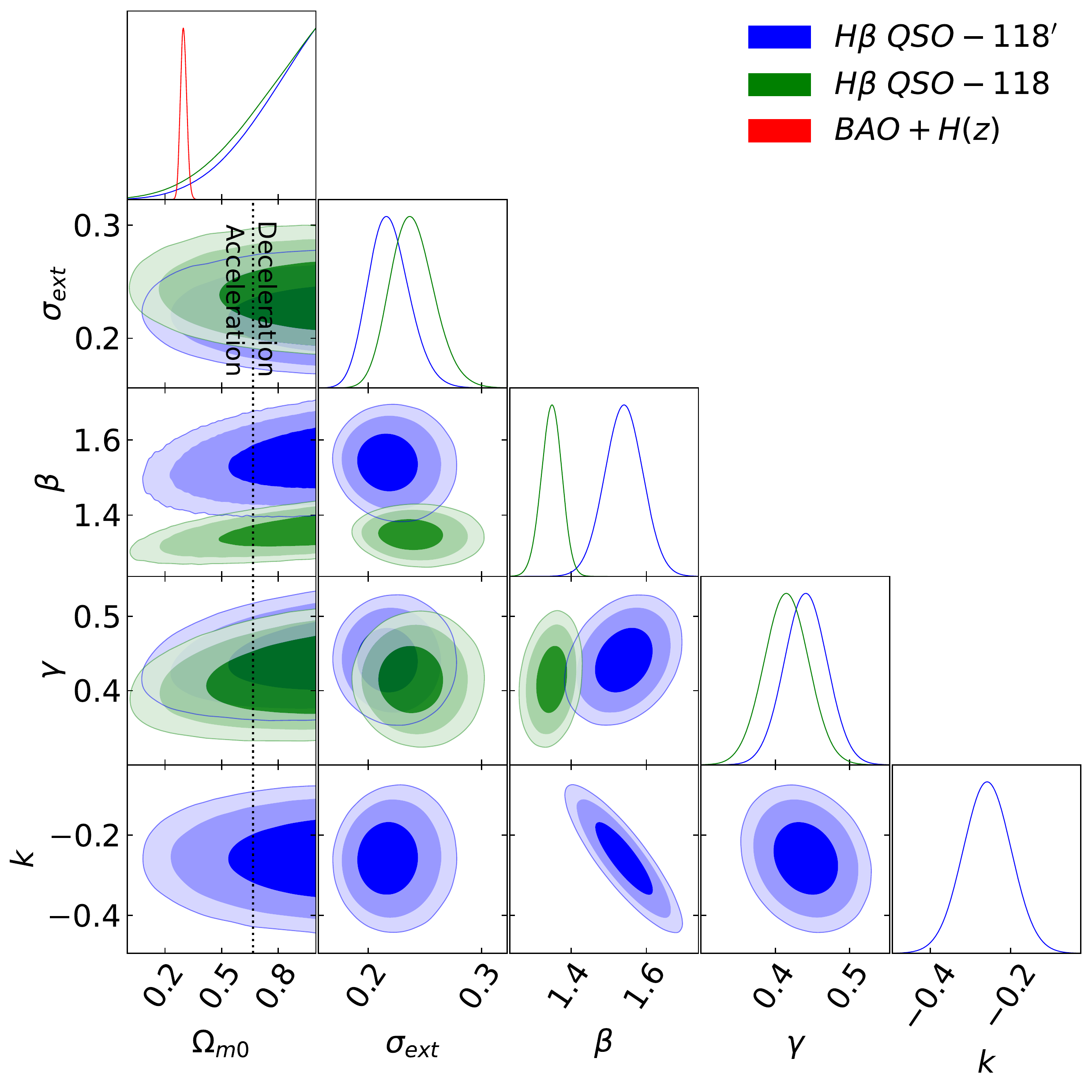}\par
    \includegraphics[width=\linewidth,height=7cm]{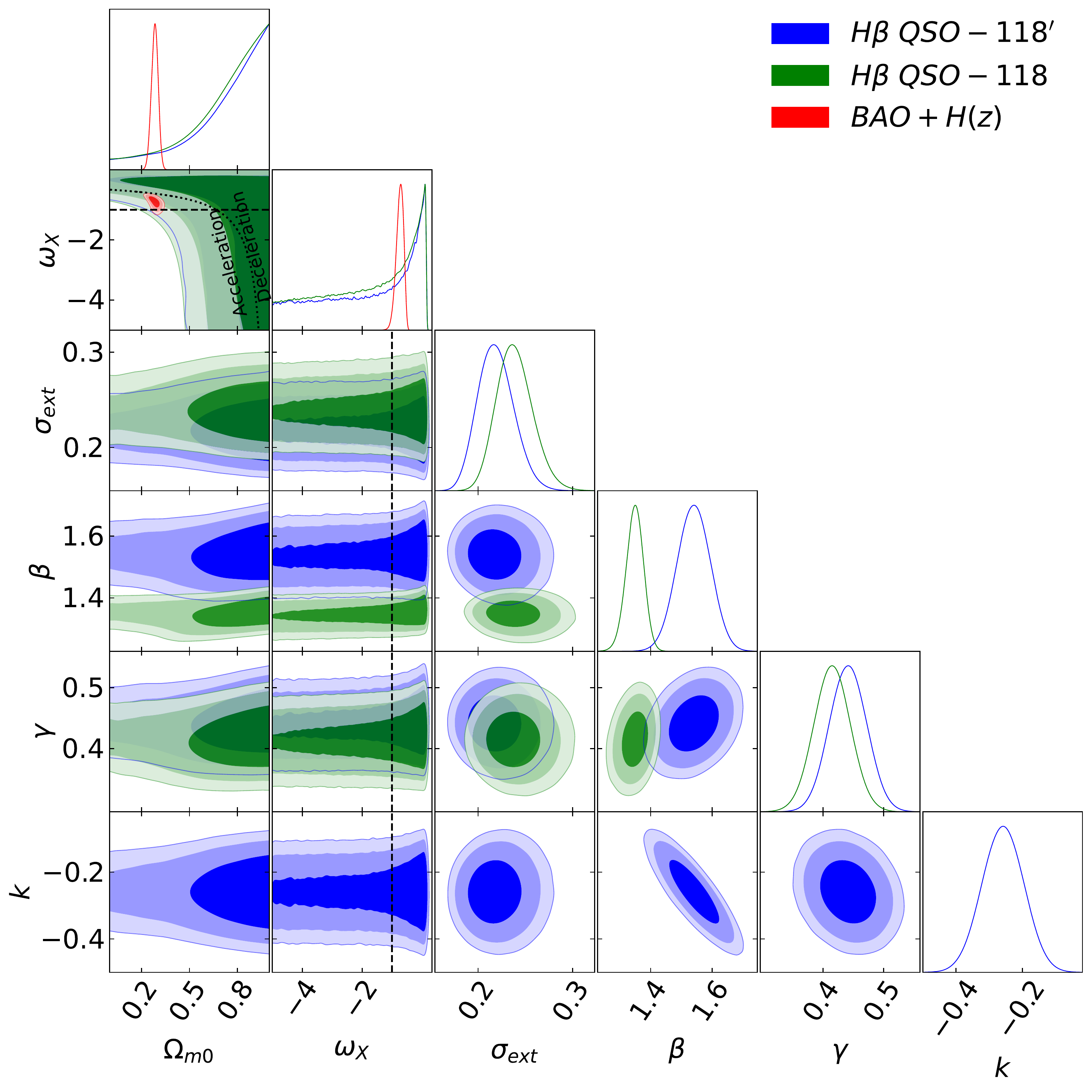}\par
    \includegraphics[width=\linewidth,height=7cm]{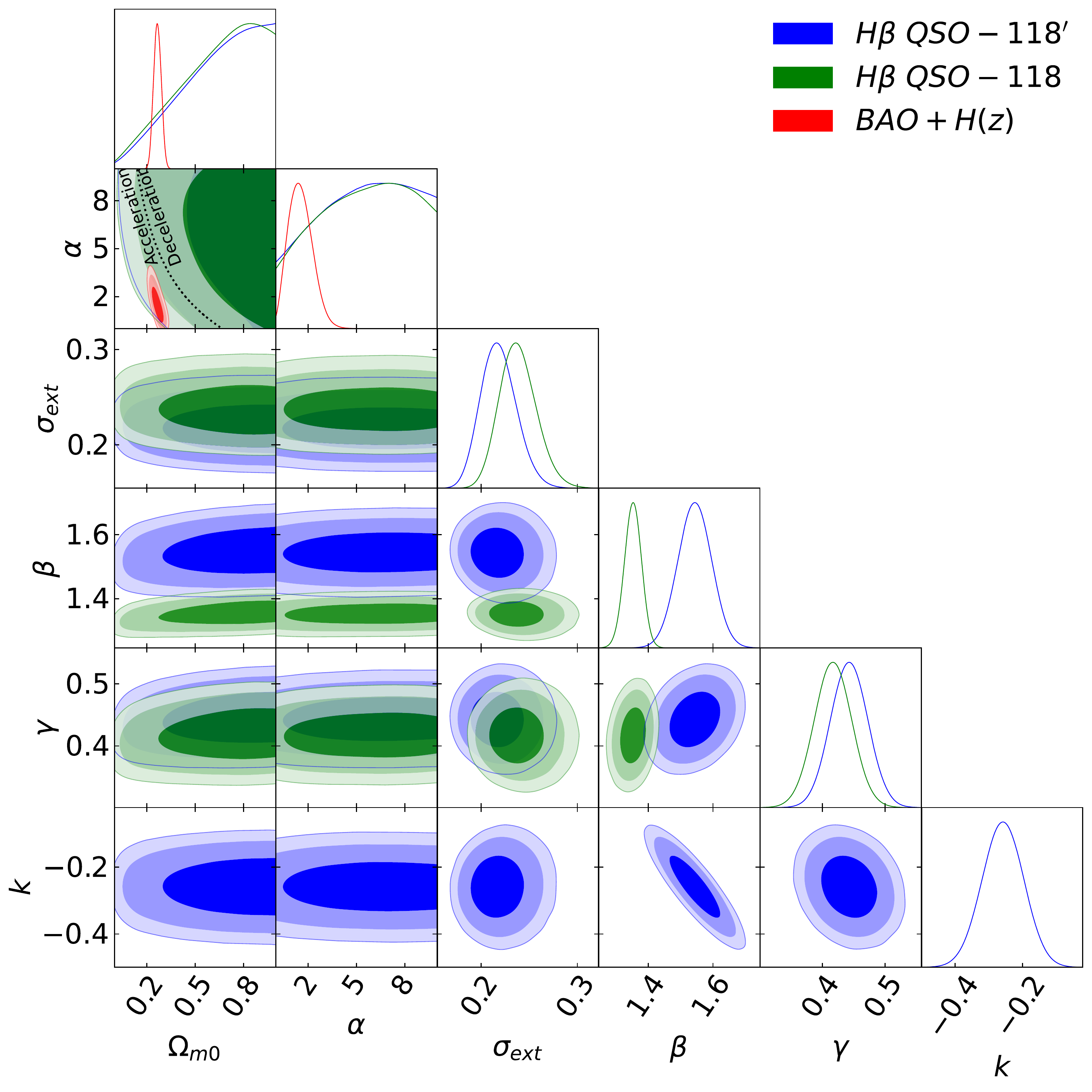}\par
    \includegraphics[width=\linewidth,height=7cm]{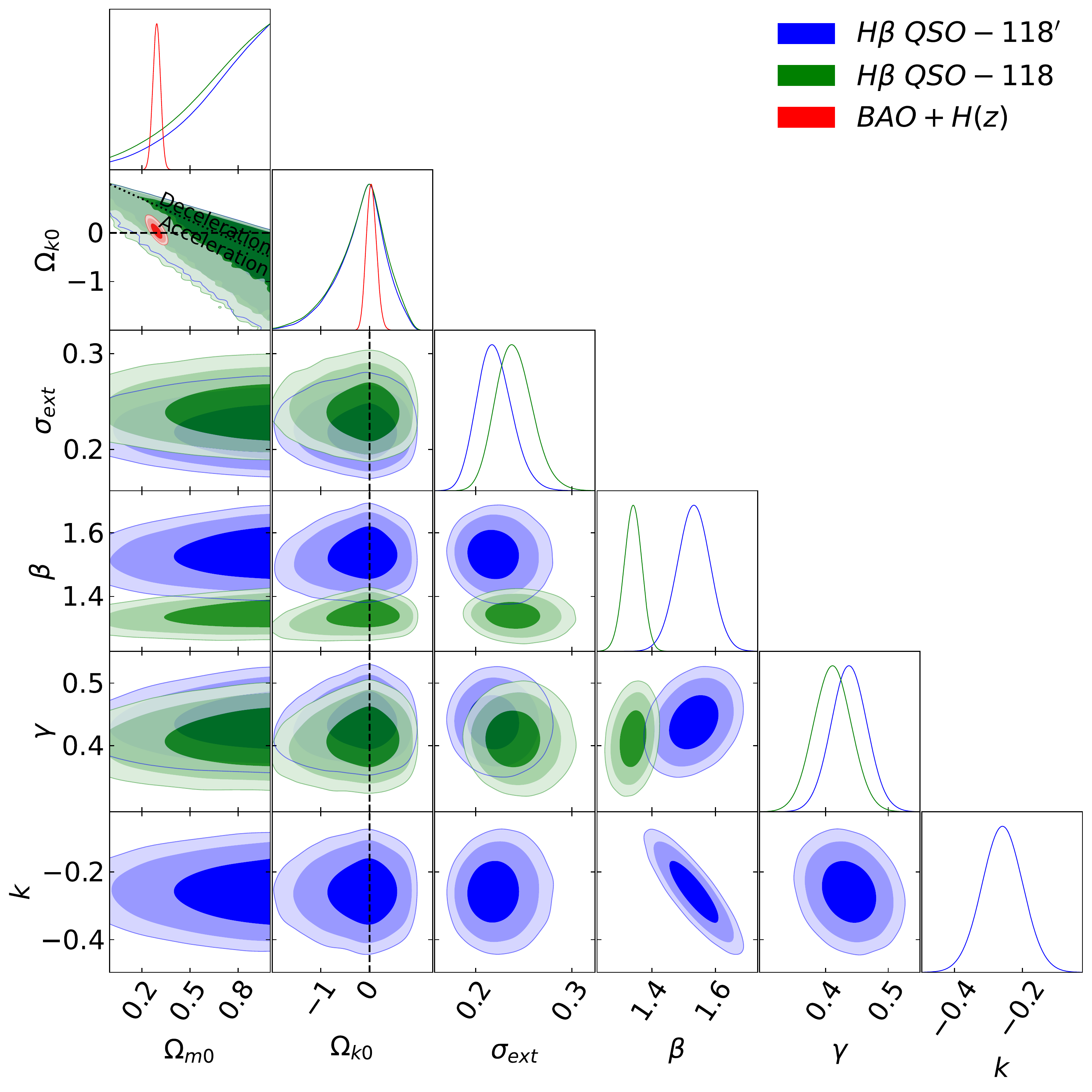}\par
    \includegraphics[width=\linewidth,height=7cm]{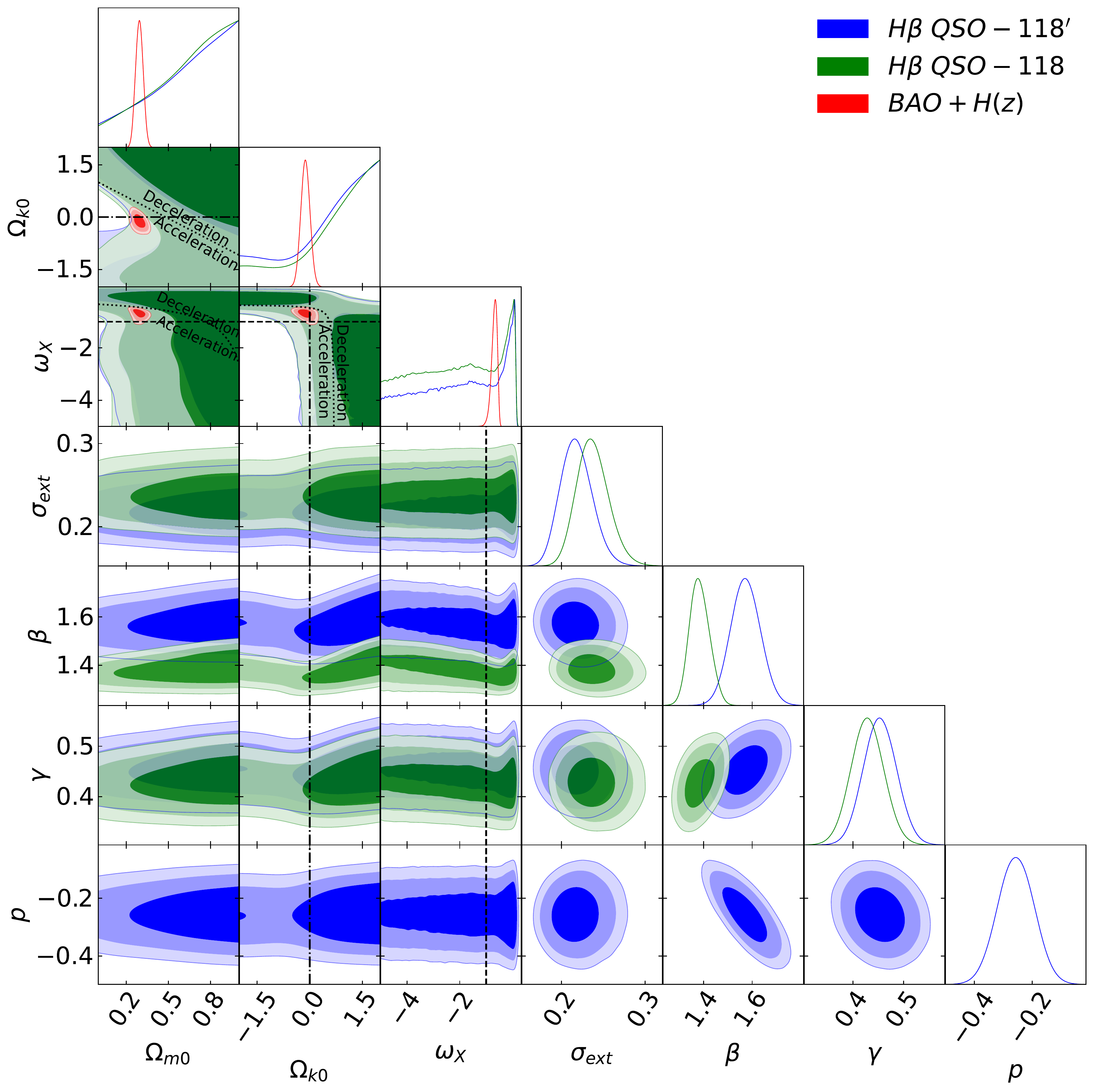}\par
    \includegraphics[width=\linewidth,height=7cm]{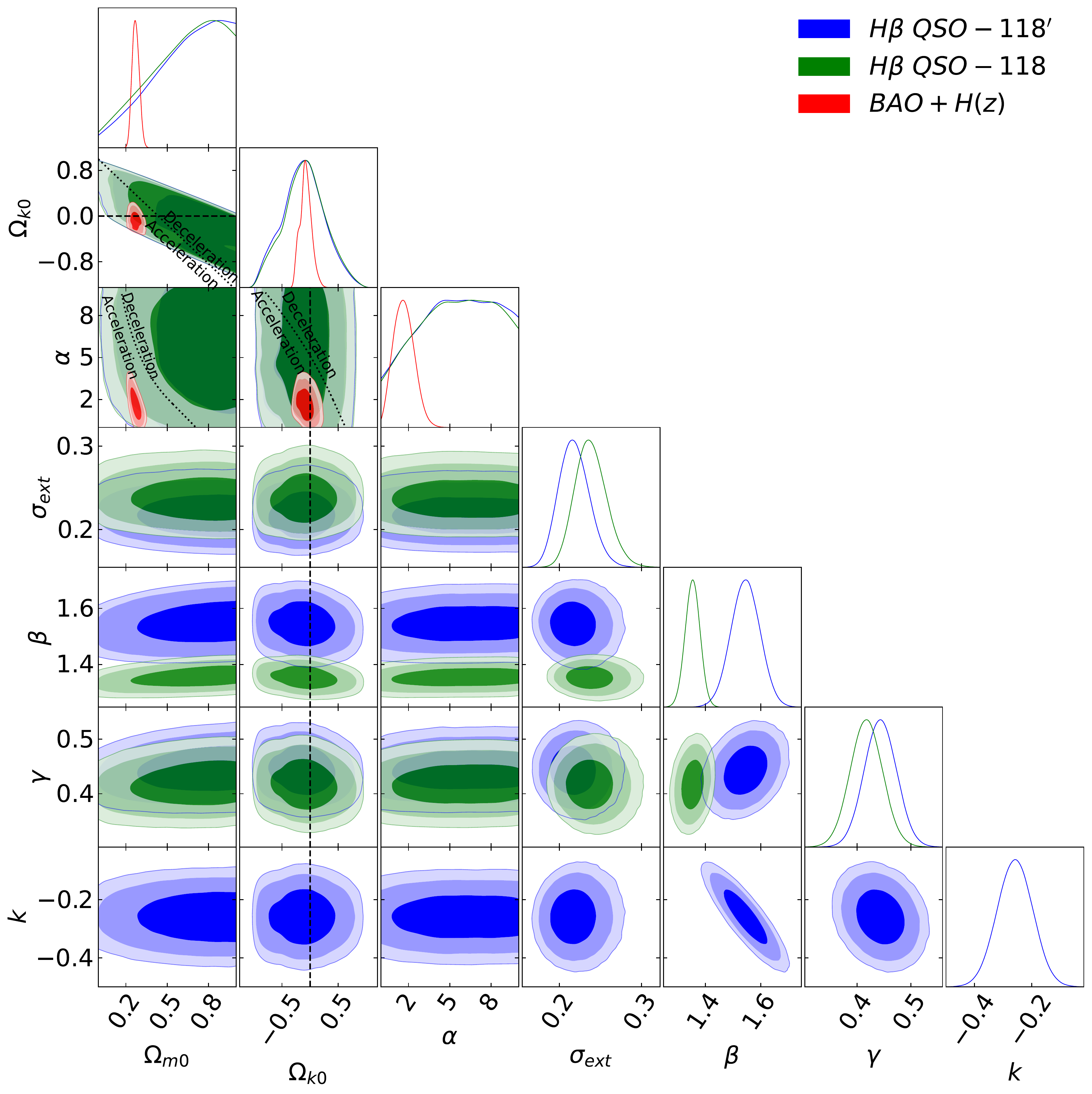}\par
\end{multicols}
\caption{One-dimensional likelihood distributions and two-dimensional likelihood contours at 1$\sigma$, 2$\sigma$, and 3$\sigma$ confidence levels using 3-parameter H$\beta$ QSO-118$^{\prime}$ (blue), 2-parameter H$\beta$ QSO-118 (green), and BAO + $H(z)$ (red) data for all free parameters. Left column shows the flat $\Lambda$CDM model, flat XCDM parametrization, and flat $\phi$CDM model respectively. The black dotted lines in all plots are the zero acceleration lines. The black dashed lines in the flat XCDM parametrization plots are the $\omega_X=-1$ lines. Right column shows the non-flat $\Lambda$CDM model, non-flat XCDM parametrization, and non-flat $\phi$CDM model respectively. Black dotted lines in all plots are the zero acceleration lines. Black dashed lines in the non-flat $\Lambda$CDM and $\phi$CDM model plots and black dotted-dashed lines in the non-flat XCDM parametrization plots correspond to $\Omega_{k0} = 0$. The black dashed lines in the non-flat XCDM parametrization plots are the $\omega_X=-1$ lines.}
\label{fig:Eiso-Ep1}
\end{figure*}

We see from Table \ref{tab:1d_BFP2}, and especially from Fig.\ \ref{fig:Eiso-Ep1}, that the most significant change in going from the 2-parameter to the 3-parameter $R-L$ relation when analyzing the full 118 sources data set is the $\sim 15$\% increase in the value of the intercept $\beta$ and an almost doubling of the $\beta$ error bars, which can be attributed to the degeneracy between $\beta$ and $k$, see Fig.~\ref{fig:Eiso-Ep1}.

A simple photoionization theory predicts a slope of $\gamma$ = 0.5 \citep{1993ApJ...404L..51N,2005ApJ...629...61K,2011A&A...525L...8C,2013peag.book.....N,2019arXiv190106507K}\footnote{The slope of $\gamma=0.5$ simply stems from the assumption that the BLR in each galaxy is characterized by the same constant ionization parameter, $U=Q(H)/[4\pi R_{\rm BLR}^2 n(H) c]$ and the BLR clouds, which are located at the distance of $R_{\rm BLR}$ from the ionizing source, have a comparable hydrogen number density $n(H)$. Since the hydrogen photon-ionizing flux $Q(H)=\int_{\nu_i}^{+\infty} L_{\rm \nu}/(h\nu)\mathrm{d}\nu\propto L_{\nu}$, then the assumption $Un(H)=\text{constant}$ leads to $R_{\rm BLR} \propto L_{\nu}^{1/2}$.} but we measure different, generally smaller values for $\gamma$. Quantitatively, in the 2-parameter $R-L$ relation case (H$\beta$ QSO-118 data) our measured values of $\gamma$ are $(2.18-2.97)\sigma$ lower than the prediction of the photoionization theory, while in the 3-parameter $R-L$ relation case (H$\beta$ QSO-118$^{\prime}$ data) our measured values of $\gamma$ are $(1.41-2.17)\sigma$ lower than the prediction of the photoionization theory. It appears that the inclusion of the third parameter $k$ in the $R-L$ relation partially corrects for the Eddington-ratio effect and increases the measured values of $\gamma$, which are then more compatible with the simple photoionization theory prediction, but there still are discrepancies which raise the question of whether a simple photoionization theory provides an adequate description of H$\beta$ emission lines, at least for the whole H$\beta$ quasar sample. The smaller slope may be due to several factors. As we show later, the smaller $\gamma$ is mostly exhibited by high-\rfe\ sources, i.e. higher accretors, which are more prone to reach ionization saturation for a given luminosity range in comparison with lower accretors \citep{guo2020}, and hence the increase in luminosity leads to a slower increase in the BLR distance. In addition, there may be additional factors determining the slope of the $R-L$ relation, namely differences in the spatial distribution of the line-emitting gas or the transfer function \citep{1995MNRAS.276..933R}, viewing angle as well as the spectral energy distribution shape  \citep{1997ApJ...490L.131W}, and occasionally, such as for NGC 5548, H$\beta$ variability is temporarily decoupled from the continuum variability --- so-called BLR holidays \citep{2019ApJ...882L..30D}.

\begin{figure*}
\begin{multicols}{2}
    \includegraphics[width=\linewidth,height=7cm]{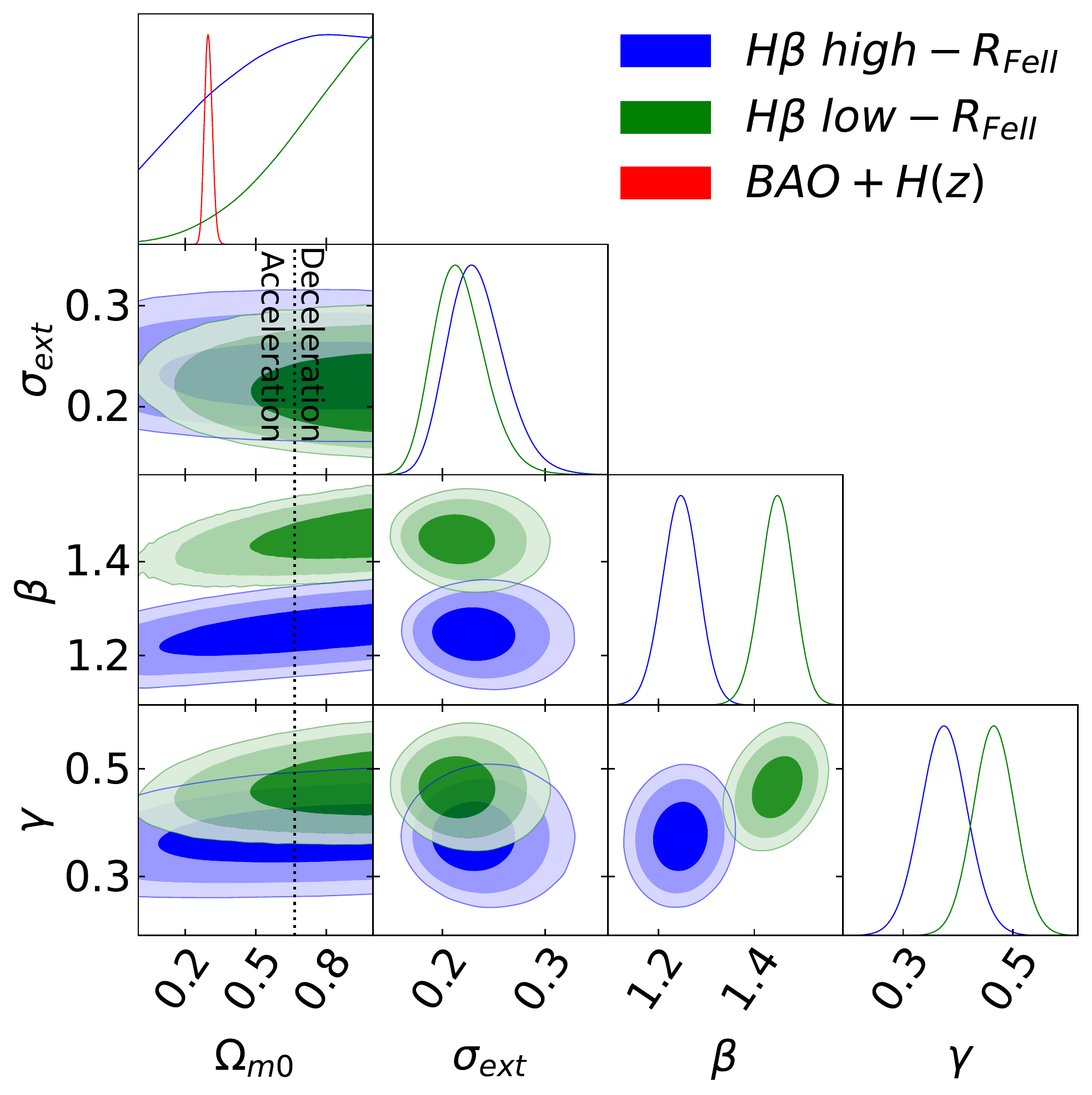}\par
    \includegraphics[width=\linewidth,height=7cm]{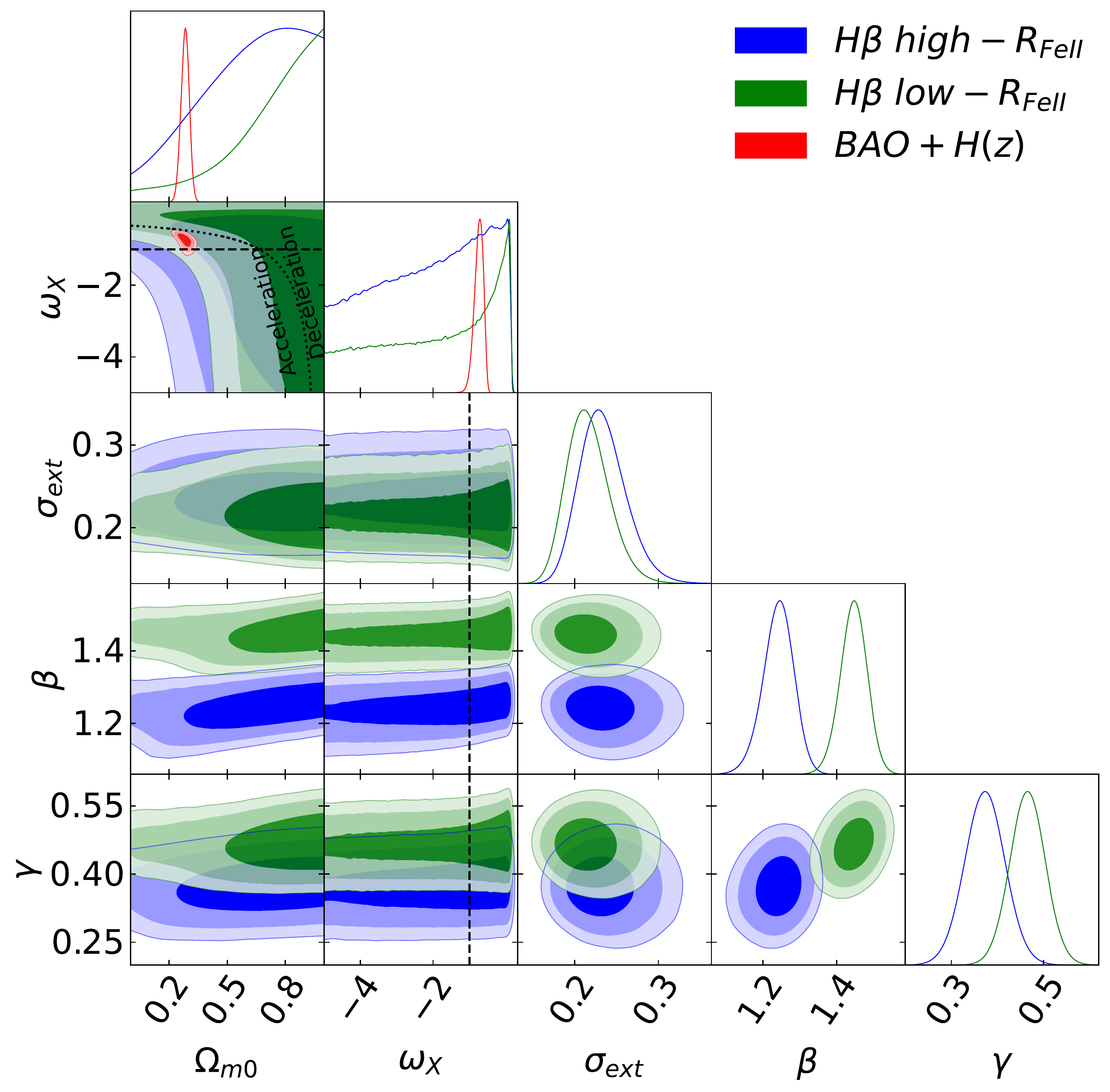}\par
    \includegraphics[width=\linewidth,height=7cm]{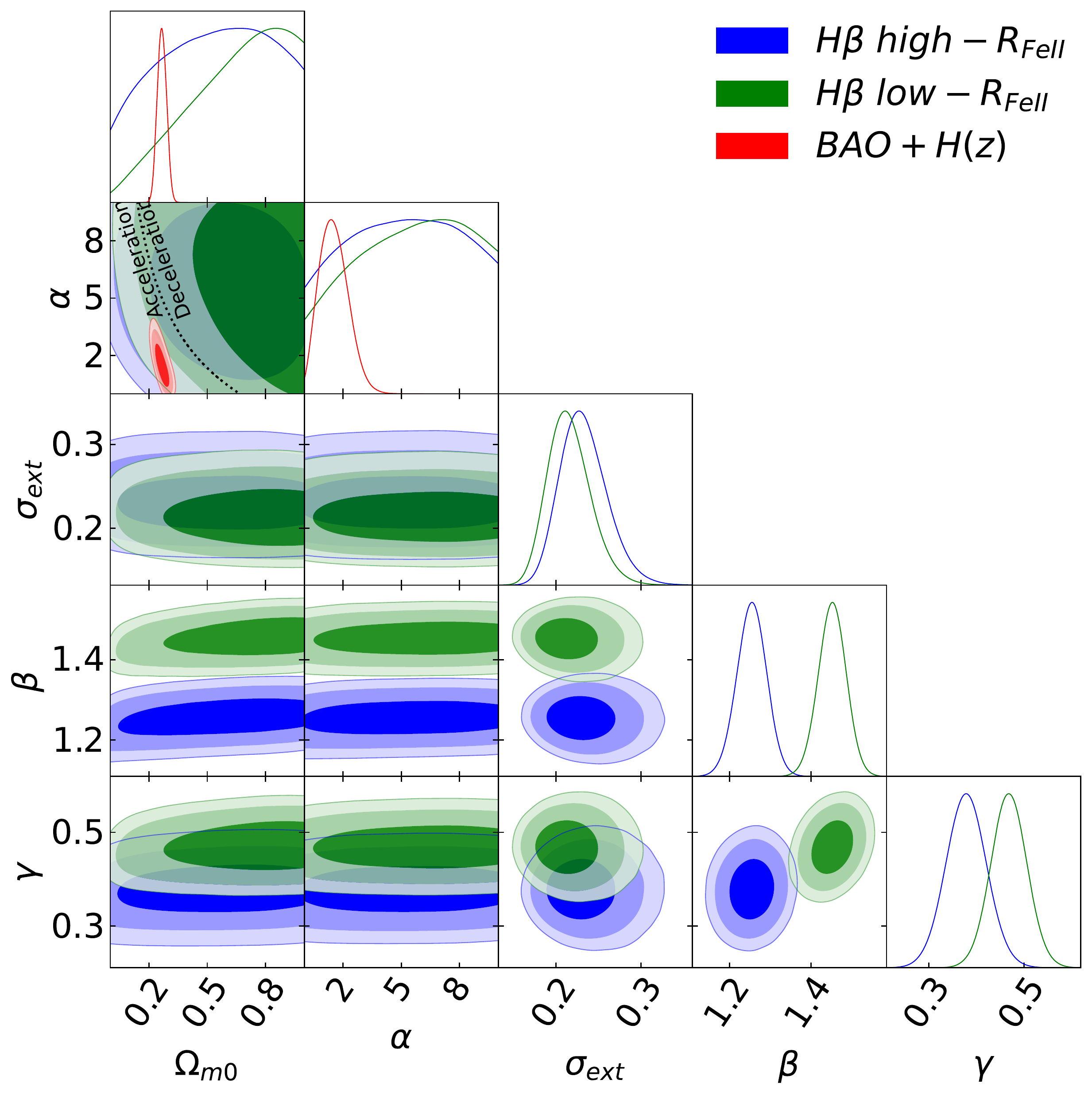}\par
    \includegraphics[width=\linewidth,height=7cm]{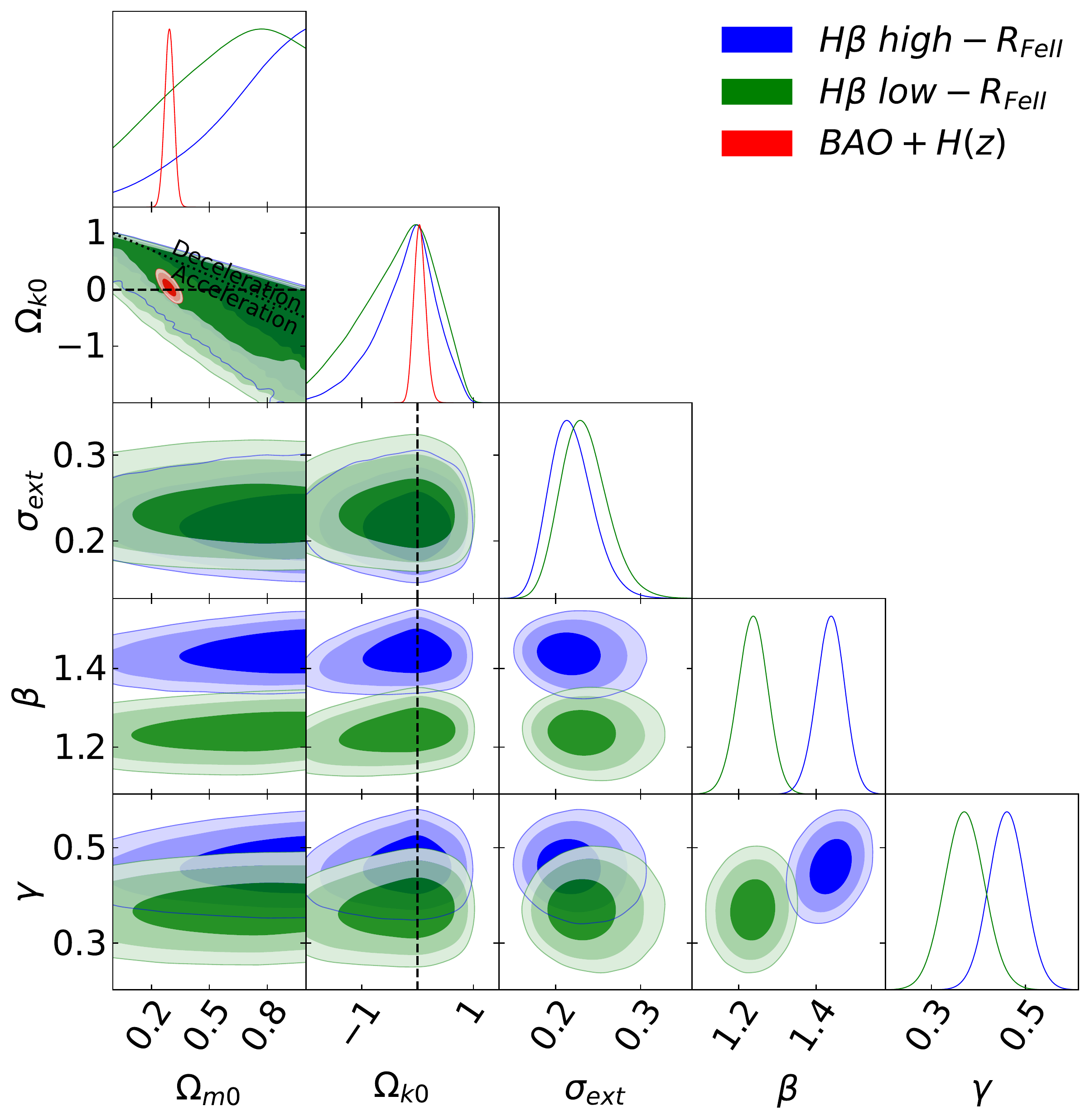}\par
    \includegraphics[width=\linewidth,height=7cm]{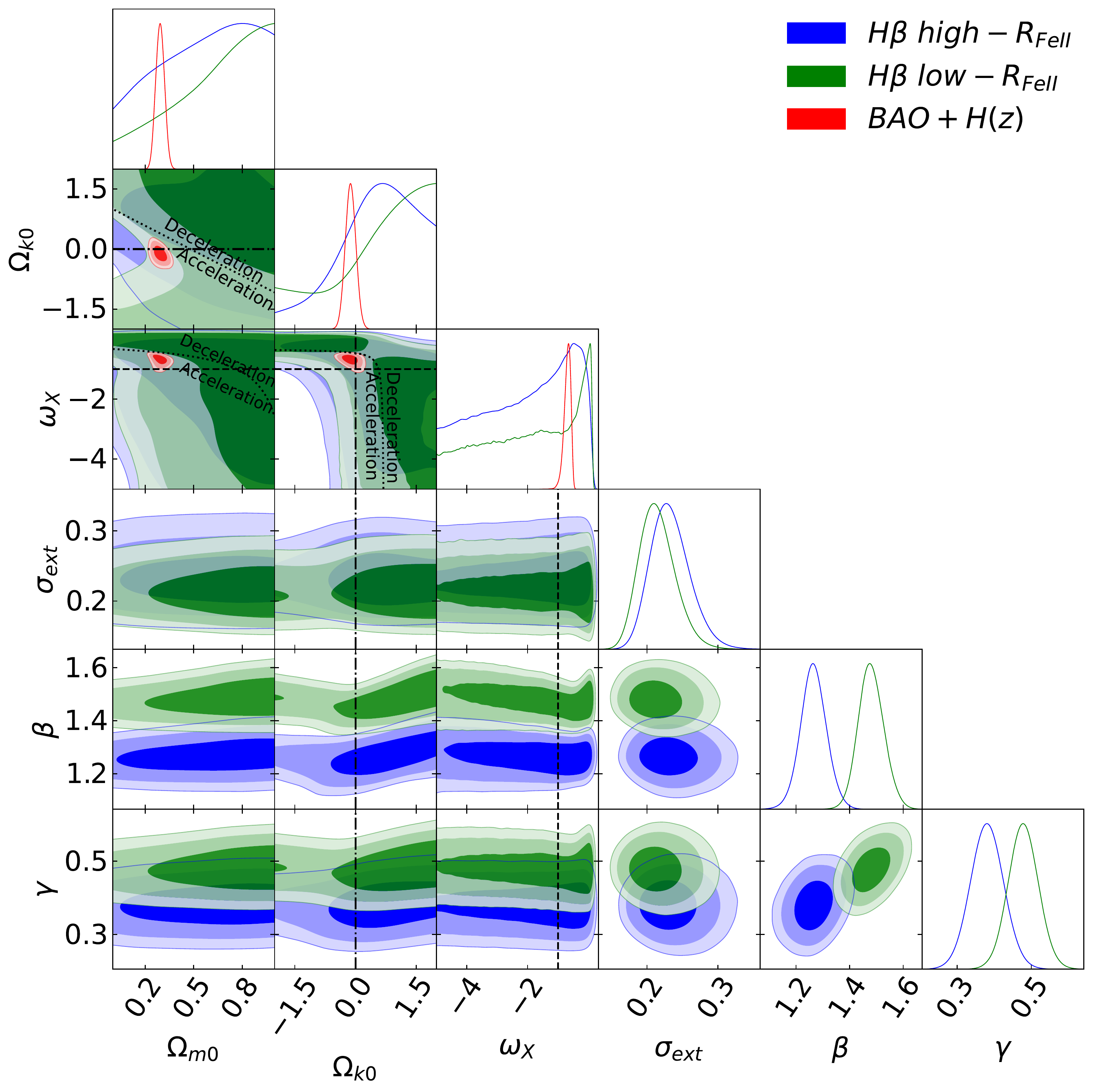}\par
    \includegraphics[width=\linewidth,height=7cm]{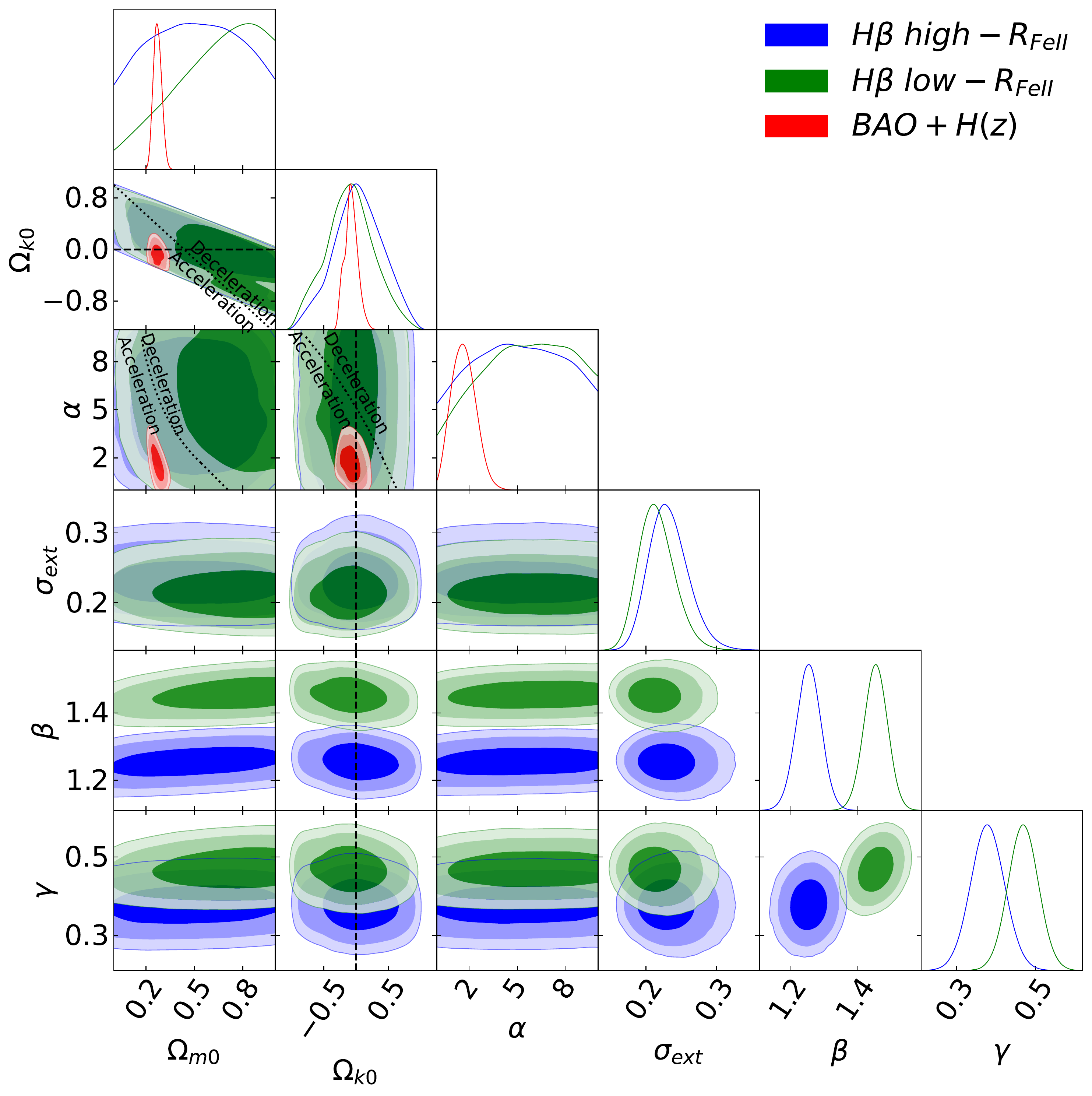}\par
\end{multicols}
\caption{One-dimensional likelihood distributions and two-dimensional likelihood contours at 1$\sigma$, 2$\sigma$, and 3$\sigma$ confidence levels using 2-parameter H$\beta$ high-\rfe\ (blue), 2-parameter H$\beta$ low-\rfe\ (green), and BAO + $H(z)$ (red) data for all free parameters. Left column shows the flat $\Lambda$CDM model, flat XCDM parametrization, and flat $\phi$CDM model respectively. The black dotted lines in all plots are the zero acceleration lines. The black dashed lines in the flat XCDM parametrization plots are the $\omega_X=-1$ lines. Right column shows the non-flat $\Lambda$CDM model, non-flat XCDM parametrization, and non-flat $\phi$CDM model respectively. Black dotted lines in all plots are the zero acceleration lines. Black dashed lines in the non-flat $\Lambda$CDM and $\phi$CDM model plots and black dotted-dashed lines in the non-flat XCDM parametrization plots correspond to $\Omega_{k0} = 0$. The black dashed lines in the non-flat XCDM parametrization plots are the $\omega_X=-1$ lines.}
\label{fig:Eiso-Ep2}
\end{figure*}

For the 2-parameter $R-L$ relation inferred using the low-\rfe\ data subset, in all cosmological models, the values of $\beta$ lie in the range $1.439^{+0.034}_{-0.037}$ to $1.475^{+0.050}_{-0.044}$ and the values of $\gamma$ lie in the range $0.460^{+0.039}_{-0.038}$ to $0.478^{+0.041}_{-0.039}$. The difference between the largest and the smallest central values of $\beta$ is 0.65$\sigma$ while this difference for $\gamma$ values is 0.33$\sigma$ and these differences are not statistically significant. For the 2-parameter $R-L$ relation obtained using the H$\beta$ high-\rfe\ data subset, in all cosmological models, the values of $\beta$ lie in the range $1.239^{+0.037}_{-0.040}$ to $1.263^{+0.050}_{-0.045}$ and the values of $\gamma$ lie in the range $0.369^{+0.043}_{-0.041}$ to $0.380^{+0.045}_{-0.044}$. The difference between the largest and the smallest central values of $\beta$ is 0.41$\sigma$ while this difference for $\gamma$ values is 0.18$\sigma$ and these differences are also not statistically significant. There are, however, differences between $\beta$ and $\gamma$ values obtained using the 2-parameter $R-L$ relation based on H$\beta$ low-\rfe\ and high-\rfe\ data subsets and these differences can be seen in Fig.\ \ref{fig:Eiso-Ep2} and they are listed in Table~\ref{tab:comp1}. From Table~\ref{tab:comp1}, in all six cosmological models, the difference in $\beta$ values $(\Delta \beta)$ lies in the range $(3.18-4.06)\sigma$ which is statistically significant and the difference in $\gamma$ values $(\Delta \gamma)$ lies in the range $(1.59-1.65)\sigma$ which could be statistically significant.\footnote{We emphasize that these computations assume that the $\beta$ and $\gamma$ (as well as $k$) values are uncorrelated, which is not correct, and so the $\Delta \beta$ and $\Delta \gamma$ (as well as $\Delta k$) values listed in Tables \ref{tab:comp1} and \ref{tab:comp2} must be viewed as qualitative indicators of the differences.} These differences show that H$\beta$ low-\rfe\ and H$\beta$ high-\rfe\ QSOs obey different 2-parameter $R-L$ correlation relations.

\begin{table}
\begin{threeparttable}
\caption{Two-parameter $R-L$ relation parameters (and $\sigma_{\rm ext}$) differences, in various cosmological models, for the $H\beta$ low-\rfe\ and $H\beta$ high-\rfe\ data sets.}
\label{tab:comp1}
\setlength{\tabcolsep}{14.5pt}
\begin{tabular}{lccc}
\hline
Model & $\Delta \sigma_{\rm ext}$  & $\Delta \gamma$  & $\Delta \beta$\\
\hline
Flat \lcdm\  & $0.44\sigma$ & $1.60\sigma$ & $3.91\sigma$\\
Non-flat \lcdm\ & $0.43\sigma$ & $1.59\sigma$ & $3.82\sigma$ \\
Flat XCDM  & $0.46\sigma$ & $1.60\sigma$ & $3.76\sigma$\\
Non-flat XCDM  & $0.46\sigma$ & $1.65\sigma$ & $3.18\sigma$\\
Flat $\phi$CDM  & $0.44\sigma$ & $1.59\sigma$ & $4.06\sigma$\\
Non-flat $\phi$CDM & $0.43\sigma$ & $1.59\sigma$ & $4.06\sigma$\\
\hline
\end{tabular}
\end{threeparttable}
\end{table}

\begin{table}
\begin{threeparttable}
\caption{Three-parameter $R-L$ relation parameters (and $\sigma_{\rm ext}$) differences, in various cosmological models, for the $H\beta^{\prime}$ low-\rfe\ and $H\beta^{\prime}$ high-\rfe\ data sets.}
\label{tab:comp2}
\setlength{\tabcolsep}{10pt}
\begin{tabular}{lcccc}
\hline
Model & $\Delta \sigma_{\rm ext}$  & $\Delta \gamma$  & $\Delta \beta$ & $\Delta k$\\
\hline
Flat \lcdm\  & $0.46\sigma$ & $1.21\sigma$ & $1.10\sigma$ & $0.64\sigma$\\
Non-flat \lcdm\ & $0.72\sigma$ & $1.23\sigma$ & $1.18\sigma$ & $0.73\sigma$ \\
Flat XCDM  & $0.47\sigma$ & $1.22\sigma$ & $1.09\sigma$ & $0.62\sigma$\\
Non-flat XCDM  & $0.44\sigma$ & $1.18\sigma$ & $0.97\sigma$ & $0.46\sigma$\\
Flat $\phi$CDM  & $0.43\sigma$ & $1.21\sigma$ & $1.06\sigma$ & $0.58\sigma$\\
Non-flat $\phi$CDM & $0.46\sigma$ & $1.21\sigma$ & $1.06\sigma$ & $0.58\sigma$\\
\hline
\end{tabular}
\end{threeparttable}
\end{table}

For the extended, 3-parameter $R-L$ relation obtained based on the low-\rfe\ data subset, in all cosmological models, the values of $\beta$ lie in the range $1.570^{+0.103}_{-0.100}$ to $1.589^{+0.102}_{-0.103}$, the values of $\gamma$ lie in the range $0.464^{+0.039}_{-0.040}$ to $0.478^{+0.043}_{-0.041}$, and the values of $k$ lie in the range $-0.322^{+0.225}_{-0.224}$ to $-0.266^{+0.223}_{-0.229}$. The difference between the largest and the smallest central values of $\beta$ is 0.13$\sigma$ while this difference for $\gamma$ and $k$ values is 0.25$\sigma$ and 0.17$\sigma$, respectively, and these differences are not statistically significant. For the 3-parameter $R-L$ relation inferred using H$\beta^{\prime}$ high-\rfe\ data subset, in all cosmological models, the values of $\beta$ lie in the range $1.378^{+0.128}_{-0.122}$ to $1.425^{+0.135}_{-0.126}$, the values of $\gamma$ lie in the range $0.387^{+0.048}_{-0.045}$ to $0.401^{+0.051}_{-0.047}$, and the values of $k$ lie in the range $-0.152^{+0.113}_{-0.109}$ to $-0.136^{+0.110}_{-0.116}$. The difference between the largest and the smallest central values of $\beta$ is 0.26$\sigma$ while this difference for $\gamma$ and $k$ values is 0.21$\sigma$ and 0.10$\sigma$, respectively, and these differences are also not statistically significant. In this 3-parameter $R-L$ relation case, the differences in corresponding $\beta$, $\gamma$, and $k$ values obtained using the H$\beta^{\prime}$ low-\rfe\ and high-\rfe\ data subsets are relatively low and are listed in Table~\ref{tab:comp2} and can be seen in Fig.\ \ref{fig:Eiso-Ep3}. In all six cosmological models, the difference in $\beta$ values $(\Delta \beta)$ lies in the range $(0.97-1.18)\sigma$ which is not very statistically significant, the difference in $\gamma$ values $(\Delta \gamma)$ lies in the range $(1.18-1.23)\sigma$ which is not very statistically significant, and the difference in $k$ values $(\Delta k)$ lies in the range $(0.46-0.73)\sigma$ which is not statistically significant. Compared to the 2-parameter $R-L$ relation case, the inclusion of $k$, the third parameter, has resulted in much larger $\beta$ parameter error bars, bringing the high-\rfe\ and low-\rfe\ $\beta$ central values in better agreement; note that the $\beta$ central values and, probably more importantly, the differences between the $\beta$ central values have changed by a much smaller factor than have the error bars. These relatively smaller $\Delta \beta$ and $\Delta \gamma$ (as well as $\Delta k$) differences indicate that the H$\beta^{\prime}$ low-\rfe\ and high-\rfe\ QSOs obey similar 3-parameter $R-L$ relations. Hence, in comparison with the 2-parameter $R-L$ relation for the two subsets, the inclusion of \rfe\ in the 3-parameter case led to a partial correction of the accretion-rate effect.

We see, from Table \ref{tab:1d_BFP2}, and especially from Figs.\ \ref{fig:Eiso-Ep4} and \ref{fig:Eiso-Ep5}, that the most significant change in going from the 2-parameter to the 3-parameter $R-L$ relation when analyzing the 59 sources low-\rfe\ (high-\rfe) data subset is the $\sim 8-9$\% ($\sim 11-13$\%) increase in the value of the intercept $\beta$ and an almost tripling of the $\beta$ error bars. Interestingly, in the 3-parameter $R-L$ analyses, $k$ is only $(1.2-1.4)\sigma$ and $(1.2-1.3)\sigma$ away from zero in the H$\beta^\prime$ low-\rfe\ and high-\rfe\ cases, while in the full H$\beta$ QSO-118$^\prime$ analyses it is $(4.1-4.2)\sigma$ away from zero. This is consistent with what we find from the $AIC$ and $BIC$ values, discussed below, which also indicate that the 3-parameter $R-L$ relation is a very significantly better fit than the 2-parameter one only for the full 118 source data set. 

From Table \ref{tab:1d_BFP2}, in the 2-parameter $R-L$ relation case, for the low-\rfe\ data subsets, the measured values of $\gamma$ are $(0.54-1.03)\sigma$ lower than the prediction of photoionization theory, which is statistically not significant, while for the high-\rfe\ data subsets, the measured values of $\gamma$ are $(2.67-3.05)\sigma$ lower than the prediction of photoionization theory, which is statistically significant. In the 3-parameter $R-L$ relation case, for the low-\rfe\ data subsets, the measured values of $\gamma$ are $(0.51-0.92)\sigma$ lower than the prediction of photoionization theory, which is statistically not significant, while for the high-\rfe\ data subsets, the measured values of $\gamma$ are $(1.94-2.35)\sigma$ lower than the prediction of photoionization theory, which is statistically significant. These differences show that the low-\rfe\ data subset is consistent with the prediction of the simple photoionization theory while the high-\rfe\ data subset shows statistically significant discrepancies with the photoionization theory, i.e. the assumption that the product of the ionization parameter and the BLR cloud density across these sources is constant is not valid. More interestingly, while the inclusion of the third parameter $k$ in the 3-parameter $R-L$ relation does bring the high-\rfe\ $\gamma$ values closer to 0.5, they are still discrepant at $\sim 2\sigma$. It appears that the high-\rfe\ subset is largely the cause for the discrepancy between the measured $\gamma$ values for the full 118 sources data and simple photoionization theory. As we mentioned above, this is linked to the higher accretion since \rfe\ and the Eddington ratio are significantly correlated, especially for high-\rfe\ sources, see Fig.~\ref{fig:delta_rfe} (left panel).  

For all data sets, in all six cosmological models, the value of the intrinsic dispersion $\sigma_{\rm ext}$ lies in the range $0.208^{+0.028}_{-0.025}$ to $0.237^{+0.020}_{-0.018}$. The minimum value is obtained in the non-flat $\phi$CDM model using H$\beta^{\prime}$ low-\rfe\ data while the maximum value is obtained in the non-flat $\phi$CDM model using the H$\beta$ QSO-118 data set. In each data set or subset, the minimum $\sigma_{\rm ext}$ value is obtained in the 3-parameter $R-L$ relation case. For the full 118 sources data set $\sigma_{\rm ext}$ is approximately 0.75$\sigma$ (of the quadrature sum of the two error bars) lower in the 3-parameter $R-L$ relation case compared to the 2-parameter case. This is not a statistically significant difference and hence, with the current data, the inclusion of one extra parameter, $k$, in the $R-L$ relation does not result in a significant reduction of the intrinsic dispersion. For the high- and the low-\rfe\ data subsets, the reductions in $\sigma_{\rm ext}$, when going from the 2-parameter $R-L$ relation to the 3-parameter one, are even less significant than for the full 118 sources data set. Taken together, this means that the extra $k$ parameter is less effective at reducing the intrinsic dispersion of source subsets that probe narrower ranges of \rfe.    

From Table \ref{tab:BFP}, for the H$\beta$ QSO-118 and H$\beta$ QSO-118$^{\prime}$ data sets, from the $\Delta AIC$ and $\Delta BIC$ values, the 3-parameter $R-L$ relation is very strongly favored over the 2-parameter $R-L$ relation. From $\Delta BIC$ values for both pairs of high- and low-\rfe\ data subsets, the 2-parameter $R-L$ relation is positively favored over the 3-parameter $R-L$ relation, except in the flat XCDM parameterization where the 3-parameter $R-L$ relation is weakly favored. On the other hand, $\Delta AIC$ values for both pairs of high- and low-\rfe\ data subsets show only weak evidence, in both directions, depending on a cosmological model. It is somewhat puzzling that while the introduction of the third parameter $k$ is strongly favored by the complete 118 source data set, it is not favored by either the 59 source high-\rfe\ data subset or by the 59 source low-\rfe\ data subset. This is likely the consequence of the moderate correlation between the offset $\Delta \tau$ and the \rfe\ parameter for the complete 118 data (See Fig.~\ref{fig:delta_rfe} right panel), while for both low- and high-\rfe\ subsets the correlation is not statistically significant. For the high-\rfe\ subset, the correlation between $\Delta \tau$ and \rfe\ appears to be stronger than the same correlation for the low-\rfe\ subset by a factor of about two in terms of Spearman's rank correlation coefficient, although just below the significance level based on the $p$-value. This is likely the reason behind the persistent small negative $\Delta AIC$ values for the 3-parameter high-\rfe\ subset case, in comparison with the 2-parameter case, for all cosmological models.

In all cases, in both the 2-parameter and 3-parameter $R-L$ relation analyses, for the full 118 source data set, and for the high- and low-\rfe\ data subsets, the $R-L$ relation parameters are largely independent of the cosmological model used in the analysis, so these H$\beta$ QSOs are standardizable through the $R-L$ relation. However, the different measured intercept $\beta$ values in the three data sets and the differences between the measured values of the slope $\gamma$ in the high- and low-\rfe\ data subsets are causes for concern.

\subsection{Cosmological model parameter measurements}

\begin{figure*}
\begin{multicols}{2}
    \includegraphics[width=\linewidth,height=7cm]{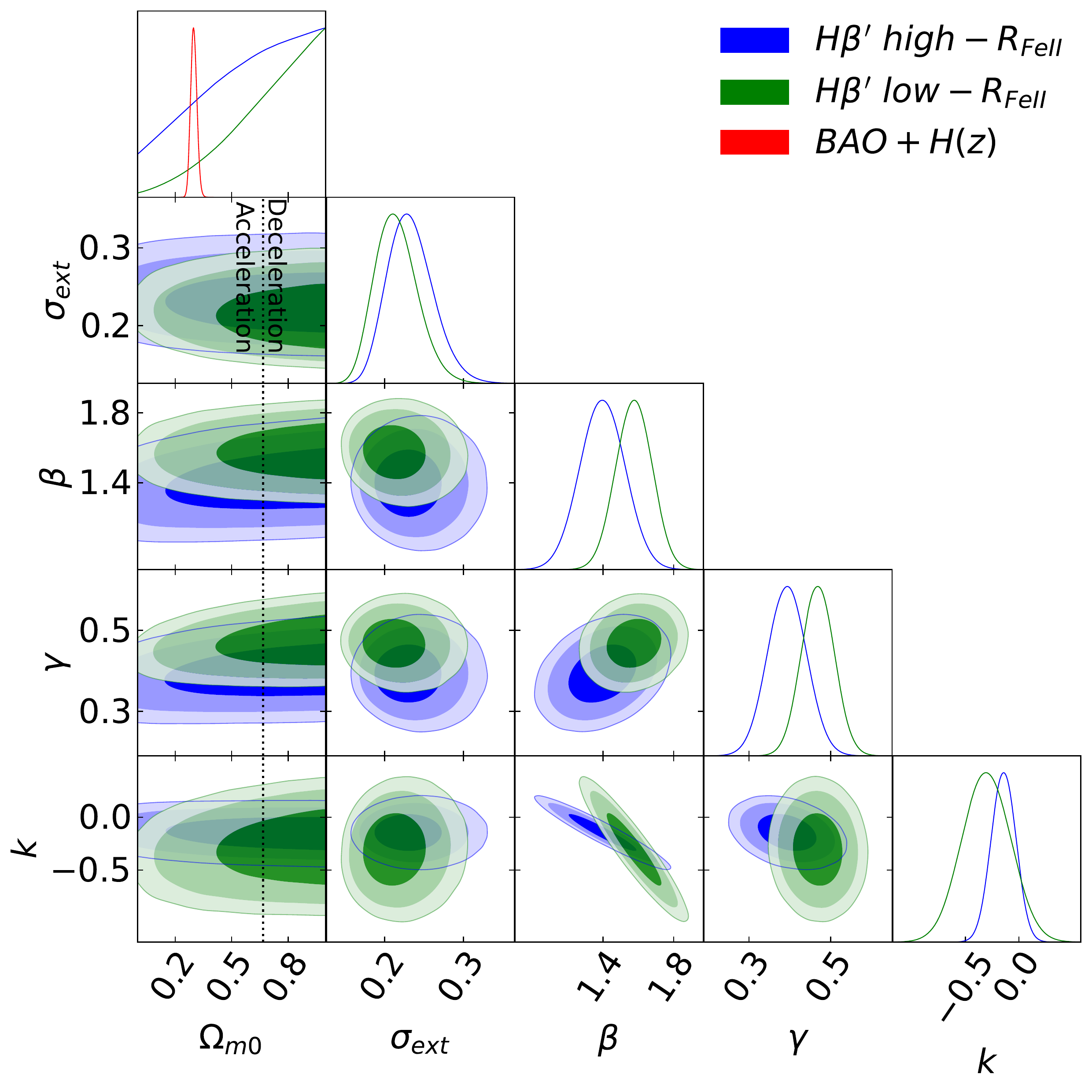}\par
    \includegraphics[width=\linewidth,height=7cm]{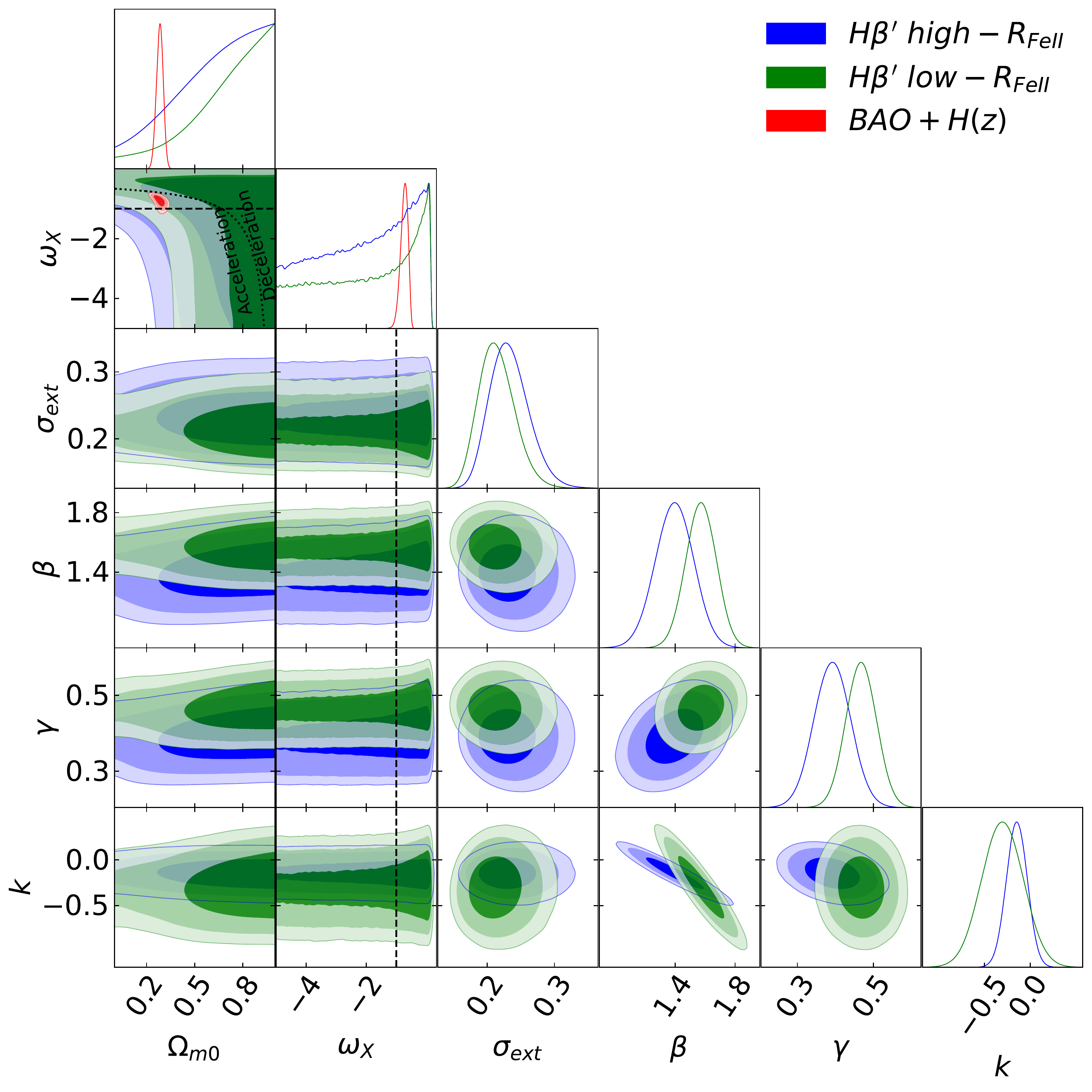}\par
    \includegraphics[width=\linewidth,height=7cm]{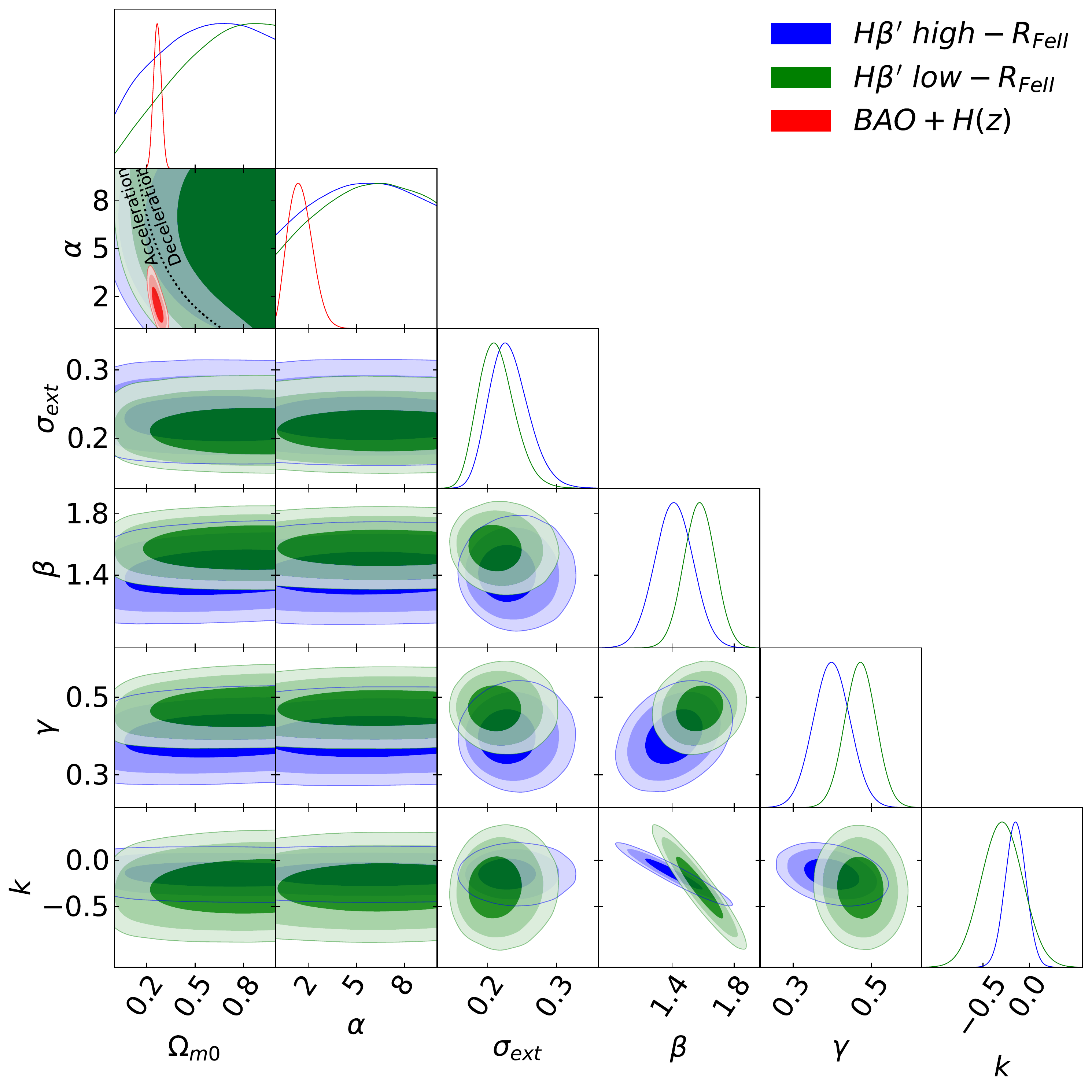}\par
    \includegraphics[width=\linewidth,height=7cm]{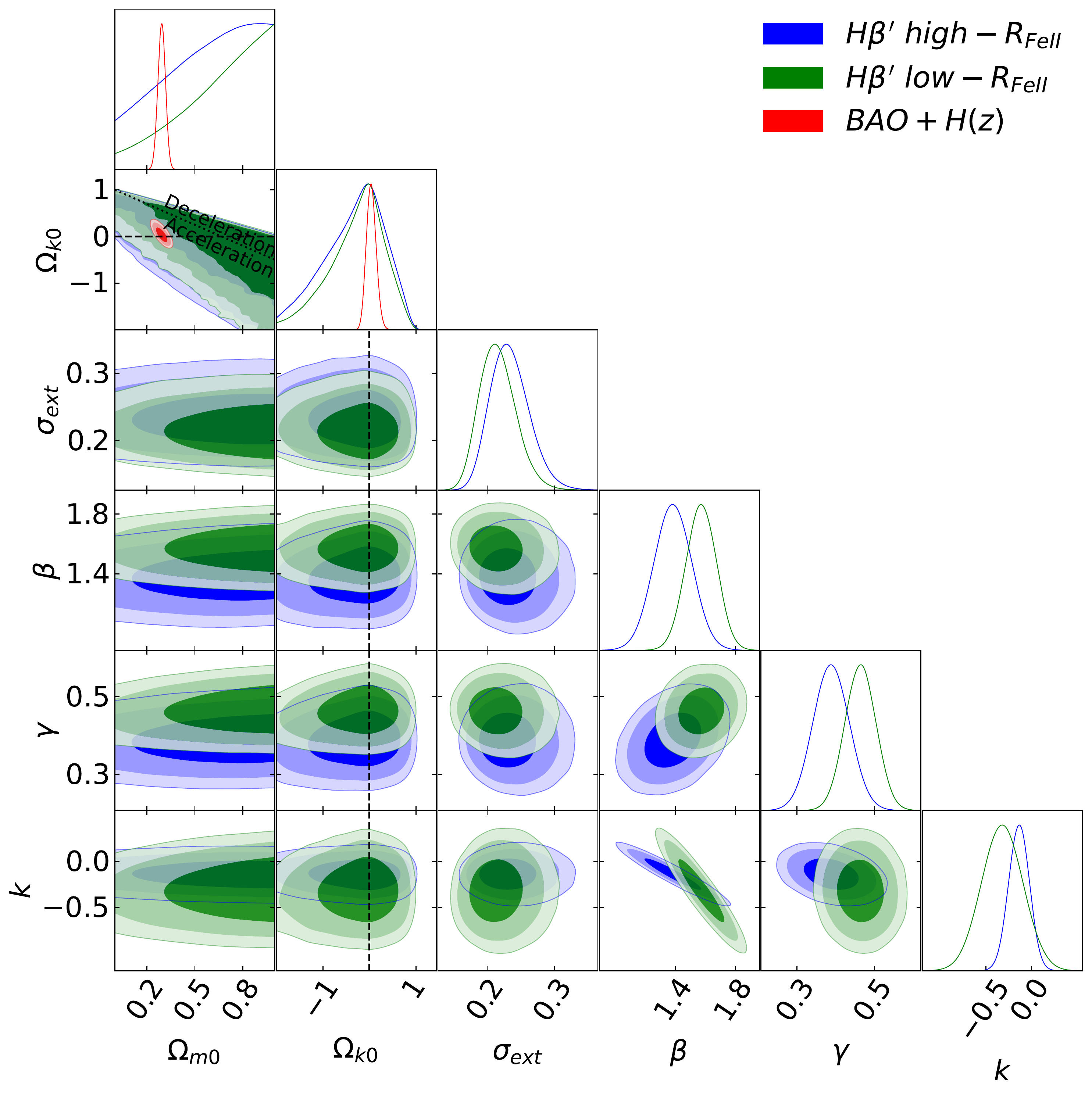}\par
    \includegraphics[width=\linewidth,height=7cm]{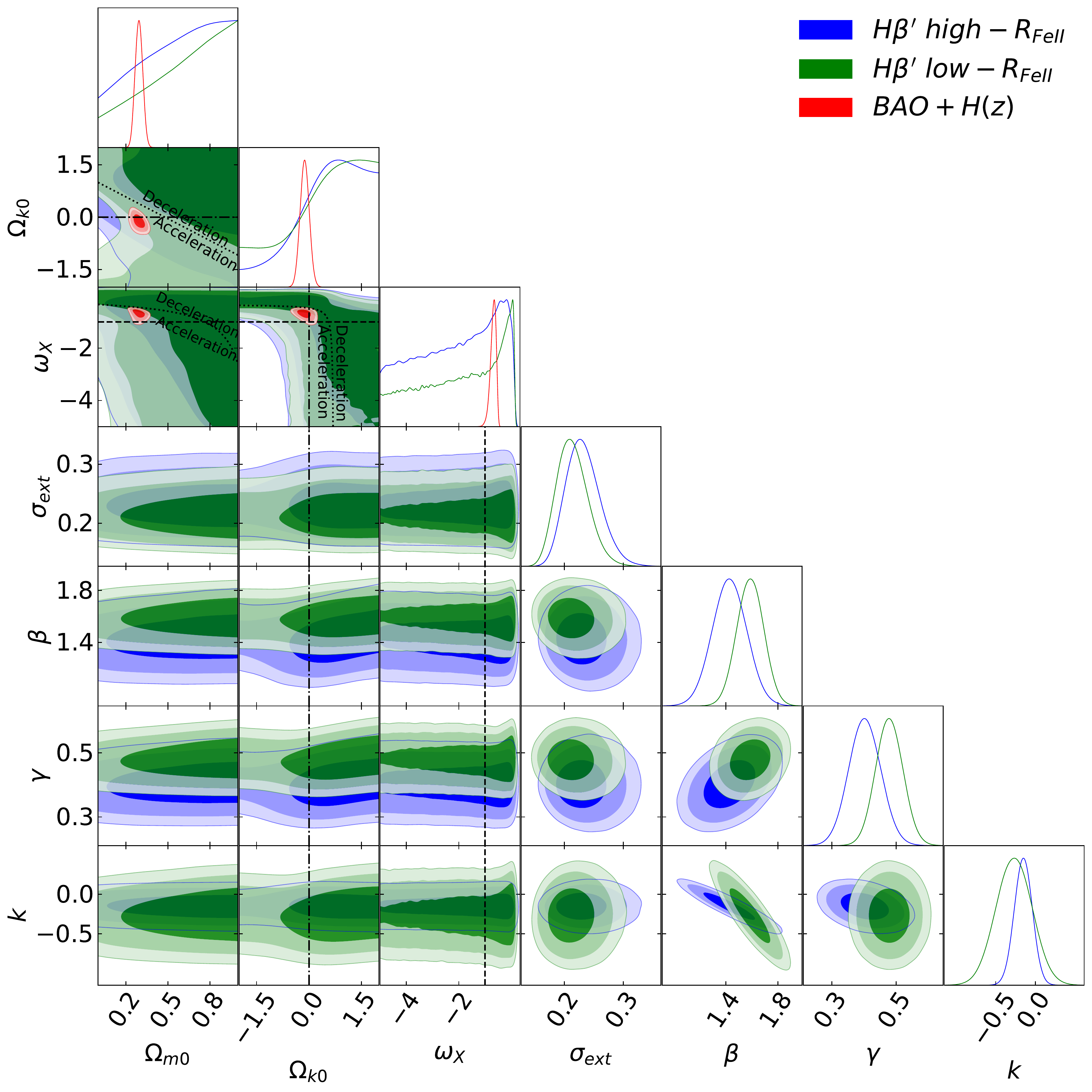}\par
    \includegraphics[width=\linewidth,height=7cm]{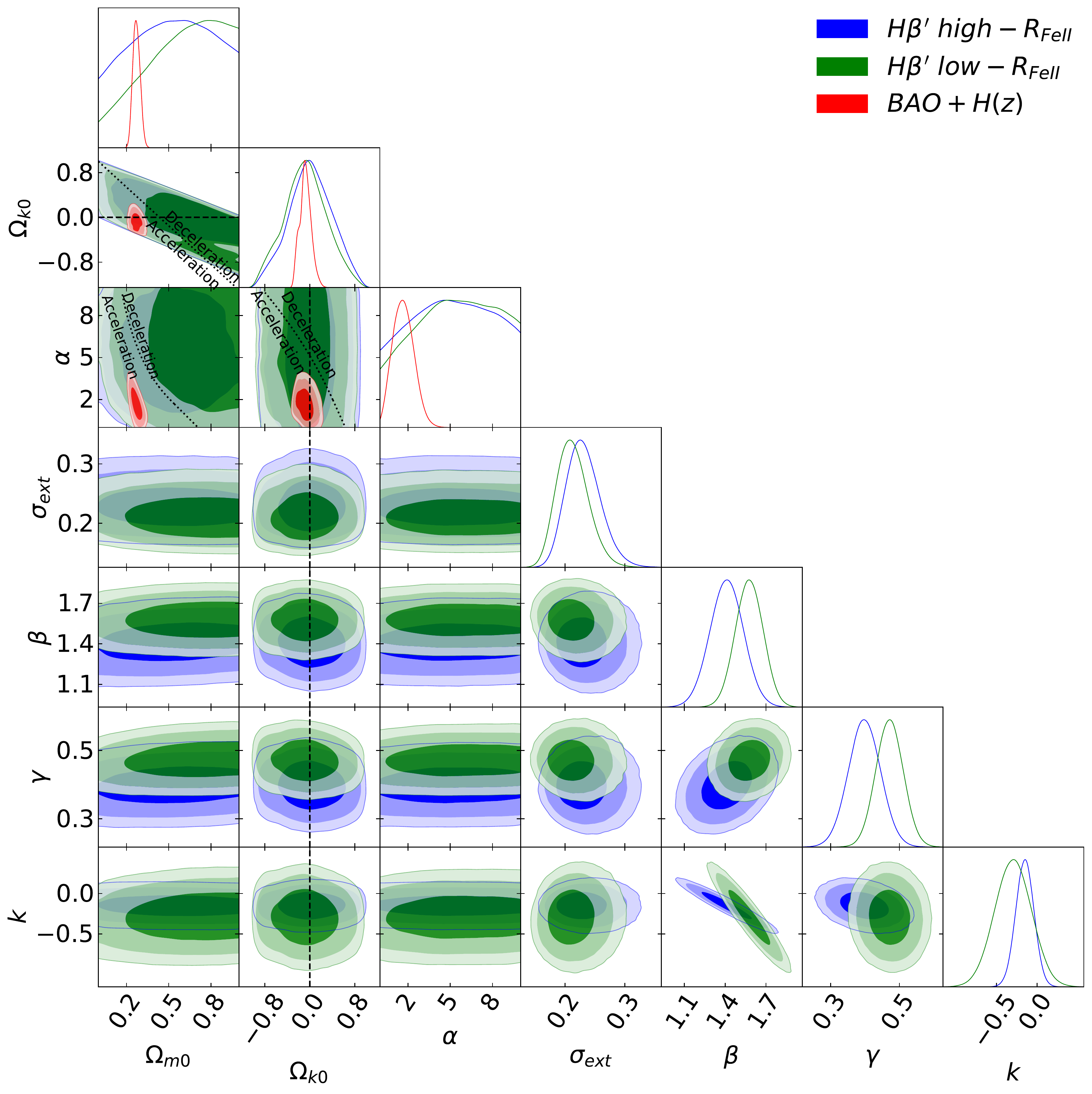}\par
\end{multicols}
\caption{One-dimensional likelihood distributions and two-dimensional likelihood contours at 1$\sigma$, 2$\sigma$, and 3$\sigma$ confidence levels using 3-parameter H$\beta^{\prime}$ high-\rfe\ (blue), 3-parameter H$\beta^{\prime}$ low-\rfe\ (green), and BAO + $H(z)$ (red) data for all free parameters. Left column shows the flat $\Lambda$CDM model, flat XCDM parametrization, and flat $\phi$CDM model respectively. The black dotted lines in all plots are the zero acceleration lines. The black dashed lines in the flat XCDM parametrization plots are the $\omega_X=-1$ lines. Right column shows the non-flat $\Lambda$CDM model, non-flat XCDM parametrization, and non-flat $\phi$CDM model respectively. Black dotted lines in all plots are the zero acceleration lines. Black dashed lines in the non-flat $\Lambda$CDM and $\phi$CDM model plots and black dotted-dashed lines in the non-flat XCDM parametrization plots correspond to $\Omega_{k0} = 0$. The black dashed lines in the non-flat XCDM parametrization plots are the $\omega_X=-1$ lines.}
\label{fig:Eiso-Ep3}
\end{figure*}

\begin{figure*}
\begin{multicols}{2}
    \includegraphics[width=\linewidth,height=7cm]{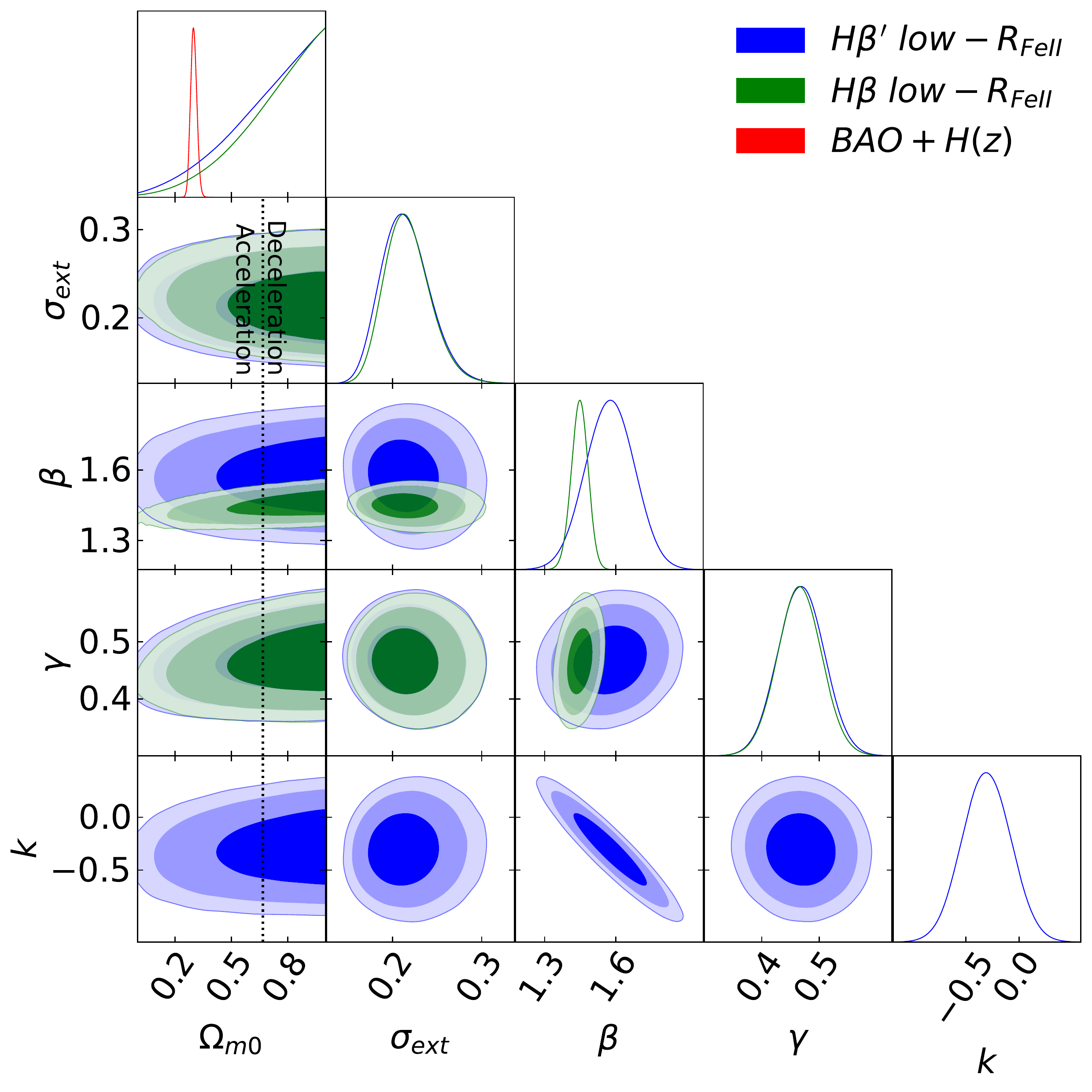}\par
    \includegraphics[width=\linewidth,height=7cm]{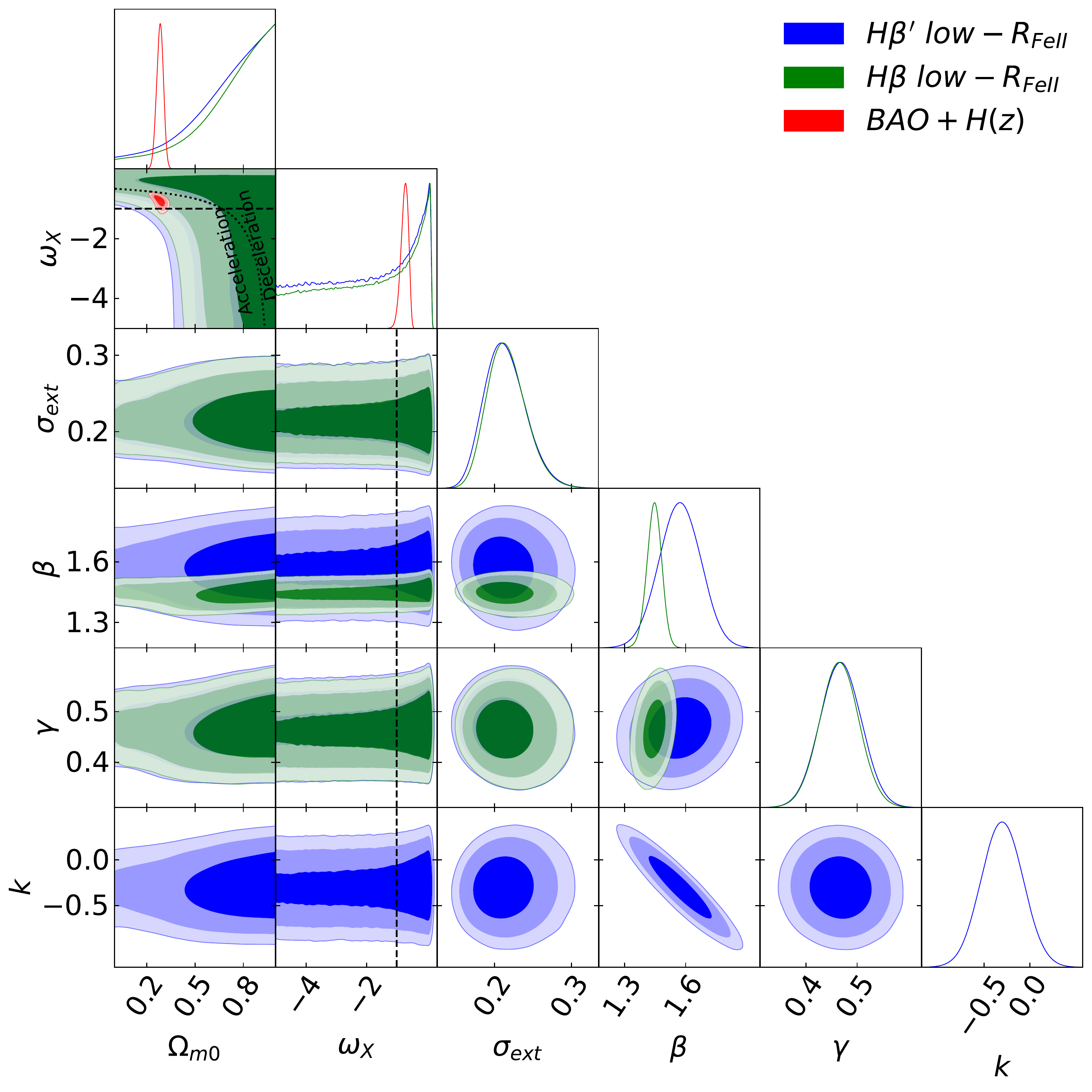}\par
    \includegraphics[width=\linewidth,height=7cm]{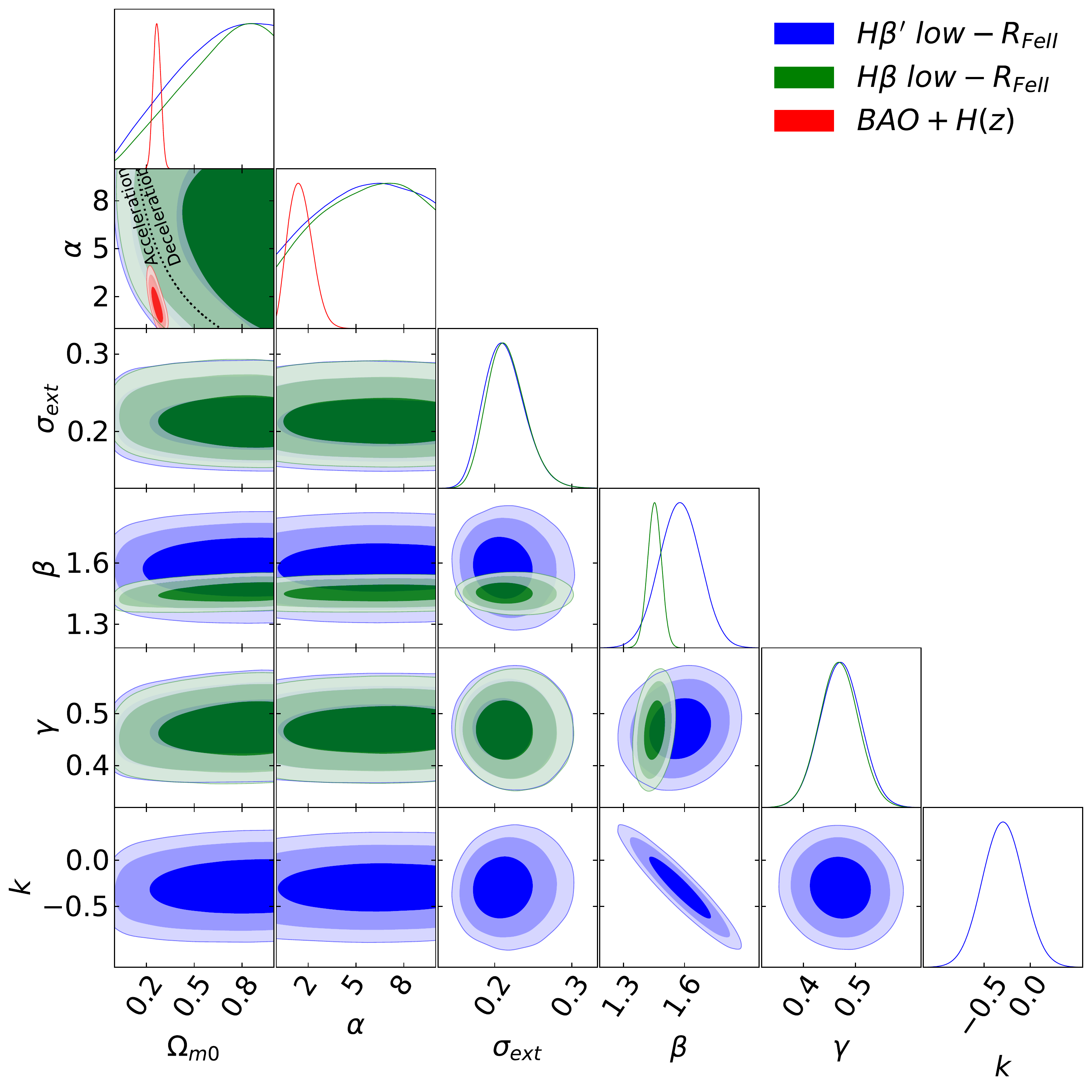}\par
    \includegraphics[width=\linewidth,height=7cm]{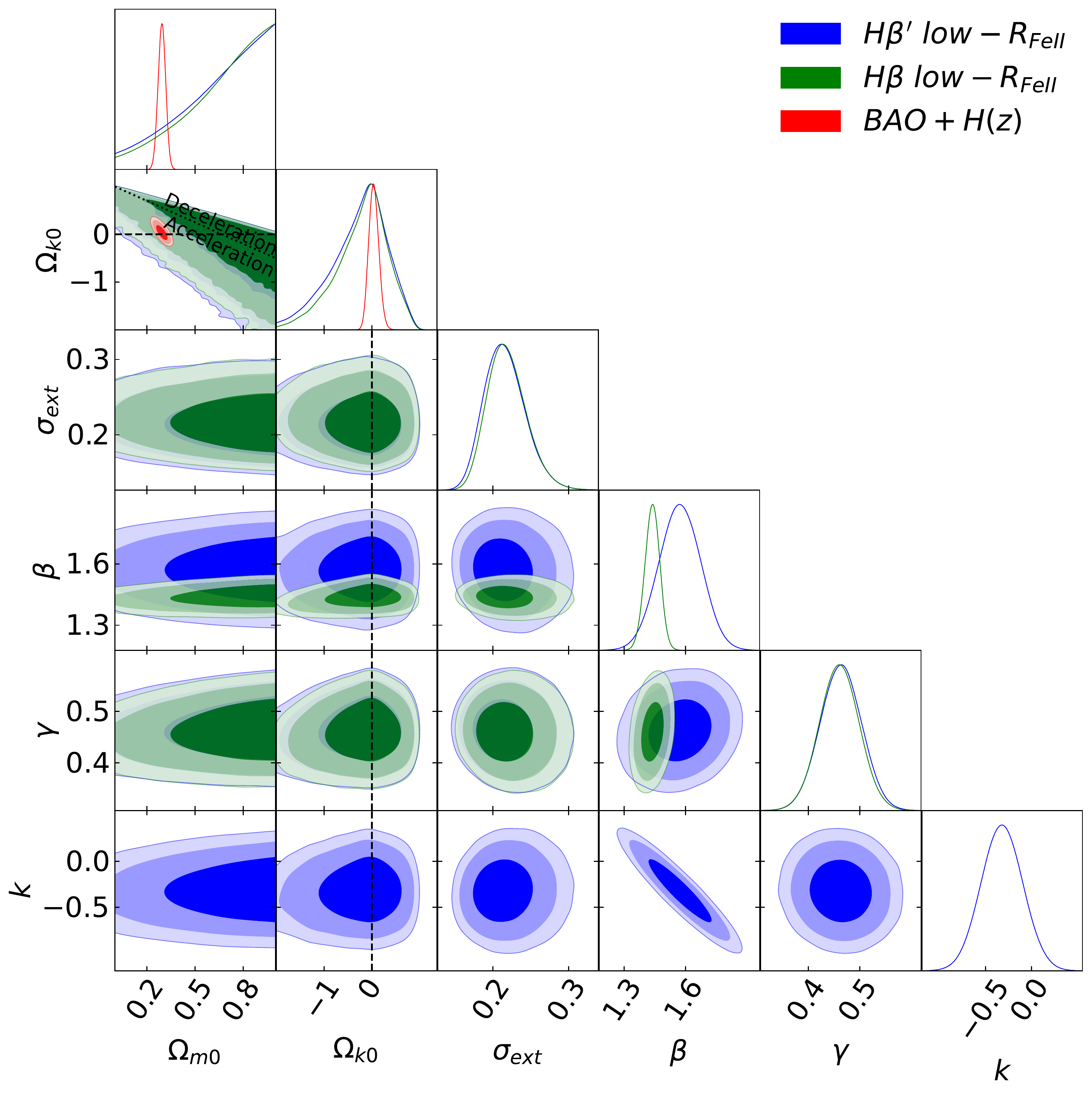}\par
    \includegraphics[width=\linewidth,height=7cm]{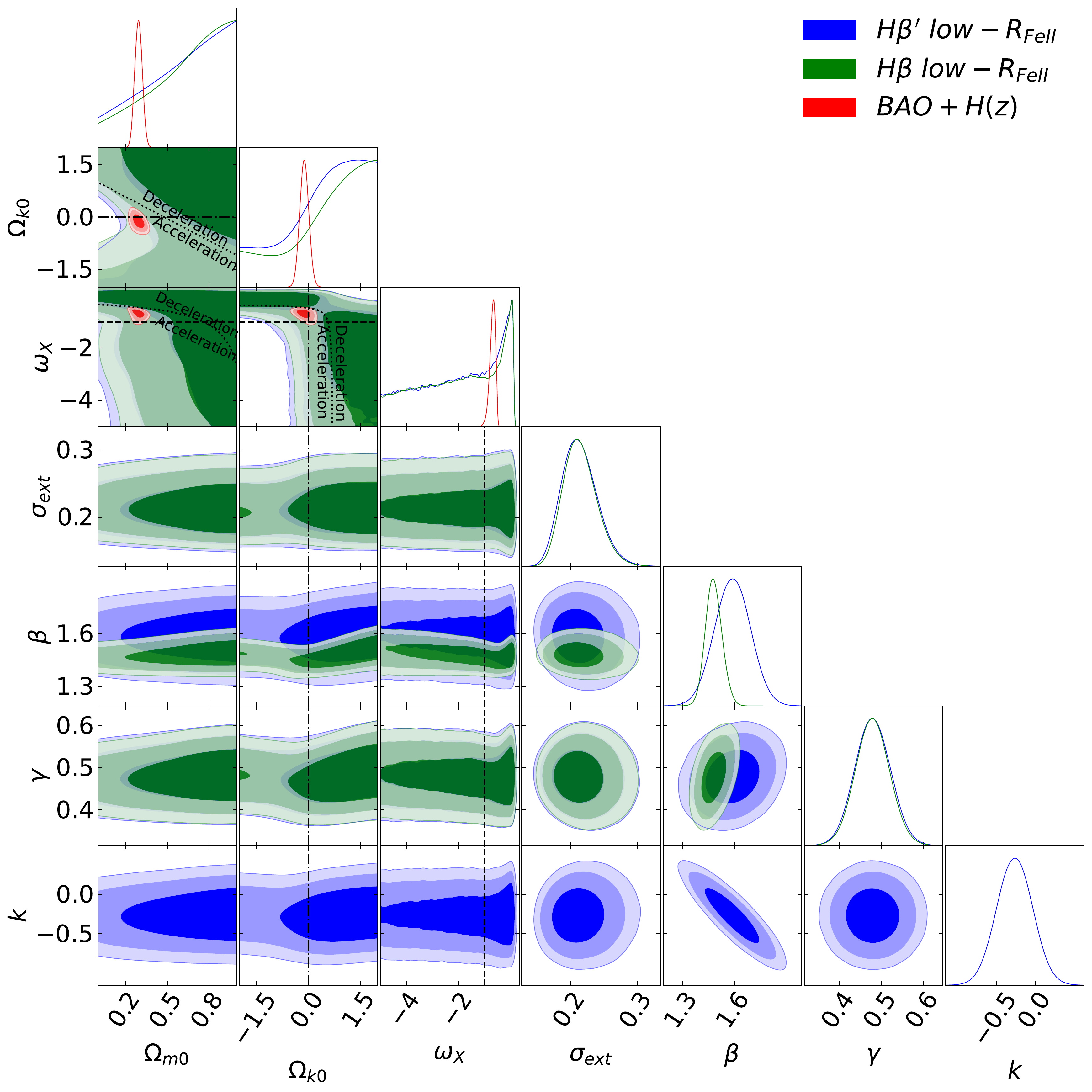}\par
    \includegraphics[width=\linewidth,height=7cm]{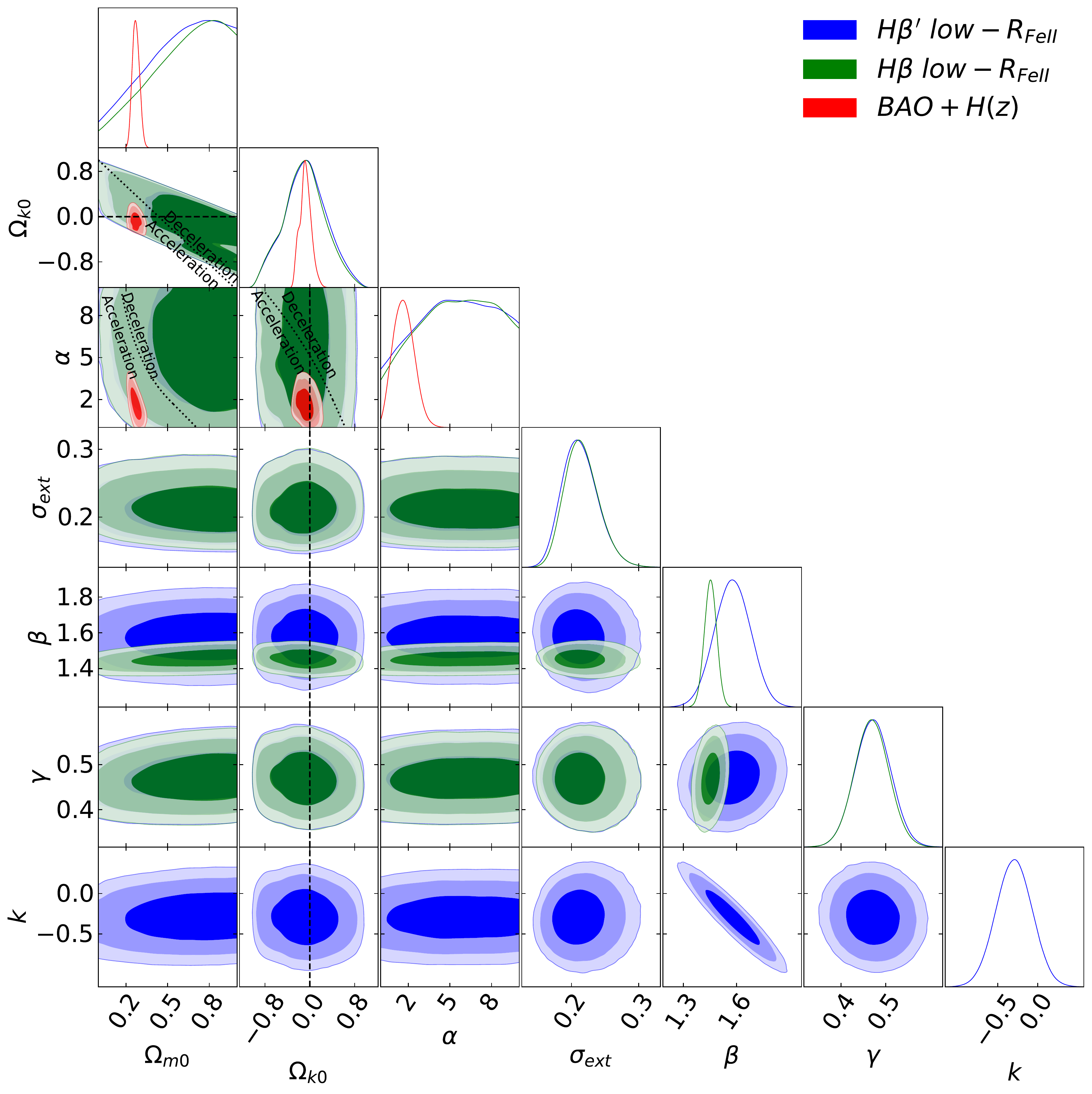}\par
\end{multicols}
\caption{One-dimensional likelihood distributions and two-dimensional likelihood contours at 1$\sigma$, 2$\sigma$, and 3$\sigma$ confidence levels using 3-parameter H$\beta^{\prime}$ low-\rfe\ (blue), 2-parameter H$\beta$ low-\rfe\ (green), and BAO + $H(z)$ (red) data for all free parameters. Left column shows the flat $\Lambda$CDM model, flat XCDM parametrization, and flat $\phi$CDM model respectively. The black dotted lines in all plots are the zero acceleration lines. The black dashed lines in the flat XCDM parametrization plots are the $\omega_X=-1$ lines. Right column shows the non-flat $\Lambda$CDM model, non-flat XCDM parametrization, and non-flat $\phi$CDM model respectively. Black dotted lines in all plots are the zero acceleration lines. Black dashed lines in the non-flat $\Lambda$CDM and $\phi$CDM model plots and black dotted-dashed lines in the non-flat XCDM parametrization plots correspond to $\Omega_{k0} = 0$. The black dashed lines in the non-flat XCDM parametrization plots are the $\omega_X=-1$ lines.}
\label{fig:Eiso-Ep4}
\end{figure*}

\begin{figure*}
\begin{multicols}{2}
    \includegraphics[width=\linewidth,height=7cm]{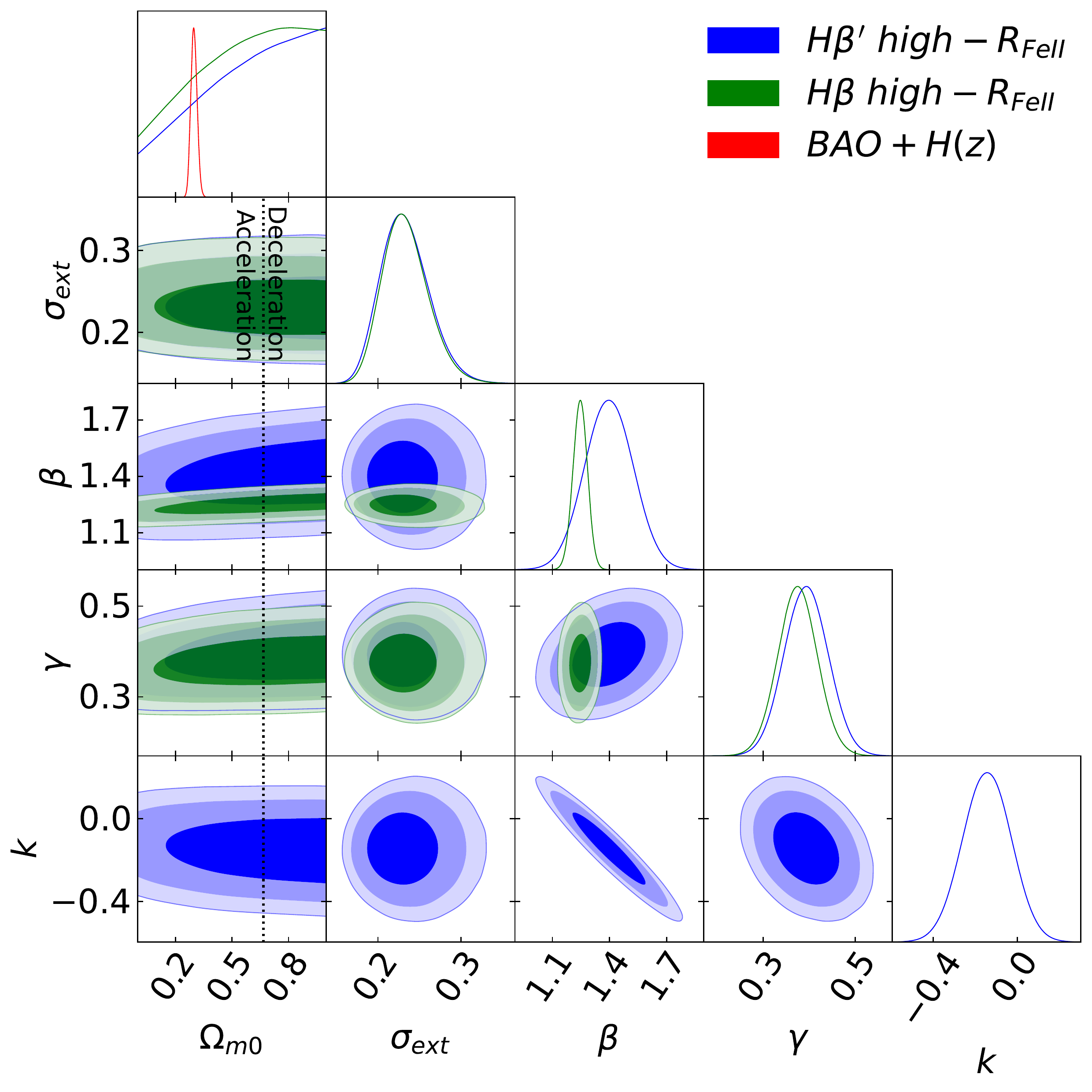}\par
    \includegraphics[width=\linewidth,height=7cm]{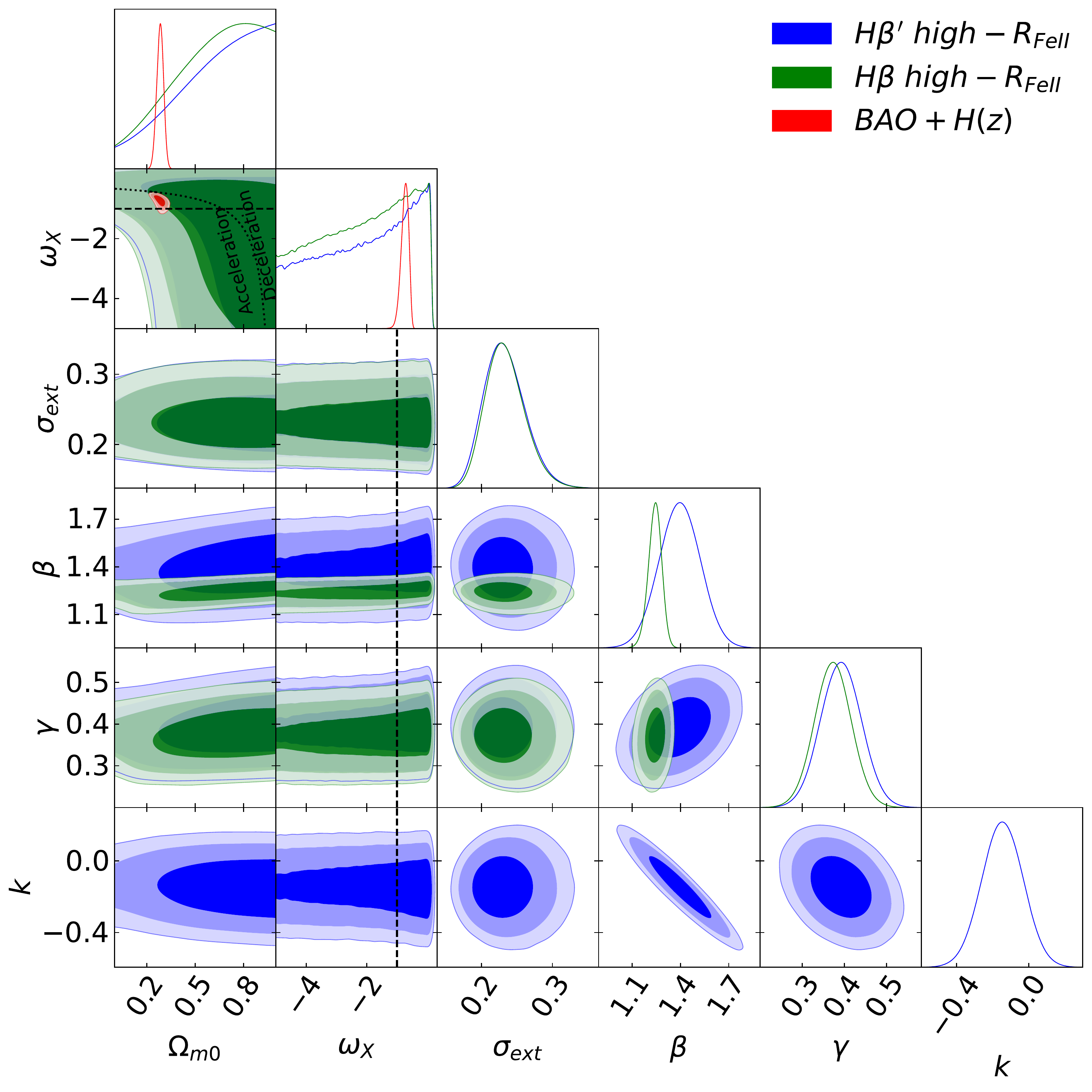}\par
    \includegraphics[width=\linewidth,height=7cm]{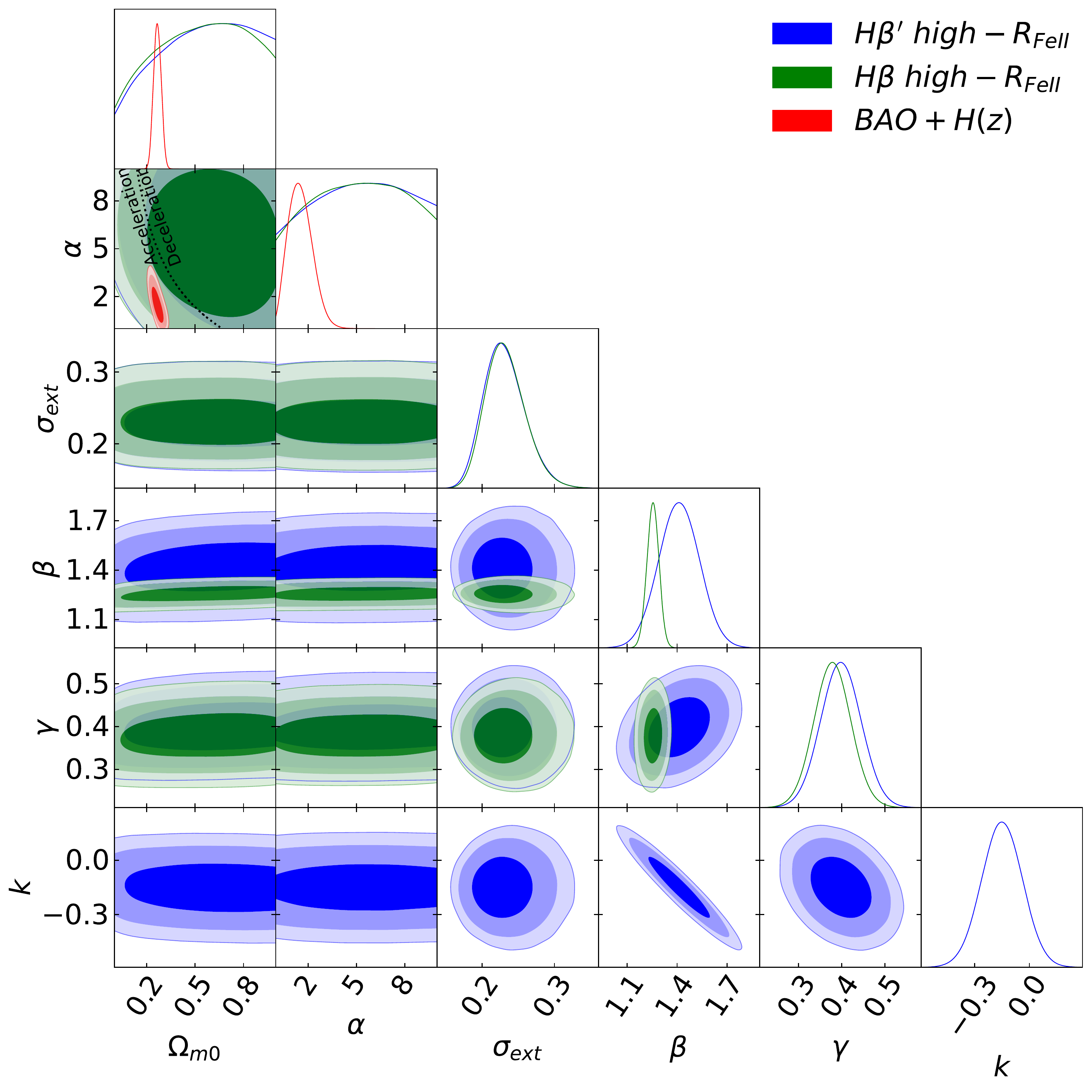}\par
    \includegraphics[width=\linewidth,height=7cm]{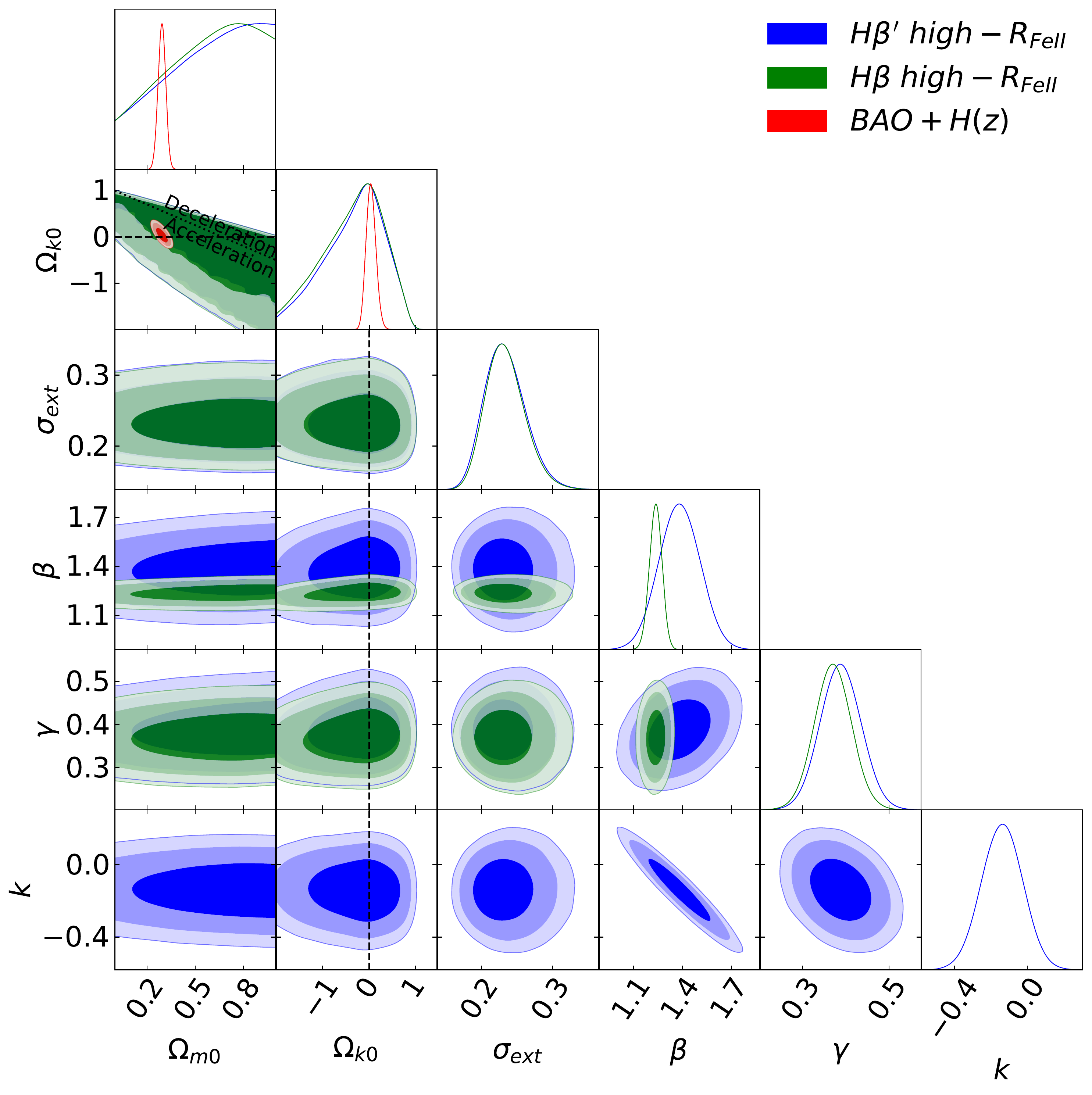}\par
    \includegraphics[width=\linewidth,height=7cm]{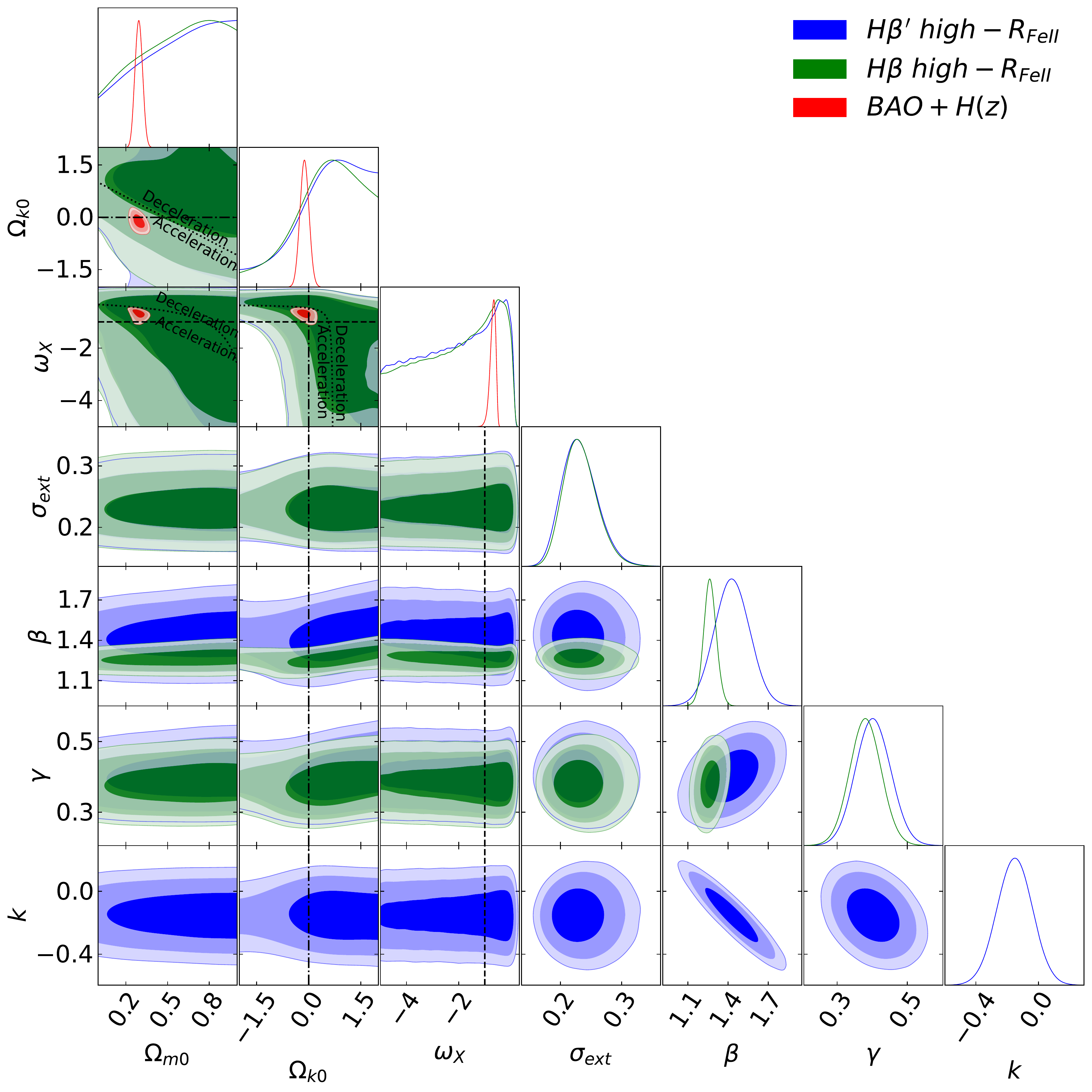}\par
    \includegraphics[width=\linewidth,height=7cm]{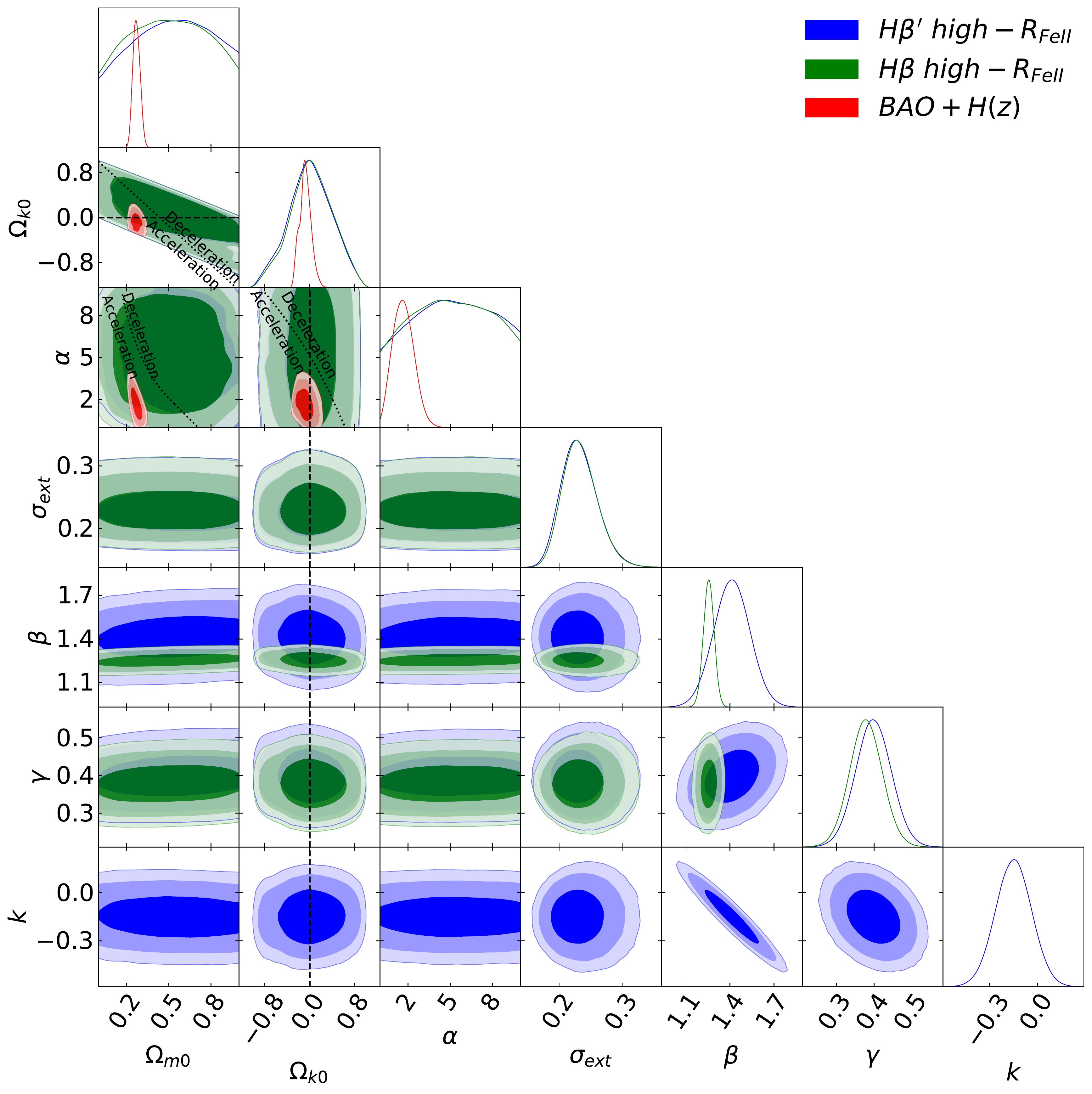}\par
\end{multicols}
\caption{One-dimensional likelihood distributions and two-dimensional likelihood contours at 1$\sigma$, 2$\sigma$, and 3$\sigma$ confidence levels using 3-parameter H$\beta^{\prime}$ high-\rfe\ (blue), 2-parameter H$\beta$ high-\rfe\ (green), and BAO + $H(z)$ (red) data for all free parameters. Left column shows the flat $\Lambda$CDM model, flat XCDM parametrization, and flat $\phi$CDM model respectively. The black dotted lines in all plots are the zero acceleration lines. The black dashed lines in the flat XCDM parametrization plots are the $\omega_X=-1$ lines. Right column shows the non-flat $\Lambda$CDM model, non-flat XCDM parametrization, and non-flat $\phi$CDM model respectively. Black dotted lines in all plots are the zero acceleration lines. Black dashed lines in the non-flat $\Lambda$CDM and $\phi$CDM model plots and black dotted-dashed lines in the non-flat XCDM parametrization plots correspond to $\Omega_{k0} = 0$. The black dashed lines in the non-flat XCDM parametrization plots are the $\omega_X=-1$ lines.}
\label{fig:Eiso-Ep5}
\end{figure*}

In Figs.\ \ref{fig:Eiso-Ep1}--\ref{fig:Eiso-Ep5}, we see that the cosmological constraints from the 2-parameter and 3-parameter $R-L$ relation analyses of the complete 118 sources data set as well as of the 59 sources low-\rfe\ data subset are more consistent with the currently decelerated cosmological expansion. On the other hand, the cosmological constraints from the 2-parameter and 3-parameter $R-L$ relation analyses of the 59 sources high-\rfe\ data subset are much less inconsistent with the currently accelerated cosmological expansion. In this context, however, it might be useful to note that the low-\rfe\ data subset is almost consistent with the simple photoionization $\gamma = 0.5$ slope prediction while the high-\rfe\ data subset $\gamma$ values are $(2-3)\sigma$ away from 0.5.

In Table \ref{tab:1d_BFP2}, we see that only H$\beta^{\prime}$ high-\rfe\ data is able to measure $\Omega_{m0}$, and only in the flat and non-flat $\phi$CDM models, resulting in  $\Omega_{m0} = 0.678^{+0.312}_{-0.289}$ and $0.571^{+0.313}_{-0.309}$, respectively. The other data sets only provide lower limits on $\Omega_{m0}$ that lie in the range  $> 0.158$ to $> 0.388$, with the minimum value obtained in the spatially-flat $\phi$CDM model using H$\beta^{\prime}$ low-\rfe\ data and the maximum value obtained in the spatially-flat $\Lambda$CDM model using H$\beta$ QSO-118$^{\prime}$ data. Mostly these $\Omega_{m0}$ results are not inconsistent with the corresponding values obtained using BAO + $H(z)$ data.

In the flat (non-flat) $\Lambda$CDM model, all H$\beta$ data sets are only able to provide upper limits on $\Omega_{\Lambda}$ and these limits lie in the range $< 0.612$ to $< 1.740$, with the minimum value in the spatially-flat model using H$\beta$ QSO-$118^{\prime}$ data and the maximum value in the spatially non-flat model using H$\beta$ high-\rfe\ data.

From Table \ref{tab:1d_BFP2}, we see that these data are more successful at constraining $\Omega_{k0}$ and, for all data sets in all cosmological models, the resulting values of $\Omega_{k0}$ lie in the range $-0.006^{+0.416}_{-0.549}$ to $0.667^{+1.126}_{-0.704}$. The minimum value is obtained in the non-flat $\Lambda$CDM model using H$\beta$ QSO-118 data and the maximum value is obtained in the non-flat XCDM parametrization using H$\beta$ high-\rfe\ data. These values are mostly consistent with flat spatial geometry.

From Table \ref{tab:1d_BFP2} we see that these data are able to only provide very weak constraints on $\omega_X$. More precisely, they only provide upper limits on $\omega_X$ which lie in the range $< 0.01$ to $< 0.20$. These data are unable to constrain the positive dark energy dynamics parameter $\alpha$ in the $\phi$CDM models.

Overall, from Figs.\ \ref{fig:Eiso-Ep1}--\ref{fig:Eiso-Ep5}, we see that the BAO + $H(z)$ data cosmological constraints lie $\sim2\sigma$ away from the peak of the H$\beta$ data constraints, so while these constraints are not too mutually inconsistent the discrepancy is a little surprising, especially since the H$\beta$ constraints are quite broad, so we do not record joint H$\beta$ and BAO + $H(z)$ data constraints here.  

In Table \ref{tab:BFP}, for the H$\beta$ QSO-118 and H$\beta$ QSO-118$^{\prime}$ data sets, from the $AIC$ and the $BIC$ values, the most favored model is the spatially-flat $\Lambda$CDM model and the least favored model is the spatially non-flat $\phi$CDM model. For the H$\beta$ low-\rfe\ data set, from the $AIC$ and the $BIC$ values, the most favored model is the spatially-flat $\Lambda$CDM model and the least favored one is the spatially-flat XCDM parametrization. For the H$\beta$ high-\rfe\, and the H$\beta^{\prime}$ low- and high-\rfe\ data sets, from the $AIC$ and the $BIC$ values, the most favored model is the spatially-flat $\Lambda$CDM model and the least favored one is the spatially non-flat $\phi$CDM model. 

\section{Discussion}
\label{sec:discussion}

The full 118 sources reverberation-measured \hb\ sample, using either a 2-parameter $R-L$ relation or a 3-parameter one that also takes into account the \rfe\ measurements, results in cosmological constraints that are somewhat {\bf $\sim 2\sigma$} inconsistent with those from better-established cosmological probes. Much more satisfactory results were obtained by \citet{khadka2021} on the basis of the \Mgii\ sample.\footnote{We note that we find that cosmological constraints based on the \hb\ high-\rfe\ data subset are mostly consistent with those from better-established cosmological probes, and that cosmological constraints based on this subset are roughly the same for both the 2-parameter and 3-parameter $R-L$ analyses. Similarly the low-\rfe\ subset cosmological constraints are also roughly similar for the 2-parameter and 3-parameter cases, but they are, however, $\sim(2-3)\sigma$ discrepant with the cosmological constraints derived using better-established cosmological probes.} 

It is possible that this discrepancy of the H$\beta$ cosmological constraints is just a statistical fluctuation. We also note that reverberation mapping of \hb\ sources were performed by several groups that used different methods to determine the time delay and different arguments to determine how reliable and significant their measurements were, see \citet{zajacek2019}. Also, for the \hb\ analyses, the spectroscopic fitting was performed using a variety of different routines and \Feii\ templates. On the other hand, 57 of the 78 \Mgii\  sources of the sample used in \citet{khadka2021} come from the Sloan Digital Sky Survey Reverberation Mapping Project (SDSS-RM) \citep{Homayouni2020} and these sources were analyzed using the same method. Hence it is also possible that the discrepancy of H$\beta$ cosmological constraints we have found is a consequence of the heterogeneity of the H$\beta$ QSO-118 sample we have used here. 

Leaving aside the \hb\ cosmological constraints discrepancy, strong evidence suggests that the accretion rate affects the $R-L$ relation for \hb\ sources, with high accretors having a smaller broad-line region size and hence a shorter time delay with respect to what is predicted by the usual 2-parameter $R-L$ relation. It has been suggested that this effect can be corrected by introducing a third parameter into the $R-L$ relation and that such an extended $R-L$ relation might possess the additional benefit of a reduced intrinsic sample scatter \citep{duwang_2019,Mary2020}. Since the \rfe\ parameter in large samples shows a strong correlation with the Eddington ratio \citep[e.g.][]{marziani2003, zamfiretal10}, and is completely independent of the time delay, we limited our sample to 118 \hb\ sources with \rfe\ measurements and also considered an extended 3-parameter $R-L$ relation with the objective of correcting the 2-parameter $R-L$ relation for the accretion-rate effect. We also consider two subsets of low and high accretors, each with 59 sources. We do not find strong evidence that this correction helps to significantly reduce the intrinsic dispersion in any of the analyzed cases.\footnote{We do find that the 3-parameter $R-L$ relation is strongly favored by the full 118-source \hb\ data set but it is not favored by either the 59-source high-\rfe\ data subset or the 59-source low-\rfe\ data subset.} The full sample shows a moderate correlation between $\eta$ (or $\Delta\tau$) with \rfe, but when high- and low-\rfe\ subsamples are considered separately only the $\eta$-\rfe\ correlation is significant, and only for the high-\rfe\ sources where the correlation is moderate and not strong. \citet{duwang_2019} originally proposed the 3-parameter $R-L$ relation through compilation of \rfe\ available values as well the estimation of new \rfe\ values. They obtained a stronger correlation between $\Delta\tau$ and \rfe\ ($\rho=-0.56$, $p$-value$=2.0\times10^{-7}$) than what we find here. In the current analysis we have also considered additional SDSS-RM sources, which increases the size of the sample but weakens the $\Delta \tau$-\rfe\ correlation.  

These results suggest  that more consistent time-delay determinations and better spectroscopic fitting, particularly better measurements of \Feii, for the full sample, might be needed to decrease the uncertainty and the scatter (Zaja\v{c}ek et al. in prep.), which is quite relevant for the cosmological analysis. Also, it will be important to apply redshift correction for the cosmic microwave background (CMB) dipole and eventually for peculiar motion for sources at $z<0.1$, such as supernova cosmology analyses do \citep[e.g.][and references therein]{davis2011}. We also note that the redshift range and the redshift distribution of sources in the \hb\ and \Mgii\ samples are very different. The Mg II sample used in \citet{khadka2021} covered a larger redshift range, and the sources were rather uniformly distributed across the redshift range. The H$\beta$ sample used here covers a much narrower redshift range and most of the \hb\ sources are located at the low redshift end. Therefore, future measurements covering redshifts $0.4\lesssim z \lesssim 0.9$ are also needed for an accurate analysis.

\section{Conclusion}
\label{con}

In this paper, we use 2- and 3-parameter $R-L$ relations to standardize 118 H$\beta$ QSOs, as well as the two 59 source high- and low-\rfe\ \hb\ subsets. We show, for the first time, that the parameters for both the 2- and 3-parameter $R-L$ relations, for all three data sets, are almost independent of cosmological model used in the analysis, indicating that these QSOs are standardizable through these $R-L$ relations. Differences in the $AIC$ and $BIC$ values show that in all cosmological models, for the 118 source data, the 3-parameter $R-L$ relation is very strongly favored over the 2-parameter one. However, for the low- and high-\rfe\ H$\beta$ QSOs data subsets, there is no significant evidence in favor of the 3-parameter $R-L$ relation in all cosmological models. We note that the 2-parameter $R-L$ relation parameters for the low- and high-\rfe\ data subsets differ significantly, but when we analyze each 59 source data subset separately these subset differences cannot significantly affect the analyses. When we analyze these subsets using the 3-parameter $R-L$ relation the parameter values change and the error bars broaden, so while the difference between the high- and low-\rfe\ subset parameter values do not change much, these differences are less significant for the 3-parameter case because of the larger error bars. In the analyses of the full 118 sources data set, the significant difference between the 2-parameter $R-L$ relation parameters for the low- and high-\rfe\ subsets plays an 
important role, and it might be the case that the very strong evidence provided by the $\Delta AIC$ and $\Delta BIC$ values in favor of the 3-parameter $R-L$ relation for the full 118 sources data set is related to the fundamentally different nature of the high and low \rfe\ H$\beta$ QSO data subsets. So it is important to be careful and not to draw a strong conclusion in favor of the 3-parameter $R-L$ relation for the 118 sources data set until the cause for this is better understood. The main motivation behind the 3-parameter $R-L$ relation was the hope that the inclusion of the third parameter $k$ in the $R-L$ relation would significantly reduce the intrinsic dispersion in the $R-L$ relation $(\sigma_{\rm ext})$, however we find only a mild reduction in the intrinsic dispersion.

We determined \hb\ constraints on cosmological parameters in six different cosmological models and found that these constraints are significantly weaker than those from BAO + $H(z)$ data. Our comparison of H$\beta$ QSO and BAO + $H(z)$ two-dimensional cosmological constraints show that typically the H$\beta$ QSO ones are $\sim 2\sigma$ discrepant with the BAO + $H(z)$ ones, which is not very significant, and also tend to more favor currently decelerating cosmological expansion.

Current H$\beta$ QSO data span a relatively narrow redshift range so we cannot test for a possible redshift evolution of the $R-L$ relation but we hope that future observations will detect more H$\beta$ QSOs over a wider redshift range and this will enable us to study the redshift evolution (if any) of the $R-L$ relation. Also, currently these data provide only weak cosmological constraints, constraints that are somewhat {\boldmath$(\sim 2\sigma)$} discrepant with those from better-established cosmological probes. We are hopeful that more H$\beta$ sources observed in the future, with more precises measurements, will help resolve this puzzle, and perhaps establish \hb\ QSOs as a new and independent cosmological probe.

\section{ACKNOWLEDGEMENTS}

This research was supported in part by US DOE grant DE-SC0011840, by the Polish Funding Agency National Science Centre, project 2017/26/A/ST9/00756 (Maestro 9), by GAČR EXPRO grant 21-13491X, and by Millenium Nucleus NCN19-058 (TITANs). Part of the computation for this project was performed on the Beocat Research Cluster at Kansas State University. We acknowledge S.-S.\ Li and J.-M.\ Wang for the extra information required for this analysis. We thank V.\ Lebedev for a comment and for pointing out a typographical error. This research has made use of the NASA/IPAC Extragalactic Database (NED), which is operated by the Jet Propulsion Laboratory, California Institute of Technology, under contract with the National Aeronautics and Space Administration.

\section*{Data availability}
The data analysed in this article are listed in Table \ref{tab:hbQSOdata} of this paper.



\bibliographystyle{mnras}
\bibliography{mybibfile}

%%%%%%%%%%%%%%%%%%%%%%%%%%%%%%%%%%%%%%%%%%%%%%%%%%

%%%%%%%%%%%%%%%%% APPENDICES %%%%%%%%%%%%%%%%%%%%%

%\appendix

\onecolumn
\begin{appendix}
\section{H$\beta$ QSO data}
\label{sec:appendix}
%\begin{landscape}
\addtolength{\tabcolsep}{0pt}
\LTcapwidth=\linewidth
\begin{longtable}{cccccccc}
\caption{Reverberation-measured \hb\ QSO samples. For each source, columns list: QSO name, RA (NED), DEC (NED), redshift, continuum flux density at 5100\,\AA, measured rest-frame time-delay, \rfe\ , and references. The SDSS-RM sources are identified with an asterisk (*) symbol. SDSS~J100402 and PG~1001+291 are the same object, but with time-delay estimated by two different groups. The flux at 5100\,\AA\ in the observed-frame was estimated from the luminosity using cosmological parameters $H_0 = 70$ km s$^{-1}$ Mpc$^{-1}$, $\Omega_{m0} = 0.3$, $\Omega_\Lambda = 0.7$. In the time-delay column the number in parenthesis corresponds to the symmetrized error, see Sec.~\ref{sec:data}. All errors are 1$\sigma$. We adopt an error of 10$\%$ for \rfe\ if the \rfe\ estimation is taken from \citet{duwang_2019}. The last column lists the literature reference for the time-delay value [1: \citet{2014ApJ...782...45D}, 2: \citet{wangSEAMBH2014}, 3: \citet{2018ApJ...856....6D}, 4: \citet{2017ApJ...851...21G}, 5: \citet{2013ApJ...767..149B}, 6: \citet{bentz2009}, 7:  \citet{bentz2014}, 8: \citet{pei2014} 9: \citet{bentz2016a}, 10:  \citet{zhang2018}, 11: \citet{lu2016}, 12: \citet{rakshit2020}, 13: \citet{du2015}, 14: \citet{du2016}, 15: \citet{barth2013}, 16: \citet{bentz2016a}, 17: \citet{fausnaugh2017}, 18: \citet{huang2019}, 19: \citet{li_2021}], the \rfe\ value [a: \citet{hu2015}, b: \citet{duwang_2019}, c: \citet{shen_2019},  d: \citet{hu2020}, e: \citet{marziani2003}, f: \citet{du2016funpl}, g: \citet{borosongreen1992}, h: \citet{marziani2003_cat}], and the set of cosmological parameters used in the original references [$\alpha$: $H_0 = 67$ km s$^{-1}$ Mpc$^{-1}$, $\Omega_{m0} = 0.32$, $\Omega_\Lambda = 0.68$; $\beta$: $H_0 = 70$ km s$^{-1}$ Mpc$^{-1}$, $\Omega_{m0} = 0.3$, $\Omega_\Lambda = 0.7$; $\gamma$: $H_0 = 72$ km s$^{-1}$ Mpc$^{-1}$, $\Omega_{m0} = 0.3$, $\Omega_\Lambda = 0.7$; $\delta$: not reported].}
\label{tab:hbQSOdata}\\
\hline\hline
Object name &  RA & DEC & $z$ &  $\log F_{5100}$  &  $\tau$ & \rfe & Ref.  \\
& (J2000.0) & (J2000.0) & & $\left({\rm erg}\,{\rm s^{-1}}{\rm cm^{-2}}\right)$ & (days) & & \\
\hline
\endfirsthead
\caption{continued.}\\
\hline\hline
Object name &  RA & DEC & $z$ &  $\log F_{5100}$  &  $\tau$ & \rfe & Ref.  \\
& (J2000.0) & (J2000.0) & & $\left({\rm erg}\,{\rm s^{-1}}{\rm cm^{-2}}\right)$ & (days) & & \\
\hline
\endhead
\hline
\endfoot
%%%LOW SUBSAMPLE
\multicolumn{8}{c}{
Low \rfe\ sources, \rfe$\leq 0.655$} \\
\hline
Mrk~335 & 00h06m19.52s & +20d12m10.5s & 0.026 & -10.422 $\pm$ 0.070 & 14.0 $_{-3.4}^{+4.6}$ ($\pm$ 3.9) & 0.620 $\pm$ 0.062 & 1,a,$\alpha$ \\
Mrk~486 & 15h36m38.36s & +54d33m33.2s & 0.039 & -10.857 $\pm$ 0.050 & 23.7 $_{-2.7}^{+7.5}$ ($\pm$ 4.2) & 0.540 $\pm$ 0.054 & 2,a,$\alpha$ \\
SDSS~J081441 & 08h14m41.92s & +21d29m18.5s & 0.163 & -11.901 $\pm$ 0.060 & 25.3 $_{-7.5}^{+10.4}$ ($\pm$ 8.8) & 0.460 $\pm$ 0.046 & 3,b,$\beta$ \\
SDSS~J141923* & 14h19m23.37s & +54d22m01.8s & 0.152 & -12.675 $\pm$ 0.010 & 11.8 $_{-1.5}^{+0.7}$ ($\pm$ 1.0) & 0.568 $\pm$ 0.017 & 4,c,$\beta$ \\
SDSS~J141625* & 14h16m25.71s & +53d54m38.6s & 0.263 & -12.365 $\pm$ 0.019 & 15.1 $_{-4.6}^{+3.2}$ ($\pm$ 3.8) & 0.329 $\pm$ 0.016 & 4,c,$\beta$ \\
SDSS~J142103* & 14h21m03.53s & +51d58m19.5s & 0.263 & -12.693 $\pm$ 0.019 & 75.2 $_{-3.3}^{+3.2}$ ($\pm$ 3.2) & 0.595 $\pm$ 0.073 & 4,c,$\beta$ \\
SDSS~J141041* & 14h10m41.24s & +53d18m49.1s & 0.359 & -12.817 $\pm$ 0.005 & 21.9 $_{-2.4}^{+4.2}$ ($\pm$ 3.1) & 0.302 $\pm$ 0.029 & 4,c,$\beta$ \\
SDSS~J141645.58* & 14h16m45.59s & +53d44m46.8s & 0.442 & -13.175 $\pm$ 0.009 & 8.5 $_{-1.4}^{+2.5}$ ($\pm$ 1.8) & 0.653 $\pm$ 0.063 & 4,c,$\beta$ \\
SDSS~J141214* & 14h12m14.20s & +53d25m46.7s & 0.458 & -12.494 $\pm$ 0.004 & 21.4 $_{-6.4}^{+4.2}$ ($\pm$ 5.1) & 0.431 $\pm$ 0.096 & 4,c,$\beta$ \\
SDSS~J140518* & 14h05m18.03s & +53d15m30.1s & 0.467 & -12.578 $\pm$ 0.004 & 41.6 $_{-8.3}^{+14.8}$ ($\pm$ 10.9) & 0.515 $\pm$ 0.021 & 4,c,$\beta$ \\
SDSS~J141018* & 14h10m18.05s & +53d29m37.5s & 0.470 & -13.334 $\pm$ 0.005 & 16.2 $_{-4.5}^{+2.9}$ ($\pm$ 3.6) & 0.598 $\pm$ 0.057 & 4,c,$\beta$ \\
SDSS~J142039* & 14h20m39.80s & +52d03m59.7s & 0.474 & -12.786 $\pm$ 0.004 & 20.7 $_{-3.0}^{+0.9}$ ($\pm$ 1.5) & 0.396 $\pm$ 0.022 & 4,c,$\beta$ \\
SDSS~J141724* & 14h17m24.60s & +52d30m24.8s & 0.482 & -12.952 $\pm$ 0.004 & 10.1 $_{-2.7}^{+12.5}$ ($\pm$ 5.1) & 0.468 $\pm$ 0.033 & 4,c,$\beta$ \\
SDSS~J141004* & 14h10m04.27s & +52d31m41.0s & 0.527 & -12.813 $\pm$ 0.003 & 53.5 $_{-4.0}^{+4.2}$ ($\pm$ 4.1) & 0.466 $\pm$ 0.081 & 4,c,$\beta$ \\
SDSS~J141706* & 14h17m06.68s & +51d43m40.1s & 0.532 & -12.861 $\pm$ 0.003 & 10.4 $_{-3.0}^{+6.3}$ ($\pm$ 4.2) & 0.566 $\pm$ 0.045 & 4,c,$\beta$ \\
SDSS~J141712* & 14h17m12.30s & +51d56m45.5s & 0.554 & -13.881 $\pm$ 0.012 & 12.5 $_{-2.6}^{+1.8}$ ($\pm$ 2.1) & 0.644 $\pm$ 0.232 & 4,c,$\beta$ \\
SDSS~J141031* & 14h10m31.32s & +52d15m33.9s & 0.608 & -13.165 $\pm$ 0.003 & 35.8 $_{-10.3}^{+1.1}$ ($\pm$ 2.7) & 0.611 $\pm$ 0.084 & 4,c,$\beta$ \\
SDSS~J141941* & 14h19m41.11s & +53d36m49.7s & 0.646 & -12.730 $\pm$ 0.017 & 30.4 $_{-8.3}^{+3.9}$ ($\pm$ 5.5) & 0.470 $\pm$ 0.086 & 4,c,$\beta$ \\
SDSS~J141147* & 14h11m47.06s & +51d56m19.8s & 0.680 & -13.281 $\pm$ 0.004 & 6.4 $_{-1.4}^{+1.5}$ ($\pm$ 1.4) & 0.449 $\pm$ 0.168 & 4,c,$\beta$ \\
SDSS~J142049* & 14h20m49.29s & +52d10m53.3s & 0.751 & -12.966 $\pm$ 0.003 & 46.0 $_{-9.5}^{+9.5}$ ($\pm$ 9.5) & 0.452 $\pm$ 0.042 & 4,c,$\beta$ \\
SDSS~J142112* & 14h21m12.29s & +52d41m47.3s & 0.843 & -13.220 $\pm$ 0.008 & 14.2 $_{-3.0}^{+3.7}$ ($\pm$ 3.3) & 0.573 $\pm$ 0.126 & 4,c,$\beta$ \\
SDSS~J141606* & 14h16m06.96s & +53d09m29.8s & 0.848 & -12.740 $\pm$ 0.003 & 32.0 $_{-15.5}^{+11.6}$ ($\pm$ 13.3) & 0.416 $\pm$ 0.123 & 4,c,$\beta$ \\
SDSS~J141952* & 14h19m52.24s & +53d13m41.0s & 0.884 & -13.340 $\pm$ 0.006 & 32.9 $_{-5.1}^{+5.6}$ ($\pm$ 5.3) & 0.277 $\pm$ 0.138 & 4,c,$\beta$ \\
PG~0026+129 & 00h29m13.70s & +13d16m04.0s & 0.142 & -10.762 $\pm$ 0.020 & 111.0 $_{-28.3}^{+24.1}$ ($\pm$ 26.1) & 0.330 $\pm$ 0.033 & 5,d,$\gamma$ \\
PG~0052+251 & 00h54m52.12s & +25d25m39.0s & 0.155 & -11.002 $\pm$ 0.030 & 89.8 $_{-24.1}^{+24.5}$ ($\pm$ 24.3) & 0.120 $\pm$ 0.012 & 5,b,$\gamma$ \\
Fairall~9 & 01h23m45.78s & -58d48m20.8s & 0.047 & -10.736 $\pm$ 0.040 & 17.4 $_{-4.3}^{+3.2}$ ($\pm$ 3.7) & 0.490 $\pm$ 0.049 & 5,b,$\gamma$ \\
Mrk~590 & 02h14m33.56s & -00d46m00.1s & 0.026 & -10.702 $\pm$ 0.210 & 25.6 $_{-5.3}^{+6.5}$ ($\pm$ 5.9) & 0.450 $\pm$ 0.045 & 5,e,$\gamma$ \\
3C~120 & 04h33m11.10s & +05d21m15.6s & 0.033 & -10.400 $\pm$ 0.100 & 26.2 $_{-6.6}^{+8.7}$ ($\pm$ 7.5) & 0.390 $\pm$ 0.039 & 5,f,$\gamma$ \\
Mrk~79 & 07h42m32.80s & +49d48m34.7s & 0.022 & -10.369 $\pm$ 0.070 & 15.6 $_{-4.9}^{+5.1}$ ($\pm$ 5.0) & 0.330 $\pm$ 0.033 & 5,b,$\gamma$ \\
PG~0804+761 & 08h10m58.60s & +76d02m42.5s & 0.100 & -10.494 $\pm$ 0.020 & 146.9 $_{-18.9}^{+18.8}$ ($\pm$ 18.8) & 0.610 $\pm$ 0.061 & 5,b,$\gamma$ \\
Mrk~110 & 09h25m12.87s & +52d17m10.5s & 0.035 & -10.800 $\pm$ 0.120 & 25.6 $_{-7.2}^{+8.9}$ ($\pm$ 8.0) & 0.140 $\pm$ 0.014 & 5,g,$\gamma$ \\
PG~0953+414 & 09h56m52.39s & +41d15m22.3s & 0.234 & -11.024 $\pm$ 0.010 & 150.1 $_{-22.6}^{+21.6}$ ($\pm$ 22.1) & 0.270 $\pm$ 0.027 & 5,g,$\gamma$ \\
NGC~3227 & 10h23m30.58s & +19d51m54.2s & 0.004 & -10.286 $\pm$ 0.110 & 3.8 $_{-0.8}^{+0.8}$ ($\pm$ 0.8) & 0.460 $\pm$ 0.046 & 5,b,$\gamma$ \\
SBS~1116+583A & 11h18m57.69s & +58d03m23.7s & 0.028 & -12.111 $\pm$ 0.230 & 2.3 $_{-0.5}^{+0.6}$ ($\pm$ 0.5) & 0.590 $\pm$ 0.059 & 5,b,$\gamma$ \\
Arp~151 & 11h25m36.17s & +54d22m57.0s & 0.021 & -11.454 $\pm$ 0.100 & 4.0 $_{-0.7}^{+0.5}$ ($\pm$ 0.6) & 0.320 $\pm$ 0.032 & 5,b,$\gamma$ \\
NGC~3783 & 11h39m01.76s & -37d44m19.2s & 0.010 & -10.761 $\pm$ 0.180 & 10.2 $_{-2.3}^{+3.3}$ ($\pm$ 2.7) & 0.040 $\pm$ 0.004 & 5,b,$\gamma$ \\
Mrk~1310 & 12h01m14.36s & -03d40m41.1s & 0.020 & -11.649 $\pm$ 0.140 & 3.7 $_{-0.6}^{+0.6}$ ($\pm$ 0.6) & 0.460 $\pm$ 0.046 & 5,b,$\gamma$ \\
NGC~4151 & 12h10m32.58s & +39d24m20.6s & 0.003 & -10.291 $\pm$ 0.210 & 6.6 $_{-0.8}^{+1.1}$ ($\pm$ 0.9) & 0.220 $\pm$ 0.022 & 5,b,$\gamma$ \\
Mrk~202 & 12h17m55.00s & +58d39m35.5s & 0.021 & -11.740 $\pm$ 0.140 & 3.0 $_{-1.1}^{+1.7}$ ($\pm$ 1.4) & 0.570 $\pm$ 0.057 & 5,b,$\gamma$ \\
PG~1307+085 & 13h09m47.00s & +08d19m48.2s & 0.155 & -10.965 $\pm$ 0.020 & 105.6 $_{-46.6}^{+36.0}$ ($\pm$ 40.8) & 0.210 $\pm$ 0.021 & 5,b,$\gamma$ \\
Mrk~279 & 13h53m03.45s & +69d18m29.6s & 0.031 & -10.620 $\pm$ 0.070 & 16.7 $_{-3.9}^{+3.9}$ ($\pm$ 3.9) & 0.550 $\pm$ 0.055 & 5,h,$\gamma$ \\
PG~1411+442 & 14h13m48.33s & +44d00m14.0s & 0.090 & -10.743 $\pm$ 0.020 & 124.3 $_{-61.7}^{+61.0}$ ($\pm$ 61.3) & 0.630 $\pm$ 0.063 & 5,b,$\gamma$ \\
PG~1426+015 & 14h29m06.59s & +01d17m06.5s & 0.087 & -10.641 $\pm$ 0.020 & 95.0 $_{-37.1}^{+29.9}$ ($\pm$ 33.2) & 0.460 $\pm$ 0.046 & 5,b,$\gamma$ \\
Mrk~290 & 15h35m52.36s & +57d54m09.2s & 0.030 & -11.133 $\pm$ 0.060 & 8.7 $_{-1.0}^{+1.2}$ ($\pm$ 1.1) & 0.290 $\pm$ 0.029 & 5,g,$\gamma$ \\
PG~1613+658 & 16h13m57.18s & +65d43m09.6s & 0.129 & -10.872 $\pm$ 0.020 & 40.1 $_{-15.2}^{+15.0}$ ($\pm$ 15.1) & 0.380 $\pm$ 0.038 & 5,g,$\gamma$ \\
3C~390.3 & 18h42m08.99s & +79d46m17.1s & 0.056 & -10.446 $\pm$ 0.580 & 44.5 $_{-17.0}^{+27.6}$ ($\pm$ 21.4) & 0.120 $\pm$ 0.012 & 5,f,$\gamma$ \\
NGC~6814 & 19h42m40.64s & -10d19m24.6s & 0.005 & -10.657 $\pm$ 0.280 & 6.6 $_{-0.9}^{+0.9}$ ($\pm$ 0.9) & 0.450 $\pm$ 0.045 & 5,b,$\gamma$ \\
Mrk~509 & 20h44m09.74s & -10d43m24.5s & 0.034 & -10.247 $\pm$ 0.050 & 79.6 $_{-5.4}^{+6.1}$ ($\pm$ 5.7) & 0.130 $\pm$ 0.013 & 5,b,$\gamma$ \\
NGC~7469 & 23h03m15.62s & +08d52m26.4s & 0.016 & -10.267 $\pm$ 0.110 & 10.8 $_{-1.3}^{+3.4}$ ($\pm$ 2.0) & 0.430 $\pm$ 0.043 & 5,b,$\gamma$ \\
PG~1211+143 & 12h14m17.67s & +14d03m13.1s & 0.081 & -10.479 $\pm$ 0.080 & 93.8 $_{-42.1}^{+25.6}$ ($\pm$ 32.3) & 0.420 $\pm$ 0.042 & 6,g,$\beta$ \\
NGC~5273 & 13h42m08.34s & +35d39m15.2s & 0.004 & -10.916 $\pm$ 0.160 & 2.2 $_{-1.6}^{+1.2}$ ($\pm$ 1.4) & 0.580 $\pm$ 0.058 & 7,b,$\delta$ \\
KA~1858-4850 & 18h58m01.10s & +48d50m23.0s & 0.078 & -11.745 $\pm$ 0.050 & 13.5 $_{-2.3}^{+2.0}$ ($\pm$ 2.1) & 0.110 $\pm$ 0.011 & 8,a,$\gamma$ \\
MCG~+08-11-011 & 05h54m53.61s & +46d26m21.6s & 0.021 & -10.648 $\pm$ 0.110 & 15.7 $_{-0.5}^{+0.5}$ ($\pm$ 0.5) & 0.290 $\pm$ 0.029 & 9,b,$\gamma$ \\
NGC~2617 & 08h35m38.79s & -04d05m17.6s & 0.014 & -10.985 $\pm$ 0.160 & 4.3 $_{-1.4}^{+1.1}$ ($\pm$ 1.2) & 0.460 $\pm$ 0.046 & 9,b,$\gamma$ \\
3C~382 & 18h35m03.39s & +32d41m46.8s & 0.058 & -11.064 $\pm$ 0.100 & 40.5 $_{-3.7}^{+8.0}$ ($\pm$ 5.3) & 0.310 $\pm$ 0.031 & 9,h,$\gamma$ \\
3C~273 & 12h29m06.70s & +02d03m08.6s & 0.158 & -9.916 $\pm$ 0.050 & 146.8 $_{-12.1}^{+8.3}$ ($\pm$ 9.9) & 0.640 $\pm$ 0.064 & 10,b,$\alpha$ \\
PG~1229+204 & 12h32m03.60s & +20d09m29.2s & 0.063 & -11.281 $\pm$ 0.050 & 37.8 $_{-15.3}^{+27.6}$ ($\pm$ 20.1) & 0.530 $\pm$ 0.053 & 5,g,$\gamma$ \\
NGC~5548 & 14h17m59.53s & +25d08m12.4s & 0.017 & -10.524 $\pm$ 0.190 & 13.9 $_{-6.2}^{+11.2}$ ($\pm$ 8.2) & 0.100 $\pm$ 0.010 & 11,b,$\alpha$ \\
PKS~1510-089 & 15h12m50.53s & -09d05m59.8s & 0.361 & -11.268 $\pm$ 0.174 & 61.1 $_{-4.0}^{+3.2}$ ($\pm$ 3.6) & 0.520 $\pm$ 0.090 & 12,$\delta$ \\

%%%% HIGH SUBSAMPLE
\hline
\multicolumn{8}{c}{High \rfe\ sources, \rfe$>0.655$} \\
\hline

Mrk~142 & 10h25m31.28s & +51d40m34.9s & 0.045 & -11.085 $\pm$ 0.040 & 6.4 $_{-3.4}^{+7.3}$ ($\pm$ 4.8) &    1.140 $\pm$ 0.114 & 1,a,$\alpha$ \\
IRAS~F12397 & 12h42m10.60s & +33d17m02.6s & 0.044 & -10.417 $\pm$ 0.050 & 9.7 $_{-1.8}^{+5.5}$ ($\pm$ 2.9) &    1.480 $\pm$ 0.148 & 1,a,$\alpha$ \\
Mrk~382 & 07h55m25.30s & +39d11m10.1s & 0.034 & -11.299 $\pm$ 0.080 & 7.5 $_{-2.0}^{+2.9}$ ($\pm$ 2.4) &    0.750 $\pm$ 0.075 & 2,a,$\alpha$ \\
IRAS~04416 & 04h44m28.78s & +12d21m11.7s & 0.089 & -10.825 $\pm$ 0.030 & 13.3 $_{-1.4}^{+13.9}$ ($\pm$ 3.5) &    1.960 $\pm$ 0.196 & 2,a,$\alpha$ \\
Mrk~493 & 15h59m09.63s & +35d01m47.5s & 0.031 & -11.243 $\pm$ 0.080 & 11.6 $_{-2.6}^{+1.2}$ ($\pm$ 1.7) &    1.130 $\pm$ 0.113 & 2,a,$\alpha$ \\
Mrk~1044 & 02h30m05.52s & -08d59m53.3s & 0.017 & -10.687 $\pm$ 0.100 & 10.5 $_{-2.7}^{+3.3}$ ($\pm$ 3.0) &    0.990 $\pm$ 0.099 & 2,b,$\alpha$ \\
SDSSJ~080101 & 08h01m01.41s & +18d48m40.8s & 0.140 & -11.446 $\pm$ 0.030 & 8.3 $_{-2.7}^{+9.7}$ ($\pm$ 4.7) &    1.010 $\pm$ 0.101 & 13,b,$\alpha$ \\
SDSSJ~081456 & 08h14m56.09s & +53d25m33.6s & 0.120 & -11.581 $\pm$ 0.040 & 24.3 $_{-16.4}^{+7.7}$ ($\pm$ 10.9) &   1.310 $\pm$ 0.131 & 13,b,$\alpha$ \\
SDSSJ~093922 & 09h39m22.90s & +37d09m44.0s & 0.186 & -11.920 $\pm$ 0.040 & 11.9 $_{-6.3}^{+2.1}$ ($\pm$ 3.4) &   1.480 $\pm$ 0.148 & 13,b,$\alpha$ \\
SDSSJ~080131 & 08h01m31.58s & +35d44m36.4s & 0.179 & -11.981 $\pm$ 0.040 & 11.5 $_{-3.7}^{+7.5}$ ($\pm$ 5.1) &   1.490 $\pm$ 0.149 & 14,b,$\alpha$ \\
SDSSJ~085946 & 08h59m46.37s & +27d45m34.7s & 0.244 & -11.844 $\pm$ 0.030 & 34.8 $_{-26.3}^{+19.2}$ ($\pm$ 22.3) &   1.390 $\pm$ 0.139 & 14,b,$\alpha$ \\
SDSSJ~102339 & 10h23m39.65s & +52d33m49.7s & 0.136 & -11.604 $\pm$ 0.030 & 24.9 $_{-3.9}^{+19.8}$ ($\pm$ 7.7) &   1.030 $\pm$ 0.103 & 14,b,$\alpha$ \\
SDSSJ~074352 & 07h43m52.04s & +27d12m39.6s & 0.252 & -10.917 $\pm$ 0.020 & 43.9 $_{-4.2}^{+5.2}$ ($\pm$ 4.7) &    1.110 $\pm$ 0.111 & 3,b,$\alpha$ \\
SDSSJ~075051 & 07h50m51.72s & +24d54m09.4s & 0.400 & -11.423 $\pm$ 0.010 & 66.6 $_{-9.9}^{+18.7}$ ($\pm$ 13.3) &    1.220 $\pm$ 0.122 & 3,b,$\alpha$ \\
SDSSJ~075101 & 07h51m01.42s & +29d14m19.2s & 0.121 & -11.401 $\pm$ 0.090 & 30.4 $_{-5.8}^{+7.3}$ ($\pm$ 6.5) &    0.970 $\pm$ 0.097 & 3,b,$\alpha$ \\
SDSSJ~075949 & 07h59m49.54s & +32d00m23.9s & 0.188 & -11.800 $\pm$ 0.030 & 43.9 $_{-19.0}^{+33.1}$ ($\pm$ 24.6) &    1.020 $\pm$ 0.102 & 3,f,$\alpha$ \\
SDSSJ~083553 & 08h35m53.46s & +05d53m17.1s & 0.205 & -11.645 $\pm$ 0.020 & 12.4 $_{-5.4}^{+5.4}$ ($\pm$ 5.4) &    1.570 $\pm$ 0.157 & 3,b,$\alpha$ \\
SDSSJ~084533 & 08h45m33.30s & +47d49m34.6s & 0.302 & -11.938 $\pm$ 0.020 & 18.1 $_{-4.7}^{+6.0}$ ($\pm$ 5.3) &    1.110 $\pm$ 0.111 & 3,b,$\alpha$ \\
SDSSJ~093302 & 09h33m02.69s & +38d52m28.0s & 0.177 & -11.634 $\pm$ 0.130 & 19.0 $_{-4.3}^{+3.8}$ ($\pm$ 4.0) &    1.440 $\pm$ 0.144 & 3,b,$\alpha$ \\
SDSSJ~100402 & 10h04m02.61s & +28d55m35.4s & 0.327 & -11.027 $\pm$ 0.010 & 32.2 $_{-4.2}^{+43.5}$ ($\pm$ 10.6) &    2.170 $\pm$ 0.217 & 3,b,$\alpha$ \\
SDSSJ~101000 & 10h10m00.69s & +30d03m21.6s & 0.256 & -11.544 $\pm$ 0.020 & 27.7 $_{-7.6}^{+23.5}$ ($\pm$ 12.4) &    1.030 $\pm$ 0.103 & 3,b,$\alpha$ \\
SDSSJ~140812* & 14h08m12.10s & +53d53m03.3s & 0.116 & -12.388 $\pm$ 0.013 & 10.5 $_{-2.2}^{+1.0}$ ($\pm$ 1.4) &    0.783 $\pm$ 0.049 & 4,c,$\beta$ \\
SDSSJ~140759* & 14h07m59.07s & +53d47m59.7s & 0.172 & -12.338 $\pm$ 0.009 & 16.3 $_{-6.6}^{+13.1}$ ($\pm$ 9.0) &    0.726 $\pm$ 0.008 & 4,c,$\beta$ \\
SDSSJ~141729* & 14h17m29.27s & +53d18m26.7s & 0.237 & -12.935 $\pm$ 0.007 & 5.5 $_{-2.1}^{+5.7}$ ($\pm$ 3.3) &    0.656 $\pm$ 0.068 & 4,c,$\beta$ \\
SDSSJ~141645.15* & 14h16m45.15s & +54d25m40.7s & 0.244 & -13.042 $\pm$ 0.007 & 5.0 $_{-1.4}^{+1.5}$ ($\pm$ 1.4) &    1.048 $\pm$ 0.021 & 4,c,$\beta$ \\
SDSSJ~142135* & 14h21m35.89s & +52d31m39.0s & 0.249 & -12.800 $\pm$ 0.007 & 3.9 $_{-0.9}^{+0.9}$ ($\pm$ 0.9) &    0.672 $\pm$ 0.025 & 4,c,$\beta$ \\
SDSSJ~142038* & 14h20m38.52s & +53d24m16.5s & 0.265 & -12.878 $\pm$ 0.006 & 25.2 $_{-5.7}^{+4.7}$ ($\pm$ 5.2) &    0.675 $\pm$ 0.030 & 4,c,$\beta$ \\
SDSSJ~142043* & 14h20m43.53s & +52d36m11.5s & 0.337 & -13.177 $\pm$ 0.005 & 5.9 $_{-0.6}^{+0.4}$ ($\pm$ 0.5) &    0.952 $\pm$ 0.053 & 4,c,$\beta$ \\
SDSSJ~141318* & 14h13m18.96s & +54d32m02.4s & 0.362 & -12.709 $\pm$ 0.005 & 20.0 $_{-3.0}^{+1.1}$ ($\pm$ 1.7) &    1.031 $\pm$ 0.018 & 4,c,$\beta$ \\
SDSSJ~141324* & 14h13m24.27s & +53d05m26.9s & 0.456 & -12.942 $\pm$ 0.004 & 25.5 $_{-5.8}^{+10.9}$ ($\pm$ 7.8) &    0.669 $\pm$ 0.022 & 4,c,$\beta$ \\
SDSSJ~141123* & 14h11m23.43s & +52d13m31.7s & 0.472 & -12.794 $\pm$ 0.004 & 13.0 $_{-0.8}^{+1.4}$ ($\pm$ 1.0) &    0.754 $\pm$ 0.023 & 4,c,$\beta$ \\
SDSSJ~142010* & 14h20m10.20s & +52d40m29.5s & 0.548 & -12.990 $\pm$ 0.003 & 12.8 $_{-4.5}^{+5.7}$ ($\pm$ 5.0) &    0.847 $\pm$ 0.028 & 4,c,$\beta$ \\
SDSSJ~141115* & 14h11m15.18s & +51d52m09.0s & 0.572 & -12.810 $\pm$ 0.003 & 49.1 $_{-2.0}^{+11.1}$ ($\pm$ 4.1) &    0.716 $\pm$ 0.016 & 4,c,$\beta$ \\
SDSSJ~141112* & 14h11m12.72s & +53d45m07.2s & 0.587 & -13.027 $\pm$ 0.003 & 20.4 $_{-2.0}^{+2.5}$ ($\pm$ 2.2) &    0.726 $\pm$ 0.054 & 4,c,$\beta$ \\
SDSSJ~141417* & 14h14m17.13s & +51d57m22.6s & 0.604 & -13.784 $\pm$ 0.013 & 15.6 $_{-5.1}^{+3.2}$ ($\pm$ 4.0) &    0.680 $\pm$ 0.139 & 4,c,$\beta$ \\
SDSSJ~141135* & 14h11m35.88s & +51d50m04.6s & 0.650 & -13.218 $\pm$ 0.004 & 17.6 $_{-7.4}^{+8.6}$ ($\pm$ 8.0) &    2.064 $\pm$ 0.079 & 4,c,$\beta$ \\
SDSSJ~140904* & 14h09m04.43s & +54d03m44.2s & 0.658 & -13.124 $\pm$ 0.004 & 11.6 $_{-4.6}^{+8.6}$ ($\pm$ 6.1) &    0.747 $\pm$ 0.105 & 4,c,$\beta$ \\
SDSSJ~142052* & 14h20m52.44s & +52d56m22.3s & 0.676 & -12.241 $\pm$ 0.003 & 11.9 $_{-1.0}^{+1.3}$ ($\pm$ 1.1) &    0.691 $\pm$ 0.006 & 4,c,$\beta$ \\
SDSSJ~141532* & 14h15m32.36s & +52d49m06.0s & 0.715 & -13.223 $\pm$ 0.004 & 26.5 $_{-8.8}^{+9.9}$ ($\pm$ 9.3) &    0.883 $\pm$ 0.054 & 4,c,$\beta$ \\
SDSSJ~142023* & 14h20m23.89s & +53d16m05.1s & 0.734 & -13.165 $\pm$ 0.006 & 8.5 $_{-3.9}^{+3.2}$ ($\pm$ 3.5) &    0.687 $\pm$ 0.073 & 4,c,$\beta$ \\
SDSSJ~141859* & 14h18m59.75s & +52d18m09.7s & 0.884 & -12.679 $\pm$ 0.003 & 20.4 $_{-7.0}^{+5.6}$ ($\pm$ 6.2) &    1.393 $\pm$ 0.047 & 4,c,$\beta$ \\
SDSSJ~142417* & 14h24m17.23s & +53d02m08.8s & 0.890 & -13.504 $\pm$ 0.060 & 36.3 $_{-5.5}^{+4.5}$ ($\pm$ 5.0) &    1.080 $\pm$ 0.078 & 4,c,$\beta$ \\
Ark120 & 05h16m11.42s & -00d08m59.4s & 0.033 & -10.522 $\pm$ 0.250 & 39.5 $_{-7.8}^{+8.5}$ ($\pm$ 8.1) &    0.830 $\pm$ 0.083 & 5,b,$\gamma$ \\
NGC3516 & 11h06m47.49s & +72d34m06.9s & 0.009 & -10.446 $\pm$ 0.200 & 11.7 $_{-1.5}^{+1.0}$ ($\pm$ 1.2) &    0.660 $\pm$ 0.066 & 5,b,$\gamma$ \\
NGC4051 & 12h03m09.61s & +44d31m52.8s & 0.002 & -10.166 $\pm$ 0.150 & 2.1 $_{-0.7}^{+0.9}$ ($\pm$ 0.8) &    1.180 $\pm$ 0.118 & 5,b,$\gamma$ \\
NGC4253 & 12h18m26.51s & +29d48m46.3s & 0.013 & -11.001 $\pm$ 0.120 & 6.2 $_{-1.2}^{+1.6}$ ($\pm$ 1.4) &    0.990 $\pm$ 0.099 & 5,b,$\gamma$ \\
NGC4593 & 12h39m39.43s & -05d20m39.3s & 0.009 & -10.636 $\pm$ 0.370 & 4.0 $_{-0.7}^{+0.8}$ ($\pm$ 0.7) &    0.890 $\pm$ 0.089 & 5,b,$\gamma$ \\
NGC4748 & 12h52m12.46s & -13d24m53.0s & 0.015 & -11.120 $\pm$ 0.120 & 5.5 $_{-2.2}^{+1.6}$ ($\pm$ 1.9) &    0.990 $\pm$ 0.099 & 5,b,$\gamma$ \\
Mrk817 & 14h36m22.07s & +58d47m39.4s & 0.032 & -10.619 $\pm$ 0.090 & 19.9 $_{-6.7}^{+9.9}$ ($\pm$ 8.1) &    0.690 $\pm$ 0.069 & 5,b,$\gamma$ \\
PG1617+175 & 16h20m11.29s & +17d24m27.7s & 0.112 & -11.123 $\pm$ 0.020 & 71.5 $_{-33.7}^{+29.6}$ ($\pm$ 31.6) &    0.740 $\pm$ 0.074 & 5,b,$\gamma$ \\
PG1700+518 & 17h01m24.80s & +51d49m20.0s & 0.292 & -10.843 $\pm$ 0.010 & 251.8 $_{-38.8}^{+45.9}$ ($\pm$ 42.1) &    1.320 $\pm$ 0.132 & 5,b,$\gamma$ \\
PG2130+099 & 21h32m27.81s & +10d08m19.5s & 0.063 & -10.661 $\pm$ 0.040 & 22.6 $_{-3.6}^{+2.7}$ ($\pm$ 3.1) &    0.960 $\pm$ 0.096 & 5,b,$\gamma$ \\
PG0844+349 & 08h47m42.47s & +34d45m04.4s & 0.064 & -10.775 $\pm$ 0.070 & 32.3 $_{-13.4}^{+13.7}$ ($\pm$ 13.5) &    0.780 $\pm$ 0.078 & 6,g,$\beta$ \\
Mrk1511 & 15h31m18.07s & +07d27m27.9s & 0.034 & -11.264 $\pm$ 0.060 & 5.7 $_{-0.8}^{+0.9}$ ($\pm$ 0.8) &    0.800 $\pm$ 0.080 & 15,b,$\delta$ \\
UGC06728 & 11h45m16.02s & +79d40m53.4s & 0.007 & -11.112 $\pm$ 0.080 & 1.4 $_{-0.8}^{+0.7}$ ($\pm$ 0.7) &    1.110 $\pm$ 0.111 & 16,b,$\delta$ \\
Mrk374 & 06h59m38.11s & +54d11m47.9s & 0.043 & -10.858 $\pm$ 0.040 & 14.8 $_{-3.3}^{+5.8}$ ($\pm$ 4.3) &    0.880 $\pm$ 0.088 & 17,b,$\gamma$ \\
IZw1 & 00h53m34.94s & +12d41m36.2s & 0.061 & -10.449 $\pm$ 0.037 & 37.2 $_{-4.5}^{+4.9}$ ($\pm$ 4.7) &    1.470 $\pm$ 0.147 & 18,$\alpha$ \\
PG0923+201 & 09h25m54.72s & +19d54m05.1s & 0.192 & -10.710 $\pm$ 0.061 & 108.2 $_{-6.6}^{+12.3}$ ($\pm$ 8.8) &    0.953 $\pm$ 0.005 & 19,$\alpha$ \\
PG1001+291 & 10h04m02.61s & +28d55m35.4s & 0.327 & -10.956 $\pm$ 0.026 & 37.3 $_{-6.9}^{+6.0}$ ($\pm$ 6.4) &    1.319 $\pm$ 0.009 & 19,$\alpha$ \\
\hline
\end{longtable}

\end{appendix}

% If you want to present additional material which would interrupt the flow of the main paper,
% it can be placed in an Appendix which appears after the list of references.

%%%%%%%%%%%%%%%%%%%%%%%%%%%%%%%%%%%%%%%%%%%%%%%%%%

% Don't change these lines
\bsp	% typesetting comment
\label{lastpage}
\end{document}